\documentclass[acmsmall]{acmart}

\acmJournal{PACMPL}
\acmVolume{1}
\acmNumber{OOPSLA} 
\acmArticle{1}
\acmYear{2022}
\acmMonth{1}
\acmDOI{} 
\startPage{1}

\setcopyright{none}

\usepackage{booktabs}   
\usepackage{subcaption} 

\usepackage{relsize}
\usepackage{mathpartir}
\usepackage{amsmath}
\usepackage{amsthm}
\usepackage{listings}
\usepackage{xspace}
\usepackage{definitions}
\usepackage{multirow,bigdelim}
\usepackage{pbox}
\usepackage{courier}

\newcommand{\kjx}[1]{{{#1}}}
\newcommand{\scd}[1]{{{#1}}}

\newcommand{\sd}[1]{{#1}}

\definecolor{amber}{rgb}{1.0, 0.75, 0.0}
\definecolor{amethyst}{rgb}{0.6, 0.4, 0.8}

\newcommand{\TODO}[1]{} 
\newcommand{\sophia}[1]{{\color{blue}#1}}
\newcommand{\toby}[1]{} 

\newcommand{\jm}[2][]{\ponders{Julian}{magenta}{#1} \textcolor{magenta}{#2}\xspace}
\renewcommand{\jm}[2][]{{#2}}

\newcommand{\sdr}[2][]{\ponders{SD:}{blue}{#1} \textcolor{blue}{#2}\xspace}
\renewcommand{\sdr}[2][]{#2}

\newcommand{\sdNr}[2][]{{#2}\xspace}
\newcommand{\sdN}[1]{{#1}\xspace}
\newcommand{\sdS}[1]{{#1}\xspace}

\newcommand{\julian}[1]{{#1}\xspace}
\newcommand{\sdM}[1]{#1\xspace}

\newcommand{\sophiaPonder}[2][]{\ponders{Sophia}{blue}{#1} \textcolor{blue}{#2}\xspace}
\renewcommand{\sophiaPonder}[2][]{#1}
\renewcommand{\sophia}[2][]

\begin{document}

\title{\Nec Specifications \sd{for Robustness}}



\author{Julian Mackay}
\orcid{0000-0003-3098-3901}             
\affiliation{
  \institution{Victoria University of Wellington}           
  \country{New Zealand}                   
}
\email{julian.mackay@ecs.vuw.ac.nz}         

\author{Sophia Drossopoulou}
\orcid{0000-0002-1993-1142}             
\affiliation{
  \institution{Meta, and Imperial College London}            
  \country{United Kingdom}                    
}
\email{scd@imperial.ac.uk}          

\author{James Noble}
\orcid{0000-0001-9036-5692}             
\affiliation{
  \institution{Creative Resaerch \& Programming}           
  \streetaddress{5 Fernlea Ave, Darkest Karori}
  \city{Wellington}
  \postcode{6012}
  \country{New Zealand}                   
}
\email{kjx@acm.org}         

\author{Susan Eisenbach}
\orcid{0000-0001-9072-6689}             
\affiliation{
  \institution{Imperial College London}           
  \country{United Kingdom}                   
}
\email{susan@imperial.ac.uk}         

\begin{abstract}
Robust modules guarantee to do \emph{only} what they are supposed to
do -- even in the presence of untrusted, malicious clients,
and considering not just the direct behaviour of individual methods,
but also the emergent behaviour from calls to more than one method. 
\Nec is a language for specifying robustness, based on novel necessity
operators capturing temporal implication, and a 
proof logic that derives explicit robustness specifications from
\funcSpecs. Soundness
and an exemplar proof
are mechanised in Coq.
\end{abstract}



\maketitle

\section{Introduction: Necessary Conditions and Robustness}
\label{s:intro}




{Software needs} to be both {\emph{correct}} ({programs do what they
  are supposed to}) and {\emph{robust}} ({programs \emph{only} do what they are supposed to}). 
 {We use the term \emph{robust} as a generalisation of \emph{robust safety}~\cite{gordonJefferyRobustSafety, Bugliesi:resource-aware,ddd}  whereby a module or process or ADT is \emph{robustly safe} if its execution preserves some safety guarantees even when run together with unknown,   unverified, potentially malicious client code. 
 The particular safety guarantees vary across the literature. 
We are interested in \emph{program-specific} safety guarantees which describe  \emph{necessary conditions}
 for some effect to take place.
In this work we propose how to specify such necessary conditions, and how to prove that modules adhere to such specifications.}   
 
 We motivate the need for  necessary conditions for effects through an example:
{Correctness is} {traditionally} specified
through \citeasnoun{Hoare69} triples: a  precondition, a code snippet, and a
 postcondition. 
 For example,  {part of the \funcSpec
   of a \prg{transfer} method for a bank module is that the source account's balance decreases:}
 \begin{quote}
   \Scorrect\ \ $\triangleq$  
 {\footnotesize{ $\{\,$\prg{pwd=src.pwd} $\,\wedge\,$ \prg{src.bal=b}$\,\}$ \prg{src.transfer(dst,pwd)} $\{$ \prg{src.bal=b-100}$\,\wedge\,\dots \}$ }} Calling \prg{transfer} on  {an account with the correct password} will transfer the money.
\end{quote}
Assuming termination, the precondition is a \emph{sufficient} condition for the {code snippet to behave correctly}: 
the precondition (\eg providing the right 
password) guarantees that
the code (\eg call the \prg{transfer} function)
will always achieve the postcondition (the money is transferred).
 
    \vspace{.05in}
 
\Scorrect  describes  {the \emph{correct use} of the  {particular function}, but is \emph{not} concerned with the {module's} \emph{robustness}.}
{For example, can I pass an account to foreign untrusted code, in the expectation of receiving a payment,
but without fear that a malicious client might use the account to steal my money \cite{ELang}?}
 A first  {attempt} to specify robustness could be:

\begin{quote}
\SrobustA\ \ $\triangleq$ \ \ An account's balance does not decrease unless \prg{transfer} was called 
with the correct password.
\end{quote}

Specification \SrobustA 
{guarantees} that it is not possible to  take money out of the account 
{without calling \prg{transfer} and} without providing the password.
  Calling \prg{transfer}   with the  correct password is 
a \emph{necessary condition} for {(the effect of)} reducing the account's  balance.

\SrobustA is  crucial, but  not   enough:
it does not take  account of the module's \emph{emergent behaviour},
{that is, does not cater for the potential interplay of several methods offered by the module.}
 What if the module provided further methods which leaked the password?  
{ While no single procedure call is capable of breaking the intent of \SrobustA, a sequence of calls might.}
{What} we really need is
 \begin{quote}
\SrobustB\ \ $\triangleq$ \ \ The balance of an account does not {\emph{ever}} decrease in the future unless some external 
object  {\emph{now}} has access to the account's current password.
\end{quote}
With \SrobustB, I can confidently pass my account to  {any, potentially untrusted context, where my password is not known;} 
 {the payment I was expecting may or may not be made,}
but I know 
 {that my money will not be stolen} \cite{ooToSecurity}. 
 Note that \SrobustB  does not mention
 the names of any functions in the module, and 
 thus can be expressed without reference to any particular API ---
 indeed \SrobustB can constrain \emph{any} API with an account, an account
 balance, and a password.

{Earlier work addressing robustness} includes object capabilities  \cite{MillerPhD, dd, threoremsFreeSep}, 
information control flow \cite{Zdancewic:Myers:01,noninteferenceOS}, 
 correspondence assertions \cite{Maffeis:aiamb:thesis00},
 sandboxing  \cite{robustSafetyPatrignani,sandbox},
robust linear temporal logic   \cite{RLTL2022} -- to name a few. 
{Most of these  
propose \emph{generic}  guarantees (\emph{e.g.} no dependencies from high values to low values),
\sdN{or preservation of module invariants},
while we work with  \emph{problem-specific}  guarantees \sdN{concerned with necessary conditions 
for specific effects}  (\emph{e.g.} no decrease in balance without access to password).}
{{{\sc{VerX}} \cite{VerX} and  \emph{Chainmail} \cite{FASE} also work on problem-specific guarantees.}
Both these approaches are able to express necessary conditions
  like \SrobustA using
  temporal logic operators and implication,
  and Chainmail is able to express \SrobustB,
  however neither have a proof logic
  \jm[removed:able]{} to prove adherence to such specifications.}

\subsection{\Nec}
\label{intro:this:work}
{In this paper we introduce \Nec,} the first approach that is able to  both express and prove
(through an inference system)
robustness specifications such as  \SrobustB.
{Developing a specification language with a proof logic that is able to prove properties such as \SrobustB 
and must tread a fine line: the language must be rich enough to express complex specifications; temporal operators are needed along with object capability style operators 
that describe \emph{permission} and \emph{provenance}, while also being simple enough that proof rules might be devised.}

\vspace{.07in}
 {
The {first main} contribution  
{is} three novel operators that merge
temporal operators and implication 
and most importantly are both expressive enough to capture the 
examples we have found in the literature and provable through an inference system.
}
%
%
%
 One such necessity operator is \\
$ 
\strut \hspace{1.7in} \onlyIf {A_{curr}} {A_{fut}} {A_{nec}}
$  
\\
This form says that  
a  {transition} from a current state satisfying assertion $A_{curr}$ to a future
state satisfying $A_{fut}$  is possible only if  the   necessary 
condition
$A_{nec}$ holds in the \emph{current} state.
Using this operator, 
we can formulate  \SrobustB  
as
\begin{lstlisting}[language = Chainmail, mathescape=true, frame=lines]
   $\text{\SrobustB}$  $\triangleq$   from   a:Account $\wedge$ a.balance==bal    to   a.balance < bal
               onlyIf  $\exists$ o.[$\external{\texttt{o}}$ $\wedge$ $\access{\texttt{o}}{\texttt{a.pwd}}$]
\end{lstlisting}
Namely, a transition from a  {current} state where an account's balance is \prg{bal}, to a  {future} state where 
it has decreased, may \emph{only} occur if  {in the current state} some {\color{blue}{\prg{external}}}, unknown client object  
has access to that account's password.
More    in \S\ref{s:bankSpecEx}.

Unlike  \emph{Chainmail}'s temporal operators, 
 the necessity operators 
 are  not first class, and may not appear in the assertions  {(\eg  ${A_{curr}}$)}. 
 This simplification enabled us to develop our proof logic. 
 Thus, we {have reached} a  sweet spot between expressiveness and 
 provability.

 \vspace{.07in}
 The second main contribution is  a logic that  enables us to prove that code 
 obeys \Nec specifications. {}
{Our insight was  that \Nec specifications are logically equivalent to the
intersection of an \emph{infinite} number of Hoare triples, \emph{i.e.,} 
$\onlyIf{A_1} {A_2} {A_3}$ is 
logically equivalent
{to}
 $\forall \prg{stmts}. \{ A_1 \wedge \neg A_3\} \prg{stmts} \{ \neg A_2 \}$.
\sdS{Note that in the above, the assertions $A_1$, $A_2$ and $A_3$ are fixed, while the code \prg(stmts) is universally quantified.
  This leaves the challenge that, usually, Hoare logics do not support such infinite quantification over the code.}}

{We addressed that  challenge} through three further insights: 
 {(1) \Nec specifications of emergent behaviour can be built up from \Nec specifications of
 single-step executions, which (2) can be built from encapsulation and \emph{finite} intersections
 of \Nec specifications of function calls, which  
 (3) in turn can be obtained from \emph{traditional} \funcSpecs.}
 %

\subsection{Contributions and Paper Organization}


%
The contributions of this {work} are:
\begin{enumerate}
\item   A language to express \Nec specifications (\S\ref{s:semantics}), including three novel \Nec operators (\S\ref{s:holistic-guarantees}) that combine implication and temporal operators.  

 \item
A logic for proving a module's adherence to its
 \Nec specifications (\S\ref{s:inference}), and a proof of soundness of the logic, (\S\ref{s:soundness}),
both mechanised in Coq \jm[]{\cite{necessityCoq2022}}. 
 \item
A proof in our logic 
  that our bank module {obeys} 
  \SrobustB (\S\ref{s:examples}),     mechanised in Coq.
  And a proof that
  a richer bank module which uses ghostfields and confined classes  obeys  \SrobustB (\S\ref{app:BankAccount}),
  also mechanised in Coq.
\item \ {Examples taken from the literature  (\S\ref{s:expressiveness} and \S\ref{s:expressiveness:appendix}) specified in \Nec  .}

\end{enumerate}


 We place \Nec into the context of 
related work (\S\ref{s:related}) and consider our overall conclusions
(\S\ref{s:conclusion}). 
The Coq proofs of 
(2) and (3) above appear in the
supplementary material, along with {a}ppendices containing expanded 
definitions and further examples.
In the next section, (\S\ref{s:outline}),  we outline our approach using a
  bank   as  a motivating example.

  {A strength of our} work 
is {that it is
parametric} with respect to assertion
satisfaction \sdS{and functional specifications} -- these questions are well covered in the literature,
{and offer several off-the-shelf solutions.}
\sdS{The current work is based on 
  a simple, imperative, typed, object oriented
language with unforgeable addresses and private fields; nevertheless,
 we believe
 that   our approach
is applicable to several programming paradigms, and 
 that   unforgeability and privacy
 can be replaced 
 by lower level mechanisms such as capability machines \cite{vanproving,davis2019cheriabi}.}
\sdS{In line  with other work in the literature, we do not --yet-- support
``callbacks'' out from internal objects (whose code gas been checked) to  external objects (ie unknown objects
whose code has not been checked).}

\section{Outline of our approach}
\label{s:outline}
 
In this Section we outline our approach:  
we  revisit  our running example, the Bank Account (\S\ref{s:bank}),  
introduce the three necessity operators (\S \ref{s:approach:necopers}),  give 
the \Nec specs (\S \ref{s:bankSpecEx}),
   outline how we model the open world (\S\ref{s:concepts}), 
give the main ideas of our proof system (\S\ref{s:approach})
and   outline 
how we use it to reason 
about adherence to \Nec specifications  (\S\ref{s:all:outline:proof}).

 \subsection{Bank Account -- three modules}
\label{s:bank}
  
Module \ModA consists of an empty 
\prg{Password} class where each instance models a unique password, and an  \prg{Account} class with a password, and a balance, an \prg{init} method to 
initialize the password, and 
a
\prg{transfer} method. 
\sdS{Note that we assume that all  fields are  ``class-private'', 
i.e., methods may read and write fields of any instance of the same class,
and  that passwords are unforgeable and not enumerable (as
in Java, albeit without reflection}.
\begin{lstlisting}[mathescape=true, language=Chainmail, frame=lines]
module $\ModA$
  class Account
    field balance:int 
    field pwd: Password
    method transfer(dest:Account, pwd':Password) -> void
      if this.pwd==pwd'
        this.balance-=100
        dest.balance+=100
     method init(pwd':Password) -> void
      if this.pwd==null
        this.pwd=pwd'
  class Password
\end{lstlisting}
\noindent 
We can capture the intended semantics of   \prg{transfer}  
through  {a}  \funcSpec with pre- and post- conditions and \prg{MODIFIES} clauses as \emph{e.g.,} in \citeauthor{Leavens-etal07,dafny13}.
The implementation of  \prg{transfer} in module  $\ModA$ meets
this specification.

\begin{lstlisting}[mathescape=true, frame=lines, language=Chainmail]
$\Sclassic$  $\triangleq$
   method transfer(dest:Account, pwd':Password) -> void  
      ENSURES:
            this.pwd$=$pwd' $\wedge$ this$\neq$dest  $\longrightarrow$  
            this.balance$_{post} =$this.balance$_{pre}$-100 $\wedge$ dest.balance$_{post} =$dest.balance$_{pre}$+100
      ENSURES:
            this.pwd$\neq$pwd' $\vee$ this$=$dest  $\longrightarrow$ 
            this.balance$_{post} =$this.balance$_{pre}$ $\wedge$ dest.balance$_{post} =$dest.balance$_{pre}$ 
      MODIFIES:  this.balance, dest.balance        
\end{lstlisting}

Now consider the following alternative implementations:
\ModB allows any client to reset an account's password at any time;
\ModC requires the existing password in order to change it.

\begin{tabular}{lll}
\begin{minipage}[b]{0.42\textwidth}
\begin{lstlisting}[mathescape=true, language=chainmail, frame=lines]
module $\ModB$
  class Account
    field balance:int 
    field pwd: Password 
    method transfer(..) ...
      ... as earlier ...
    method init(...) ...
       ... as earlier ...
    method set(pwd': Password)
      this.pwd=pwd'
      
  class Password
\end{lstlisting}
\end{minipage}
&\ \ \  \ \   &%
\begin{minipage}[b]{0.45\textwidth}
\begin{lstlisting}[mathescape=true, language=chainmail, frame=lines]
module $\ModC$
  class Account
    field balance:int 
    field pwd: Password 
    method transfer(..) 
      ... as earlier ...
    
    
    method set(pwd',pwd'': Password)
      if (this.pwd==pwd') 
        this.pwd=pwd''
  class Password
\end{lstlisting}
\end{minipage} 
\end{tabular}

Although the \prg{transfer} method is the same in
all three alternatives, and each one satisfies \Sclassic,
code  {such as}
\\ 
$\ \strut \hspace{.2in} $ \prg{an\_account.set(42); an\_account.transfer(rogue\_account,42)}
\\ 
is enough to drain  \prg{an\_account} in \ModB without knowing the password.

\sdS{This example also demonstrates the importance of field privacy: $\ModA$ and $\ModC$ would not be any
more robust than $\ModB$ if the underlying programming language did not restrict access to fields. Without 
such a restriction,
any external object would have been able to directly manipulate the fields \prg{balance} and \prg{pwd}}.

\subsection{The three necessity operators}
\label{s:approach:necopers}

{We need  a specification that rules out \ModB while permitting \ModA and
\ModC.  For this, we will be using  one of the  three necessity operators mentioned in  \S \ref{intro:this:work}. These operators are:}
\\
$ \strut \hspace{1.7in} \onlyIf {A_{curr}} {A_{fut}} {A_{nec}} $ \\
$ \strut \hspace{1.7in}   \onlyIfSingle {A_{curr}} {A_{fut}} {A_{nec}} $ \\
$ \strut \hspace{1.7in}   \onlyThrough {A_{curr}} {A_{fut}} {A_{intrm}} $
\\
The first operator was already introduced in \S \ref{intro:this:work}: it says that  
a  {transition} from a current state satisfying assertion $A_{curr}$ to a future
state satisfying $A_{fut}$  is possible only if  the   necessary 
condition
$A_{nec}$ holds in the \emph{current} state.
{The  second operator says    that 
a  \emph{one-step} {transition} from a current state satisfying assertion $A_{curr}$ to a future
state satisfying $A_{fut}$  
is possible only if 
$A_{nec}$ holds in the \emph{current} state.   
The   third operator   says that a change from 
 $A_{curr}$ to  $A_{fut}$  may happen only if 
 $A_{intrm}$ holds in some \emph{intermediate} state.}

%

{
Our assertions $A$, also allow for the use of capability operators, such as 
1) having \prg{access} to an object ($\access{\prg{o}}{\prg{o'}}$) which means that $\prg{o}$ has a reference to $\prg{o'}$,
 or 2) \prg{calling} a method with on receiver with certain arguments, 
  ($\calls{\prg{o}}{\prg{o'}}{\prg{m}}{\prg{args}} $), or 3) an object being \prg{external},
  where  $\external{\prg{o}}$ means that \prg{o} belongs to a class that is not defined in the current module,
  and thus its behaviour is unrestricted.}
 These are the capability operators that we have 
adopted from Chainmail.

 \subsection{Bank Account -- the right specification}
\label{s:bankSpecEx}

We now {return to our quest for} a specification that rules out \ModB while permitting \ModA and
\ModC. The catch is that the vulnerability present in \ModB is the result
of  \emph{emergent} behaviour from the interactions of the \prg{set}
and \prg{transfer} methods --- even though \ModC also has a
\prg{set} method, it does not exhibit the unwanted interaction.
This is exactly where a necessary condition can help:
we want to avoid transferring money
(or more generally, reducing an account's balance)
\textit{without} the existing account password.  Phrasing the same condition
the other way around 
rules out the theft: that money \textit{can only} be
transferred when the account's password is known.

In \Nec  syntax, and {recalling \S \ref{intro:this:work}, and \ref{s:approach:necopers},}

%
%
%
%
%
%
\begin{lstlisting}[language = Chainmail, mathescape=true, frame=lines]
   $\text{\SrobustA}$  $\triangleq$   from   a:Account $\wedge$ a.balance==bal    next    a.balance < bal
                onlyIf $\exists$ o,a'. [$\external{\prg{o}}$ $\wedge$ $\calls{\prg{o}}{\prg{a}}{\prg{transfer}}{\prg{a'},\prg{a.pwd}} $]    
                
   $\text{\SrobustB}$  $\triangleq$   from   a:Account $\wedge$ a.balance==bal    to    a.balance < bal
                onlyIf $\exists$ o.[$\external{\prg{o}}$ $\wedge$ $\access{\prg{o}}{\prg{a.pwd}}$]    
           
\end{lstlisting}
%
%
%
 %

{\SrobustA does not fit the bill: all three modules   satisfy  it.
 But  \SrobustB does fit the bill: \ModA and \ModC satisfy \SrobustB, while \ModB does not.}
 
A critical point of \SrobustB 
is that it is
expressed in terms of observable effects (the account's balance is
reduced: \prg{a.balance < bal}) and the shape of the heap 
(external access to the password:
$\external{\prg{o}}\ \wedge\ \access{\prg{o}}{\prg{a.pwd}}$) 
rather than in terms of individual methods such as
\prg{set} and \prg{transfer}.
This gives our specifications the
vital advantage that they can be used to constrain
\jm[typo]{\textit{implementations}} of a bank account with a balance and a
password, irrespective of the API it
offers, the services it exports, or the dependencies on other parts of
the system.

 This example also demonstrates that 
adherence to   \Nec specifications is not monotonic:
adding a method to a module does not necessarily preserve adherence to
a specification, 
and while separate methods may adhere to a  specification, their combination does
not necessarily do so. 
{For example, \ModA satisfies \SrobustB, while \ModB does not.}
This is why we say that \Nec   specifications capture a module's \emph{emergent behaviour}.

\subsubsection{How  useful is  \SrobustB?}
\label{sec:how}

{One might think that \SrobustB was not useful: normally, there will exist somewhere in the heap
at least one external object  
with access to the password --  if no such object existed, then \sdN{nobody} would be able to use the money of
the account.
And if such an object did exist, \sdN{then the premise of \SrobustB would not hold, and thus}
the guarantee given by \SrobustB might seem vacuous.}

{
This is \emph{not} so: 
\sdN{in scopes   from which such external objects with access to the password
are not (transitively) reachable, 
\SrobustB  guarantees that the balance of the account will not decrease.
}
We illustrate this through the following  code snippet:

\begin{lstlisting}[mathescape=true, language=chainmail, frame=lines]
module $\ModParam{1}$
     ...
    method cautious(untrusted:Object)
        a = new Account
        p = new Password
        a.set(null,p)
        ...
        untrusted.make_payment(a)
        ...
\end{lstlisting}

{The method \prg{cautious} has as  argument an external object \prg{untrusted}, of unknown provenance.
{It} creates a new \prg{Account} and initializes its password. 
In the scope of  this method,  external objects with access to the password are reachable:
thus,  during execution of  line 7, or   line {9} the balance may decrease.
}

{Assume that class \prg{Account} is from a module which satisfies \SrobustB. 
Assume also that the code in line 7 does not leak the password to \prg{untrusted}. Then no external object
reachable from the scope of execution of \prg{make\_payment} at line 8 has access to the password.
Therefore, 
even though we are calling   an untrusted object, \SrobustB guarantees that \prg{untrusted}
 will not be able to take any money out of  \prg{a}.
 }
 
\jm[]{
A  proof sketch of the  safety provided by \SrobustB appears in Appendix \ref{app:safety}.
}
\sdN{Note that in this example, we have (at least) three modules: the internal module 
which defines class \prg{Account} adhering to  \SrobustB, the  
external module $\ModParam{1}$, and the external module which contains the class definition for \prg{untrusted}.
Our methodology allows the external module, $\ModParam{1}$ to reason about its own code, and thus 
pass \prg{a} to code from the second external module, without fear of losing money.}
\sdN{In further work we want to make such arguments more generally applicable, and 
extend Hoare logics to encompass such proof steps.}

\subsection{Internal and external modules, objects, and calls}
\label{s:concepts}

Our work concentrates on guarantees made in an \emph{open} setting; that is, a given module
$M$ must be programmed so that 
execution of $M$ together with \emph{any} \externalM 
module $M'$ will uphold these guarantees. In the tradition of
visible states semantics, we are  only interested in upholding the guarantees while 
$M'$, the  \emph{\externalM} module, is executing. A module can
temporarily break its own invariants,
so long as the broken invariants are never visible externally.
   
We therefore distinguish between  \emph{\internalO}
objects --- instances of classes defined in $M$ ---
and \emph{\externalO} objects defined in any other module.
We also distinguish between
  \emph{\internalC} calls  (from either an internal or an external object)  made 
 to \internalO objects and \emph{\externalC} calls made 
 to \externalC objects. 
{Looking at the code snippet from \S \ref{sec:how}, the call to \prg{set} on line 6 is an 
 internal call, while the call to \prg{make\_payment} is an external call -- from the external 
 object  \prg{this}  to the external object \prg{untrusted}.}
 
%
%
Because we only require guarantees while 
the  \externalM module  is executing,
we develop an \emph{external states} semantics, where
 any internal calls are executed in one, large, step.
With external steps semantics,  the executing object (\prg{this}) is always   external. 
  In line  with other work in the literature \cite{Permenev, Grossman, Albert}, we currently forbid 
  calls from internal to  external objects
  -- further details on call-backs in \S\ref{s:related}. 

{For the purposes of the current work we are only interested in one internal, and one external module.
But the interested reader might ask: what if there is more than one external module?
The answer is that from the internal module's viewpoint, all external modules are considered as one;
for this we provide a module linking operator with  the expected semantics -- more details in Def. \ref{def:pair-reduce} and \S \ref{app:loo}. 
But from the external module's viewpoint, there may be more than one external module: for example, in \S \ref{sec:how}, 
module $\ModParam{1}$ is external to the module   implementing class \prg{Account}, and the module 
implementing the class of \prg{untrusted} is external to
$\ModParam{1}$.
}

%

\newcommand{\vertsp} {\vspace{.05in}} 
 
\subsection{Reasoning about \Nec}
\label{s:approach}

{We will now outline the key ingredients of  our logic with which we prove
that modules obey \Nec specifications. 
We will use the auxiliary concept that  
an assertion $A$ is \emph{encapsulated} by
a module $M$, if  $A$  can only be invalidated through 
a call to a method from $M$ -- more   in \S \ref{s:encaps-proof}.}

%

The \Nec logic is based on the {insight} that the specification\\
\strut $\hspace{1in}$  $\onlyIf{A_1} {A_2} {A_3}$
\\
 is  logically equivalent {to} \\
 \strut $\hspace{1in}$  $\forall \prg{stmts}. \{ A_1 \wedge \neg A_3\} \prg{stmts} \{ \neg A_2 \}$\\
 -- that is,
 with an \emph{infinite} conjunction  of Hoare triples, \sdS{where the three assertions are fixed, but the code,
 \prg{stmts}, is universally quantified.
This leaves the challenge that \sdr[changed "no Hoare logics support"]{usually, Hoare logics do
 not support} such infinite conjunctions over code.}
 Three \sdr[dropped breakthroughs]{ideas} helped us address   that challenge: 

 \begin{description}
 \item
 [From Hoare triples to per-call specs] 
  The Hoare triple 
$ \{ A_1 \wedge \neg A_3 \} \ \prg{x.m(ys)}\  \{ \neg A_2 \}$ is logically equivalent 
 {to} the specification
$ \onlyIfSingle {(A_1 \wedge  \calls{\_}{\prg{x}}{\prg{m}}{\prg{ys}} )} {A_2} {A_3}$.  
 
 \item 
 [From per-call specs to per-step specs] 
 If an assertion $A_2$  is \emph{encapsulated} by a module -- and thus the only way from a 
 state that satisfies $A_2$ to a state that does not, is through a call to a method in that module -- then
{the
\emph{finite conjunction}
that all methods of that module {$ \onlyIfSingle {(\, A_1 \wedge A_2 \wedge {\calls{\_}{\prg{x}}{\prg{m}}{\prg{ys}}}\, )} {\neg  A_2} {A_3}$}
  is logically equivalent 
 {to}
{$ \onlyIfSingle {A_1 \wedge A_2} {\neg A_2} {A_3}$. }}
 
  \item [Proof logic  for emergent behaviour] 
  combines several specifications to reason about the
  emergent behaviour, \emph{e.g.,} 
   $ \onlyThrough  {A_1} {A_2} {A_3}$  and $ \onlyIf  {A_1} {A_3} {A_4}$ implies 
   $ \onlyIf  {A_1} {A_2} {A_4}$.
 \end{description}



  Thus, our system consists of four parts (five including \funcSpecs):
(\textbf{Part 1}) assertion encapsulation, (\textbf{Part 2}) {per-method} specifications, 
(\textbf{Part 3}) per-step specifications, and (\textbf{Part 4}) specifications of emergent behaviour.
The structure of the system, and the dependency of each part on preceding parts is given in Fig. \ref{fig:dependency}.
 \FuncSpecs are used to prove \jm[]{per-method} specifications, which coupled with 
assertion encapsulation is used to prove per-step specifications, which is used to 
prove specifications of emergent behaviour.
\begin{figure}[t]
\includegraphics[width=\textwidth]{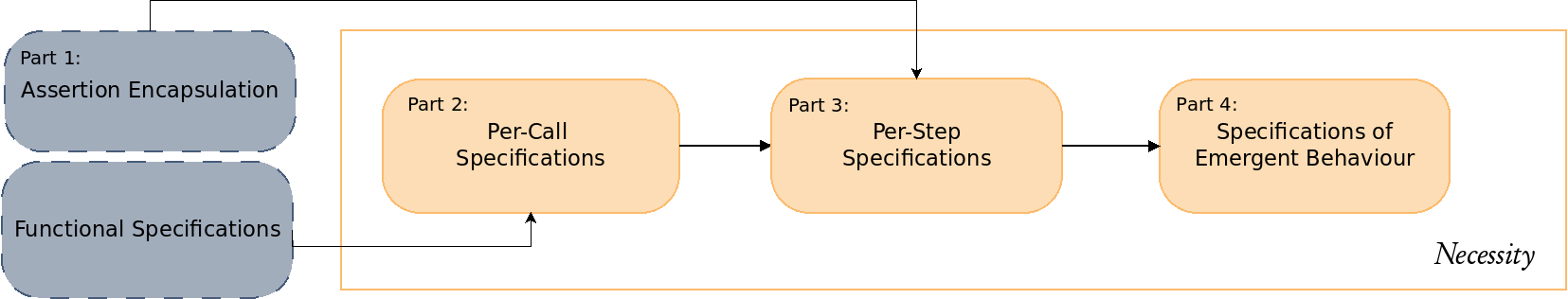}
   \caption{
   Parts of \Nec Logic {and their Dependencies}. 
   Note that gray parts with a dashed border indicate
   parts that are not part of \Nec, and on which \Nec 
   is parametric.}

   \label{fig:dependency}
 \end{figure}

Our \Nec logic  is parametric with respect to  {the way we ascertain whether an assertion
is encapsulated and the way we obtain \funcSpecs.}
As a result we can leverage  results from many different approaches.
Further, our proofs of \Nec do not inspect method
bodies: we rely on simple annotations to infer encapsulation, and
{on} pre and post-conditions  to infer per-method conditions.

 \subsection{Outline of the proof that  \ModC obeys \SrobustB}
 \label{s:all:outline:proof}
 
For illustration, we  outline  a proof that  \ModC adheres to \SrobustB. 
note that for illustration purposes, in this paper we show how assertion encapsulation can be proven
based on simple annotations inspired by confinement types
\cite{confined}; we could just as easily rely  on other language mechanisms,  \emph{e.g.,} ownership types,
{or even develop custom  logics.}
 
\begin{description}
\item[Part 1: Assertion Encapsulation.]  \
\vertsp\\
{\fbox{\parbox{\linewidth}{
We begin by proving that \ModC encapsulates: 
\begin{description}
\item[(A)] The balance
\item[(B)] The password
\item[(C)] External accessibility to an account's password -- 
that is, the property that no external object has access to the password may only 
be invalidated by calls to \ModC.
\end{description}
}}}
\vertsp
\item[Part 2: Per-Method Specifications]   \
\vertsp\\
{\fbox{\parbox{\linewidth}{
{We prove} that the call of any  method {from} \ModC (\prg{set} and \prg{transfer}) 
satisfies: 
\begin{description}
\item[(D)] If the balance {decreases}, then 
 \prg{transfer} was called with the correct password
\item[(E)] If the password changes, then the method called was
 \prg{set} with the correct password
\item[(F)] It will not provide external accessibility to the password.
\end{description}
}}}
\vertsp
\item[Part 3: Per-step Specifications]  \
\vertsp \\ 
{\fbox{\parbox{\linewidth}{
We then raise our results of Parts 1 and 2 to reason about arbitrary \emph{single-step} executions:
\begin{description}
\item[(F)]
By \textbf{(A)} and \textbf{(D)} only \prg{transfer} and external access to the password may decrease the balance.
\item[(G)]
By \textbf{(B)} and \textbf{(E)} only \prg{set} and external access to the password may change the password.
\item[(H)]
By \textbf{(C)} and \textbf{(F)} no step may grant external accessibility to an account's password.
\end{description}
}}}
%
%
\vertsp
\item[Part 4: Specifications of Emergent Behaviour] \
\vertsp \\
{\fbox{\parbox{\linewidth}{
We then raise our necessary conditions of Part 3 to reason about \emph{arbitrary} executions:
\begin{description}
\item [(I)]
A decrease in balance over any number of steps implies that some single intermediate step 
reduced the account's balance.
\item [(J)] 
By \textbf{(F)} we know that step must be a call to \prg{transfer} with the correct password.
\item [(K)]
When \prg{transfer} was called, either
\begin{description}
\item[(K1)]
The password used was the current password, and thus by \textbf{(H)} we know that 
the current password must be externally known, satisfying \SrobustB, or
\item[(K2)] 
The password had been changed, and thus by \textbf{(G)} some intermediate step must have been
a call to \prg{set} with the current password. Thus, by \textbf{(H)} we know that
the current password must be externally known, satisfying \SrobustB.
\end{description}
\end{description}
}}}
\end{description}

\noindent

\section{The Meaning of Necessity}
\label{s:semantics}

In this section we define {the}  \Nec specification language.  
We first define an underlying programming language, \Loo (\S \ref{sub:Loo}).
We then define an assertion language, \SpecO, which can talk about the
contents of the state, as well as about provenance, permission and
control (\S \ref{sub:SpecO}).  Finally, we define the syntax and
semantics of our full language for writing \Nec
specifications (\S \ref{s:holistic-guarantees}).

\subsection{\Loo}
\label{sub:Loo} 
 \Loo is a \jm[removed: formal model of a]{} {small}, imperative, sequential, 
class based, typed, object-oriented language, whose
\sdS{fields are private to the class where they are defined.}
\Loo is straightforward:
Appendix \ref{app:loo} contains 
the full definitions.
\Loo is based on \LangOO 
\cite{FASE}, with some small variations, as well as 
the addition of  
 a simple type system -- more in \ref{types}.
%
%
A \Loo state $\sigma$ consists of a 
heap $\chi$, and a  {stack $\psi$ which is a sequence of frames}.
A frame $\phi$ consists of
local variable map, and a continuation, \ie a sequence of statements to be executed.
 A statement may assign to variables, create new objects and push them to the heap, 
perform field reads and writes on objects,  or
 call methods on those objects. 

 Modules are mappings
from class names to class definitions. 
Execution 
is in the context of  a module $M$ and   a state $\sigma$,
 defined via unsurprising small-step semantics of the form \ \ 
   $M, \sigma \leadsto \sigma'$.
The   top frame's continuation contains the statement to be 
executed next.
 
As discussed in \S \ref{s:approach}, 
{open world specifications need to be able to provide}
guarantees which hold
during execution of an internal, 
known, trusted module $M$ when linked together with any
unknown, untrusted, module $M'$. These guarantees need only hold 
when the external module is executing; we are not concerned if they are
temporarily broken by the internal module. Therefore, we are only interested in states where the
executing object (\prg{this}) is an external object. 
To express our focus on external states, we define the  \emph{external states semantics}, of the form 
$\reduction{M'}{M}{\sigma}{\sigma'}$, where $M'$ is the external
module, and $M$ is the internal module, and where we
collapse all internal steps into one single step.

\begin{definition}[External States Semantics]
\label{def:pair-reduce}
For  
  modules $M$,  $M'$, and 
   states $\sigma$, $\sigma'$, 
we say that $\ \ \ \ \ \ \ \ \reduction{M'}{M}{\sigma}{\sigma'}\ \ \ \ \ \ \ \ $ if and only if there exist 
$n\in\mathbb{N}$, and states $\sigma_0$,...$\sigma_n$, such that
\begin{itemize}
\item
$\sigma$=$\sigma_1$, and  $\sigma'$=$\sigma_n$,
\item
$M' \circ M, \sigma_i \leadsto \sigma_{i+1}$  \ \ \ for all $i\in [0..n)$,
\item
$\class{\sigma}{\scd{\prg{this}}}, \class{\sigma'}{\scd{\prg{this}}}\in M'$,
\item
$\class{\sigma_i}{\scd{\prg{this}}} \in M$\ \ \ for all $i\in (1..n)$.
\end{itemize} 
\end{definition}
The function $\class{\sigma}{\_}$ is overloaded:
  applied to a variable, 
$\class{\sigma}{x}$  looks up the variable $x$ in the top frame of $\sigma$, and returns the 
class of the corresponding object in the  heap of $\sigma$;
applied to an address, $\class{\sigma}{\alpha}$  returns
the class of   the object referred by address $\alpha$ in the heap of $\sigma$.
 The module linking operator $\circ$, applied to two modules, $M'\circ M$, 
 combines the two modules into one module in the obvious way, provided their
domains are disjoint.
Full details in  Appendix \ref{app:loo}.
\begin{figure}[htb]
\begin{minipage}{\textwidth}
\begin{tabular}{cc}
(A) &
\begin{tikzpicture}[->,>=to,shorten >=1pt,auto,node distance=7mm,
                    thick,
                    external node/.style={circle,draw,minimum size=7mm,font=\sffamily\Large\bfseries, color=hotpink, fill = hotpink, text = black, fill opacity = 0.5},
                    internal node/.style={circle,draw,minimum size=7mm,font=\sffamily\Large\bfseries, color=lightseagreen, fill = lightseagreen, text = black, fill opacity = 0.5},
                    decoration = snake]
    
	\node[internal node] (a) {$\sigma_1$};
	\node[external node] (b) [right = of a] {$\sigma_2$};
	\node[internal node] (c) [right = of b] {$\sigma_3$};
	\node[internal node] (d) [right = of c] {$\sigma_4$};
	\node[external node] (e) [right = of d] {$\sigma_5$};
	\node[external node] (f) [right = of e] {$\sigma_6$};
	\node[internal node] (g) [right = of f] {$\sigma_7$};
	\node[internal node] (h) [right = of g] {$\sigma_8$};
	\node[external node] (i) [right = of h] {$\sigma_9$}; 


	\draw [decorate, ->]
	(a) -- (b);
	\draw [decorate, ->]
	(b) -- (c);
	\draw [decorate, ->]
	(c) -- (d);
	\draw [decorate, ->]
	(d) -- (e);
	\draw [decorate, ->]
	(e) -- (f);
	\draw [decorate, ->]
	(f) -- (g);
	\draw [decorate, ->]
	(g) -- (h);
	\draw [decorate, ->]
	(h) -- (i);
\end{tikzpicture} \\
(B) &
\begin{tikzpicture}[->,>=to,shorten >=1pt,auto,node distance=7mm,
                    thick,
                    external node/.style={circle,draw,minimum size=7mm,font=\sffamily\Large\bfseries, color=hotpink, fill = hotpink, text = black, fill opacity = 0.5},
                    internal node/.style={circle,draw,minimum size=7mm,font=\sffamily\Large\bfseries, color=lightseagreen, fill = lightseagreen, text = black, fill opacity = 0.2, draw opacity = 0.5},
                    decoration = snake,]
    
	\node[internal node] (a) {$\sigma_1$};
	\node[external node] (b) [right = of a] {$\sigma_2$};
	\node[internal node] (c) [right = of b] {$\sigma_3$};
	\node[internal node] (d) [right = of c] {$\sigma_4$};
	\node[external node] (e) [right = of d] {$\sigma_5$};
	\node[external node] (f) [right = of e] {$\sigma_6$};
	\node[internal node] (g) [right = of f] {$\sigma_7$};
	\node[internal node] (h) [right = of g] {$\sigma_8$};
	\node[external node] (i) [right = of h] {$\sigma_9$}; 


	\draw [decorate, ->]
	(b) -- (e);
	\draw [decorate, ->]
	(e) -- (f);
	\draw [decorate, ->]
	(f) -- (i);
\end{tikzpicture}
\end{tabular}
\end{minipage}
   \caption{External States Semantics
     (Def. \ref{def:pair-reduce}),  %
     (A) $\exec{{\color{hotpink}M'} \circ {\color{lightseagreen}M}}{\sigma_1}{\ldots}\leadsto \sigma_9$ \ \ \ and \ \ \ 
     (B) $\reduction{{\color{hotpink}M'}}{{\color{lightseagreen}M}}{\sigma_2}{\ldots}\leadsto \sigma_9$, \ \ \ 
     \\
     where $\class{{\color{lightseagreen}\sigma_1}}{\scd{\prg{this}}}$,$\class{{\color{lightseagreen}\sigma_3}}{\scd{\prg{this}}}$,$\class{{\color{lightseagreen}\sigma_4}}{\scd{\prg{this}}}$,$\class{{\color{lightseagreen}\sigma_7}}{\scd{\prg{this}}}$,$\class{{\color{lightseagreen}\sigma_8}}{\scd{\prg{this}}}\in {\color{lightseagreen}M}$,\\
     and where $\class{{\color{hotpink}\sigma_2}}{\scd{\prg{this}}},\class{{\color{hotpink}\sigma_5}}{\scd{\prg{this}}} 
     \class{{\color{hotpink}\sigma_6}}{\scd{\prg{this}}},\class{{\color{hotpink}\sigma_9}}{\scd{\prg{this}}}\in {\color{hotpink}M'}$.
    }
   \label{fig:VisibleStates}
 \end{figure}
 
Fig. \ref{fig:VisibleStates} inspired by \citeasnoun{FASE} provides a simple graphical description of 
our external states semantics: (A) is the ``normal'' execution after 
linking two modules into one: \ $M' \circ M, ... \leadsto ...$ whereas (B) is the
 external states execution when $M'$ is external,\   $\reduction{M'}{M}{...}{...}$.
Note that whether a module is external or internal depends on 
perspective -- nothing in a module itself renders it internal or external. For example, in
 $\reduction{M_1}{M_2}{...}{...}$ the external module is $M_1$,
  while in  $\reduction{M_2}{M_1}{...}{...}$  the external module is $M_2$.

We  use the notation\ \  $\reductions{M'}{M}{\sigma}{\sigma'}$ \ 
to denote zero or more  steps starting at state $\sigma$ and ending at state $\sigma'$, in the context of internal module 
$M$ and external module $M'$.
We are \jm[]{not} concerned with internal states or states that can never arise.
{A state $\sigma$ is \emph{arising},}  written $\arising{M'}{M}{\sigma}$, {if it  may arise by external states} execution
starting at some initial configuration:

\begin{definition}[Arising  States]
\label{def:arising}
For   modules $M$ and  $M'$, a 
 state $\sigma$ is 
called an \emph{arising} state, formally \ \ \ $\arising{M'}{M}{\sigma}$,\ \ \ 
if and only if there exists some $\sigma_0$ such that $\initial{\sigma_0}$ and
$\reductions{M'}{M}{\sigma_0}{\sigma}$.
\end{definition}

An \emph{Initial} state's heap
contains a single object of class \prg{Object}, and
its  stack   consists of a single frame, whose local variable map is a
mapping from \prg{this} to the single object, and whose continuation is  any statement.
(See Definitions \ref{def:initial} and \ref{def:arising}).

\paragraph{Applicability} 
{While our work is based on 
  a simple, imperative, typed, object oriented}
language with unforgeable addresses and private fields, we believe
 that 
 it is applicable to several programming paradigms, and 
 that   unforgeability and privacy
 can be replaced 
 by lower level mechanisms such as capability machines \cite{vanproving,davis2019cheriabi}.

\subsection{\SpecO}
\label{sub:SpecO}

\SpecO is \jm[removed: a subset of the \emph{Chainmail} assertions language, \ie]{}
a basic assertion language extended with
object-capability assertions.

\subsubsection{Syntax of \SpecO}
The syntax of \SpecO   is given in
Definition \ref{f:chainmail-syntax}.
An assertion may be an expression,   a query of the defining class of
  an object, the usual connectives and quantifiers, along 
with three non-standard assertion forms:
(1) \emph{Permission} and (2) \emph{Provenance}, inspired by the capabilities literature, and
(3) \emph{Control} which allows tighter  characterisation of the cause of effects --  
useful for the specification of large APIs.
\begin{itemize}
\item
\emph{Permission} ($\access{x}{y}$):  
  $x$ has access to $y$.
\item
{\emph{Provenance}} ($\internal{x}$ and $\external{y}$):   $x$ is an internal \jm[]{(i.e. trusted) object}, and $y$ is an external \jm[]{(i.e. untrusted) object}.
\item
\emph{Control} ($\calls{x}{y}{m}{\overline{z}}$): 
$x$ calls method $m$ on object $y$ with arguments $\overline{z}$.
\end{itemize}

\begin{definition}
Assertions ($A$) in
\SpecO are defined as follows:

\label{f:chainmail-syntax}
 \[
\begin{syntax}
\syntaxElement{A}{}
		{
		\syntaxline
				{e}
				{e : C}
				{\neg A}
				{A\ \wedge\ A}
				{A\ \vee\ A}
				{\all{x}{A}}
				{\ex{x}{A}}
		\endsyntaxline
		}
		{
		\syntaxline
				{\access{x}{y}}
				{\internal{x}}
				{\external{x}}
				{\calls{x}{y}{m}{\overline{z}}}
		\endsyntaxline
		}
\endSyntaxElement\\
\end{syntax}
\]

\end{definition}

\subsubsection{Semantics of \SpecO}
The semantics of \SpecO   
is given in Definition \ref{def:chainmail-semantics}. 
We   use the evaluation relation, $\eval{M}{\sigma}{e}{v}$,
which says that the expression $e$ evaluates
to value $v$ in the context of state $\sigma$ and module $M$.
Note that expressions in \Loo may be recursively defined, and thus evaluation 
need not \jm[removed: always]{} 
 terminate. Nevertheless, the logic of $A$ remains classical because recursion is restricted
to expressions, and not generally to assertions.
We have taken this approach from \citeasnoun{FASE}, which also contains a mechanized Coq proof that assertions are classical \cite{coqFASE}.
The semantics of $\hookrightarrow$ \jm[]{is} unsurprising (see Fig.\ref{f:evaluation}).

Shorthands: 
 $\interpret{\phi}{x} = v$  means that $x$ maps to
value $v$ in the local variable map of frame $\phi$, $\interpret{\sigma}{x} = v$ means that $x$ 
maps to $v$ in the top most frame of $\sigma$'s stack, and $\interpret{\sigma}{x.f} = v$
has the obvious meaning. The terms $\sigma.\prg{stack}$,  
$\sigma.\prg{contn}$, 
$\sigma.\prg{heap}$     mean the stack, 
the continuation at the
top frame of $\sigma$, 
and the heap of $\sigma$.
The term $\alpha\!\in\!\sigma.\prg{heap}$ means that $\alpha$ is in the domain of the heap of $\sigma$, and \emph{$x$ fresh in $\sigma$} means that 
$x$ isn't in the variable map of the top frame of $\sigma$, 
while the substitution  $\sigma[x \mapsto \alpha]$ is applied to the top frame of $\sigma$.
\jm[added extra space]{\ }$C\in M$ means that class $C$ is in the domain of module $M$. 

\begin{definition}[Satisfaction 
of Assertions by a module and a state] 
\label{def:chainmail-semantics}
We define satisfaction of an assertion $A$ by a 
state $\sigma$ with 
 module $M$ as:
\begin{enumerate}
\item
\label{cExpr}
$\satisfiesA{M}{\sigma}{e}$ \ \ \ iff \ \ \  $\eval{M}{\sigma}{e}{\true}$
\item
\label{cClass}
$\satisfiesA{M}{\sigma}{e : C}$ \ \ \ iff \ \ \  $\eval{M}{\sigma}{e}{\alpha}$ \textit{and} $\class{\sigma}{\alpha} = C$
\item
$\satisfiesA{M}{\sigma}{\neg A}$ \ \ \ iff \ \ \  ${M},{\sigma}\nvDash{A}$
\item
$\satisfiesA{M}{\sigma}{A_1\ \wedge\ A_2}$ \ \ \ iff \ \ \  $\satisfiesA{M}{\sigma}{A_1}$ and 
$\satisfiesA{M}{\sigma}{A_2}$
\item
$\satisfiesA{M}{\sigma}{A_1\ \vee\ A_2}$ \ \ \ iff \ \ \  $\satisfiesA{M}{\sigma}{A_1}$ or 
$\satisfiesA{M}{\sigma}{A_2}$
\item
\label{quant1}
$\satisfiesA{M}{\sigma}{\all{x}{A}}$ \ \ \ iff \ \ \  
$\satisfiesA{M}{\sigma[x \mapsto \alpha]}{A}$, \ 
\ \ \ for some $x$ fresh in $\sigma$, and for all $\alpha\!\in\!\sigma.\prg{heap}$.
\item
\label{quant2}
$\satisfiesA{M}{\sigma}{\ex{x}{A}}$ \ \ \ iff \ \ \  
$\satisfiesA{M}{\sigma[x \mapsto \alpha]}{A}$, \ 
\ \ for some $x$ fresh in $\sigma$, and for some $ \alpha\!\in\!\sigma.\prg{heap}$. 
\item
\label{cAccess}
$\satisfiesA{M}{\sigma}{\access{x}{y}}$ \ \ \ iff \ \ \  
\begin{enumerate}
\item
\label{c1}
$\interpret{\sigma}{x.f}={\interpret{\sigma}{y}}$ for some $f$, \\
  or
\item
\label{c2}
{$\interpret{\sigma}{x}=\interpret{\phi}{\prg{this}}$}, {$\interpret{\sigma}{y}=\interpret{\phi}{z}$}, and $z\ \in\ \phi.\prg{contn}$\ \ \ \
for some variable $z$, and some frame $\phi$ in $\sigma.
\prg{stack}$.
\end{enumerate}
\item
\label{cInternal}
$\satisfiesA{M}{\sigma}{\internal{x}}$ \ \ \ iff \ \ \  
$\textit{classOf}(\sigma,x) \in M$
\item
\label{cExternal}
$\satisfiesA{M}{\sigma}{\external{x}}$ \ \ \ iff \ \ \  
$\textit{classOf}(\sigma,x) \not\in M$
\item
\label{cCall}
$\satisfiesA{M}{\sigma}{\calls{x}{y}{m}{z_1, \ldots, z_n}}$ \ \ \ iff \ \ \ 
\begin{enumerate}
\item
$\sigma.\prg{contn} = (w := y'.m(z'_1,\ldots,z'_n)\scd{; s})$,\ \ for some 
variable $w$, and some statement $s$,
\item
$\satisfiesA{M}{\sigma}{x = \prg{this}}$
\ \ and \ \ 
$\satisfiesA{M}{\sigma}{y = y'}$,
\item
$\satisfiesA{M}{\sigma}{z_i = z'_i}$\ \ \ for all $1\!\leq i\!\leq n$
\end{enumerate}
\end{enumerate}
\end{definition}

\julian{Quantification (defined in \ref{quant1} and \ref{quant2}) is done over all objects on the heap.
We do not include quantification over primitive types such as integers as \Loo is too simple. The 
Coq mechanisation does include primitive types.}
 
The assertion ${\access{x}{y}}$ (defined in  \ref{cAccess})
requires  that $x$ has access to $y$
either through a field of $x$ (case \ref{c1}),
or through some call in the stack, where $x$ is the receiver and $y$ is one of the
arguments (case \ref{c2}).
{Note that access is not deep, and only refers to objects that 
an object has direct access to via a field or within the context of a current scope. 
 The restricted form of access used in \Nec specifically captures a crucial property of robust programs in the open world: access to an object does not imply access to that object's internal data. For example, an object may have access to an account \prg{a}, but a safe implementation of the account would never allow that object to leverage that access to gain direct access to {\prg{a.pwd}}}.
 
 The assertion  
 ${\calls{x}{y}{m}{z_1, \ldots, z_n}}$  (defined in \ref{cCall})
\sdM{describes the current innermost active call}. It
requires that the current receiver (\prg{this}) is $x$, and that it calls the method $m$ on $y$ with
 arguments $z_1$, ... $z_n$ -- It does \emph{not} mean  that somewhere in the 
 call stack there exists a call from $x$ to $y.m(...)$. 
 Note that in most cases, satisfaction of an assertion not only depends on the state $\sigma$, but 
also depends on the module in the case of expressions (\ref{cExpr}), class membership
(\ref{cClass}), and internal or external provenance (\ref{cInternal} and \ref{cExternal}).

We now define what it means for a module to satisfy an assertion:
 $M$ satisfies  $A$ if any state arising from external steps execution of that
module with any other external module  satisfies $A$. 
 
\begin{definition} [Satisfaction 
of Assertions
by a module] 
\label{def:mdl-sat}
For a module $M$ and assertion $A$, we say that\ \  $\satisfies{M}{A}$ \ \ if and only if 
for all modules $M'$, and all $\sigma$, if $\arising{M'}{M}{\sigma}$, then $\satisfiesA{M}{\sigma}{A}$.
\end{definition}

In the current work we assume the existence of a proof system that judges
$\proves{M}{A}$, to prove  satisfaction of assertions. 
 We will not define such a judgement, but will rely on its existence {later on for} Theorem \ref{thm:soundness}.
We define soundness of such a judgement in the usual way:

\begin{definition}[Soundness of \SpecO Provability]
\label{ax:specW-prove-soundness}
A judgement of the form {$\proves{M}{A}$} is \emph{sound}, if for all
 modules $M$ and assertions $A$, \ if $\proves{M}{A}$ then $\satisfies{M}{A}$.
\end{definition}

\subsubsection{Inside}

We define
a final shorthand 
predicate $\wrapped{\prg{o}}$ which states 
that only \internalO objects have access to \prg{o}.
The object \prg{o} may be either \internalO or \externalO.
\begin{definition}[Inside]
$\wrapped{o}\ \triangleq\ \all{x}{\access{x}{o}\ \Rightarrow\ \internal{x}} $ 
\end{definition}

\inside is a very useful concept. For example, the balance of an account whose
  password is \inside  will not decrease in the next step.
  Often, API implementations contain objects whose capabilities, while  crucial for the implementation, if exposed,
would break the intended guarantees of the API. Such objects need to remain \inside - see
such an example in Section \ref{s:examples}.

\subsection {\Nec operators}
\label{s:holistic-guarantees}

\subsubsection{Syntax of \Nec Specifications}
The \Nec specification language extends \SpecO with our three novel 
 \jm[]{\Nec} \emph{operators}:

\begin{description}
\item 
[$\onlyIfSingle{A_1}{A_2}{A}$]: If an arising  
  state satisfies $A_1$, and \sd{a single execution step reaches}  a state satisfying $A_2$, 
then the original  
state must have also satisfied $A$.

\item 
[$\onlyIf{A_1}{A_2}{A}$]: If an arising  
  state satisfies $A_1$ and \sd{a number of execution steps reach} a state   satisfying $A_2$, 
then the original  
state must have also satisfied $A$.

\item 
[$\onlyThrough{A_1}{A_2}{A}$]: If an arising  
 state satisfies $A_1$, and \sd{a number of execution steps reach} a state  
 satisfying $A_2$,  then  execution must have passed through some \emph{intermediate} state satisfying $A$.
\end{description}

\noindent
The syntax of  \Nec specifications is given below

\begin{definition}  

\noindent
{\emph{\sd{Syntax of \Nec Specifications}}}

\label{f:holistic-syntax}
\footnotesize
\[
\begin{syntax}
\syntaxElement{S}{}
		{
		\syntaxline
				{A}
				{\onlyIf{A_1}{A_2}{A_3}}
				{\onlyThrough{A_1}{A_2}{A_3}}
				 {\onlyIfSingle{A_1}{A_2}{A_3}}
		\endsyntaxline
		}
\endSyntaxElement\\
\end{syntax}
\]
\normalsize
\end{definition}

\label{sec:adapt:motivate}

\noindent
As an example, we consider the following three  specifications:

\begin{lstlisting}[language = Chainmail, mathescape=true, frame=lines]
$\text{\SRobustNextAcc}$   $\triangleq$  from a:Account $\wedge$ a.balance==bal  next a.balance < bal
                        onlyIf $\exists$ o.[$\external{\texttt{o}}$ $\wedge$ $\access{\prg{o}}{\prg{a.pwd}}$]
$\text{\SRobustIfAcc}$   $\triangleq$  from a:Account $\wedge$ a.balance==bal  to a.balance < bal
                        onlyIf $\exists$ o.[$\external{\texttt{o}}$ $\wedge$ $\access{\prg{o}}{\prg{a.pwd}}$]
$\text{\SRobustThroughAcc}$   $\triangleq$  from a:Account $\wedge$ a.balance==bal  next a.balance < bal
                        onlyThrough $\exists$ o.[$\external{\texttt{o}}$ $\wedge$ $\access{\prg{o}}{\prg{a.pwd}}$]                                   
\end{lstlisting}

\noindent
\SRobustNextAcc  requires that an account's balance may decrease in \emph{one step} (go from a state where the balance is \prg{bal}
to a state where it is less than \prg{bal}) only if the password is accessible to an external object (in the original state an external object had access to the password).
\SRobustIfAcc  requires that an account's balance may decrease in \emph{any number of steps}    only if the password is accessible to an external object.
\SRobustThroughAcc requires that an account's balance may decrease in \emph{any number of steps}    only if in \emph{some intermediate state} the password was accessible to an external object --  the   \emph{intermediate} state  where the password is accessible to the external object might be the \emph{starting}  
state, the \emph{final} state, or any state in between.

\subsubsection{Semantics of \Nec Specifications}
We now  define what it means for  a module  $M$ to satisfy specification  $S$, written as $M \vDash S$. The
 Definition~\ref{def:necessity-semantics} below is straightforward, apart from  
the use of the $\adapt  {\sigma'}{\sigma}$  (best read as ``$\sigma'$ seen
from $\sigma$'')
to deal with the fact that execution might  change the bindings in local variables.
We explain this in detail in   \S \ref{sub:adapt:full}, but for now, the reader may ignore the applications of that operator and
read $\sigma' \triangleleft \sigma$ as $\sigma'$,
and also read ${\sigma_k \triangleleft \sigma_1}$ as  $\sigma_k$.
We illustrate the meaning of the three operators in 
Fig.~\ref{fig:Operators}.

\begin{figure}[htbp]
\newcommand{\mathsmall}[1]{\substack{\scalebox{0.8}{$#1$}}}
\begin{minipage}{\textwidth}
$\onlyIf{A_1}{A_2}{A}$:\\\\
\begin{tikzpicture}[->,>=to,shorten >=1pt,auto,node distance=5.5mm,
                    thick,
                    state/.style={circle,draw,minimum size=5mm,font=\sffamily\bfseries, color=hotpink, fill = hotpink, text = black, fill opacity = 0.5, scale=0.9},
                    dots/.style={
                    minimum size=7mm,
                    font=\sffamily\Large\bfseries, 
                    color=lightseagreen, text = black, fill opacity = 0.5},
                    space/.style={
                    minimum size=7mm,
                    font=\sffamily\Large\bfseries, 
                    color=lightseagreen, text = black, fill opacity = 0.5},
                    arrow/.style={
                    minimum size=7mm,
                    font=\sffamily\Large\bfseries, 
                    color=lightseagreen, text = black, fill opacity = 0.5},
                    models/.style={
                    minimum size=7mm,
                    font=\sffamily\Large\bfseries, 
                    color=lightseagreen, text = black, fill opacity = 0.5},
                    decoration = snake]
    
	\node[state, label={270:$\mathsmall{\vDash A_1}$}] (la) {$\sigma_1$};
	\node[space] (s1) [right = of la] {};
	\node[dots] (lb) [right = of s1] {$\ldots$};
	\node[space] (s2) [right = of lb] {};
	\node[state, label={270:$\mathsmall{\vDash A_2}$}] (lc) [right = of s2] {$\sigma_n$};
	\draw [decorate, ->]
	(la) -- (s1);
	\draw [decorate, ->]
	(s2) -- (lc);

	\node[arrow] (c) [right = of lc] {$\Longrightarrow$};
    
	\node[state, label={270:$\mathsmall{\vDash A_1 \color{hotpink}{\mathbf{\wedge A}}}$}] (ra) [right = of c] {$\sigma_1$};
	\node[space] (s3) [right = of ra] {};
	\node[dots] (rb) [right = of s3] {$\ldots$};
	\node[space] (s4) [right = of rb] {};
	\node[state, label={270:$\mathsmall{\vDash A_2}$}] (rc) [right = of s4] {$\sigma_n$};

	\draw [decorate, ->]
	(ra) -- (s3);
	\draw [decorate, ->]
	(s4) -- (rc);
\end{tikzpicture}\\\\
$\onlyIfSingle{A_1}{A_2}{A}$:\\\\
\begin{tikzpicture}[->,>=to,shorten >=1pt,auto,node distance=5.5mm,
                    thick,
                    state/.style={circle,draw,minimum size=7mm,font=\sffamily\bfseries, color=hotpink, fill = hotpink, text = black, fill opacity = 0.5, scale=0.9},
                    dots/.style={
                    minimum size=7mm,
                    font=\sffamily\Large\bfseries, 
                    color=lightseagreen, 
                    text = black, 
                    fill opacity = 0.5},
                    space/.style={
                    minimum size=7mm,
                    font=\sffamily\Large\bfseries, 
                    color=lightseagreen, 
                    text = black, 
                    fill opacity = 0.5},
                    arrow/.style={
                    minimum size=7mm,
                    font=\sffamily\Large\bfseries, 
                    color=lightseagreen, 
                    text = black, 
                    fill opacity = 0.5},
                    models/.style={
                    minimum size=7mm,
                    font=\sffamily\Large\bfseries, 
                    color=lightseagreen, 
                    text = black, 
                    fill opacity = 0.5},
                    decoration = snake]
    
    \node[space] (s) at (0,0) {};
	\node[state, label={270:$\mathsmall{\vDash A_1}$}] (a) at (3.85,0) {$\sigma_1$};
	\node[state, label={270:{$\mathsmall{\vDash A_2}$}}] (c) [right = of a] {$\sigma_n$};
	\draw [decorate, ->]
	(a) -- (c);

	\node[arrow] (d) [right = of c] {$\Longrightarrow$};
    
	\node[state, label={270:$\mathsmall{\vDash A_1 \color{hotpink}{\mathbf{\wedge A}}}$}] (e) [right = of d] {$\sigma_1$};
	\node[state, label={270:$\mathsmall{\vDash A_2}$}] (g) [right = of e] {$\sigma_n$};

	\draw [decorate, ->]
	(e) -- (g);
\end{tikzpicture}\\\\
$\onlyThrough{A_1}{A_2}{A}$:\\\\
\begin{tikzpicture}[->,>=to,shorten >=1pt,auto,node distance=5.5mm,
                    thick,
                    state/.style={circle,draw,minimum size=7mm,font=\sffamily\bfseries, color=hotpink, fill = hotpink, text = black, fill opacity = 0.5, scale=0.9},
                    dots/.style={
                    minimum size=7mm,
                    font=\sffamily\Large\bfseries, color=lightseagreen, text = black, fill opacity = 0.5},
                    space/.style={
                    minimum size=7mm,
                    font=\sffamily\Large\bfseries, color=lightseagreen, text = black, fill opacity = 0.5},
                    arrow/.style={
                    minimum size=7mm,
                    font=\sffamily\Large\bfseries, color=lightseagreen, text = black, fill opacity = 0.5},
                    models/.style={
                    minimum size=7mm,
                    font=\sffamily\Large\bfseries, color=lightseagreen, text = black, fill opacity = 0.5},
                    decoration = snake]
    
	\node[state, label={270:$\mathsmall{\vDash A_1}$}] (la) {$\sigma_1$};
	\node[space] (s1) [right = of la] {};
	\node[dots] (lb) [right = of s1] {$\ldots$};
	\node[space] (s2) [right = of lb] {};
	\node[state, label={270:$\mathsmall{\vDash A_2}$}] (lc) [right = of s2] {$\sigma_n$};
	\draw [decorate, ->]
	(la) -- (s1);
	\draw [decorate, ->]
	(s2) -- (lc);

	\node[arrow] (c) [right = of lc] {$\Longrightarrow$};
    
	\node[state, label={270:$\mathsmall{\vDash A_1}$}] (ra) [right = of c] {$\sigma_1$};
	\node[dots] (d1) [right = of ra] {$\ldots$};
	\node[state, label={270:$\mathsmall{\mathbf{\color{hotpink}{\vDash A}}}$}] (rb) [right = of d1] {$\sigma_k$};
	\node[dots] (d2) [right = of rb] {$\ldots$};
	\node[state, label={270:$\mathsmall{\vDash A_2}$}] (rc) [right = of d2] {$\sigma_n$};

	\draw [decorate, ->]
	(ra) -- (d1);
	\draw [decorate, ->]
	(d1) -- (rb);
	\draw [decorate, ->]
	(rb) -- (d2);
	\draw [decorate, ->]
	(d2) -- (rc);
\end{tikzpicture}
\end{minipage}
\caption{Illustrating the three \Nec operators}
\label{fig:Operators}
\end{figure}

\begin{definition}[Semantics of \Nec Specifications]
We define $\satisfies{M}{{S}}$ by cases over the four possible syntactic forms.
For any assertions   $A_1$, $A_2$, and $A$: \\

\label{def:necessity-semantics}

$\bullet$ \ $\satisfies{M}{{A}}$ \ \ \ iff\ \ \ for all $M'$, $\sigma$,\ if $\arising{M'}{M}{\sigma}$, then $\satisfiesA{M}{\sigma}{A}$. (see Def. \ref{def:mdl-sat})\\


$\bullet$ \ $\satisfies{M}{\onlyIf {A_1}{A_2}{A}}$ \ \ iff\ \  for all $M'$, $\sigma$, $\sigma'$, such that $\arising{M'}{M}{\sigma}$: \\  

\begin{tabular}{lr}
$\;\;\;\;$- $\satisfiesA{M}{\sigma}{A_1}$  & \rdelim\}{3}{3mm}[$\;\;\;\Rightarrow\;\;\;$  $\satisfiesA{M}{\sigma}{A}$] \\
$\;\;\;\;$- $\satisfiesA{M}{\sigma' \triangleleft \sigma}{A_2}$   \\
$\;\;\;\;$- $\reductions{M'}{M}{\sigma}{\sigma'}$   \\
\end{tabular}\\ 

$\bullet$ \  $\satisfies{M}{\onlyIfSingle {A_1}{A_2}{A}}$\ \ iff\ \   for all $M'$, $\sigma$,   $\sigma'$, such that $\arising{M'}{M}{\sigma}$: \\

\begin{tabular}{lr}
$\;\;\;\;$- $\satisfiesA{M}{\sigma}{A_1}$  & \rdelim\}{3}{3mm}[$\;\;\;\Rightarrow\;\;\;$  $\satisfiesA{M}{\sigma}{A}$] \\
$\;\;\;\;$- $\satisfiesA{M}{\sigma' \triangleleft \sigma}{A_2}$   \\
$\;\;\;\;$- $\reduction{M'}{M}{\sigma}{\sigma'}$   \\
\end{tabular}\\ 
  
$\bullet$ \  $\satisfies{M}{\onlyThrough {A_1}{A_2}{A}}$ \ \ iff\ \  for all $M'$, $\sigma_1$,  \sd{$\sigma_2$,  ....} $\sigma_n$, such that $\arising{M'}{M}{\sigma_1}$: \\

\begin{tabular}{lr}
$\;\;\;\;$- $\satisfiesA{M}{\sigma_1}{A_1}$  & \rdelim\}{3}{3mm}[$\;\;\;\Rightarrow\;\;\;$  $\exists k. \ 1\leq k \leq n \ \wedge \ \satisfiesA{M}{\sigma_k \triangleleft \sigma_1}{A}$]   \\
$\;\;\;\;$- $\satisfiesA{M}{\sigma_n\triangleleft \sigma\sd{_1}}{A_2}$   \\
$\;\;\;\;$- $\forall i\!\in\![1..n).\ \reduction{M'}{M}{\sigma_i}{\sigma_{i+1}}$   \\
\end{tabular} 

\end{definition} 

Revisiting the examples from the previous subsection, we obtain that all three modules satisfy 
$\SRobustNextAcc$. But $\ModB$ does not satisfy $\SRobustIfAcc$: as  already discussed in \S \ref{s:bank}, 
with \prg{a} of class \prg{Account} implemented as in $\ModB$,
starting in a state where no external object has access to \prg{a}'s password, and executing 
\prg{a.set(42); a.transfer(rogue\_account,42)} leads to a state where the balance has decreased.
All three modules satisfy $\SRobustThroughAcc$: namely, in all cases, the balance can only decrease if 
there was a call to \prg{a.transfer(\_,p)} where $\prg{p}=\prg{a.pwd}$, and since that call can only be made from an external object,
\prg{p} is externally known at the time of that call.

 $\begin{array}{llll}
  \   & \ModA \vDash \SRobustNextAcc  \   & \ModB \vDash \SRobustNextAcc \  
  & \ModC \vDash \SRobustNextAcc \\
  \   & \ModA \vDash \SRobustIfAcc  \   & \ModB \nvDash \SRobustIfAcc \  
  & \ModC \vDash \SRobustIfAcc \\
   \   & \ModA \vDash \SRobustThroughAcc  \   & \ModB \vDash \SRobustThroughAcc \  
  & \ModC \vDash \SRobustThroughAcc \\ 
  \end{array}$

\subsubsection{Adaptation}
\label{sub:adapt:full}
We  now discuss  the adaptation operator.  To see the need, 
consider  specification
\begin{lstlisting}[language = Chainmail, mathescape=true, frame=lines]
$\text{\Sadapt}$   $\triangleq$  from a:Account $\wedge$ a.balance==350  next a.balance == 250
                   onlyIf $\exists$ o.[$\external{\texttt{o}}$ $\wedge$ $\calls{\prg{o}}{\prg{a}}{\prg{transfer}}{\prg{\_, \_, \_}}$]
\end{lstlisting}

\sd{Without adaptation,} the semantics of  \Sadapt would be: If
$..,\sigma  \models \prg{a.balance==350}$, 
and  $.., \sigma  \leadsto^* \sigma'$ and $\sigma' \models \prg{a.balance==250}$,
then 
between $\sigma$ and $\sigma'$ there must be
call to   \prg{a.transfer}. 
But if $\sigma$ happened to have another account \prg{a1} with balance
\prg{350}, and if we reach $\sigma'$ from $\sigma$ by executing \prg{a1.transfer(}$..,..$\prg{); a=a1}, then we would 
reach a $\sigma'$ 
\emph{without} \prg{a.transfer} having been called: indeed, without the
account \prg{a} from $\sigma$  having changed at all.
\sd{In fact, with such a semantics,  a module would satisfy $\Sadapt$ only if it did not support decrease of \jm[]{the} balance by $100$, or
\jm[]{if} states where an account's balance is \prg{350} were unreachable!}

This is the remit of the adaptation operator: when we consider the
future state, we must ``see it from'' the perspective of the current
state; the binding for variables such as \prg{a} must be from the
current state, even though we may have assigned to them in the mean
time.  Thus, $\adapt {\sigma'} {\sigma}$ keeps the heap from $\sigma'$,
and renames the variables
in the top stack frame of  $\sigma'$ so that all variables defined in $\sigma$ have the same 
bindings
as in   $\sigma$; the continuation must be adapted similarly (see
Fig.~\ref{fig:Adaptation}).
%
%
\newcommand{\mathsmall}[1]{\substack{\scalebox{0.8}{$#1$}}}
\begin{figure}[tbp]
\begin{tabular}{clclc}
 \begin{minipage}{0.27\textwidth}
 $\sigma:$\\
 \includegraphics[width=\linewidth]{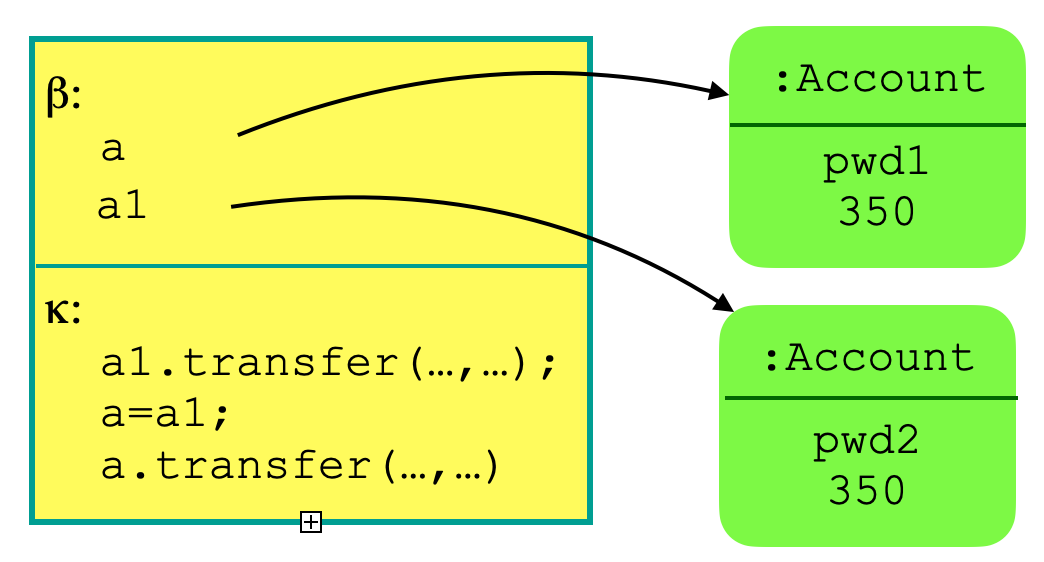}
   \end{minipage}
 & \ \ \ &
 \begin{minipage}{0.27\textwidth}
  $\sigma':$\\
  \includegraphics[width=\linewidth]{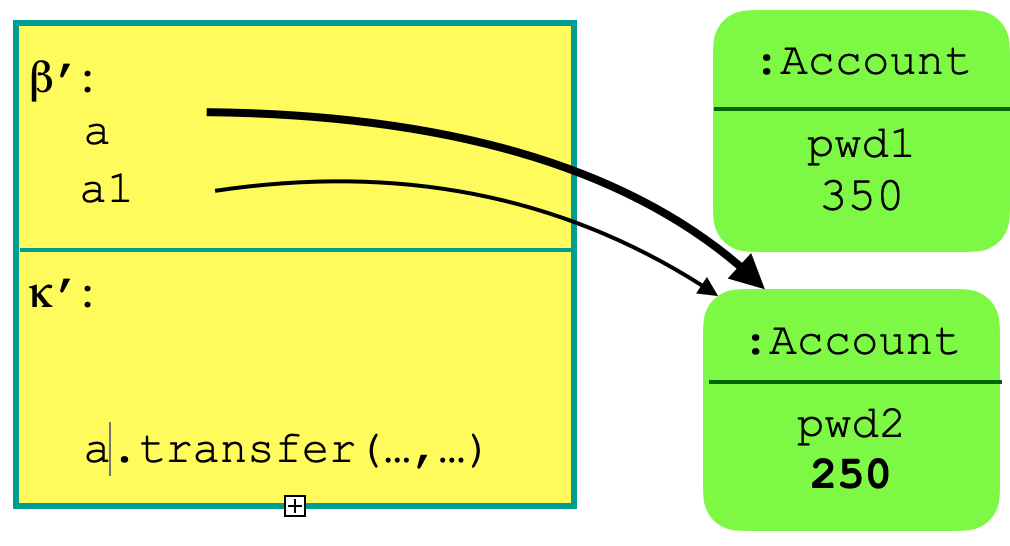}
   \end{minipage}
   & \ \ \  &
    \begin{minipage}{0.27\textwidth}
$\adapt {\sigma'}{\sigma}:$\\
  \includegraphics[width=\linewidth]{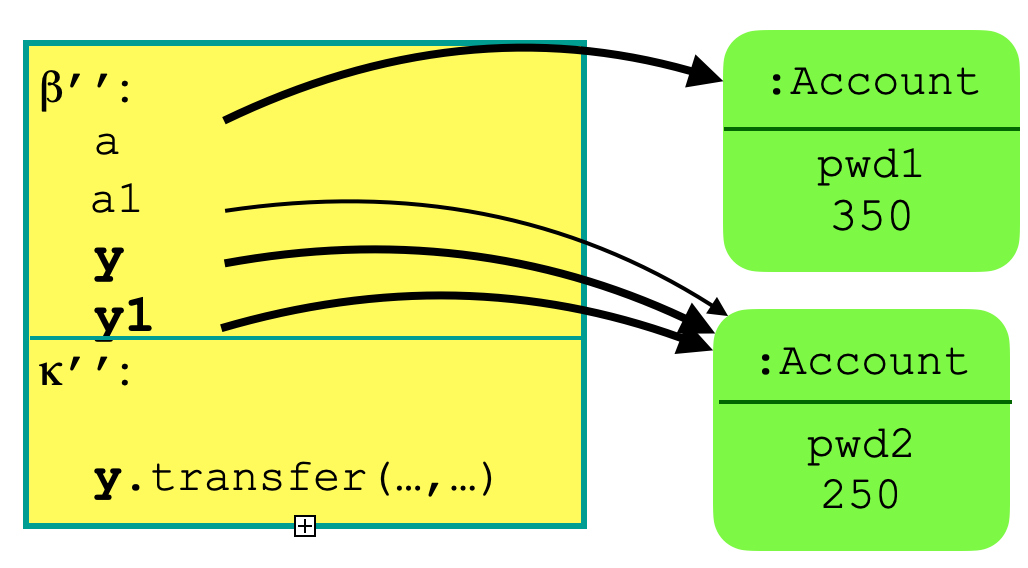}
   \end{minipage}
\end{tabular}

%
\caption{Illustrating adaptation
}
\label{fig:Adaptation}
\end{figure}

Under adaptation, the semantics of \Sadapt is:\  if $..,\sigma \models \prg{a.balance==350}$,
and  $.., \sigma \leadsto^* \sigma'$ and $..., \boldsymbol{\adapt {\sigma'}{\sigma}} \models \prg{a.balance==250}$,
then {some intermediate state's continuation must contain  a call to \prg{a.transfer}};
\sd{where,  all \sd{variables bound in the initial state, $\sigma$,}
have the same bindings in $\adapt {\sigma'}{\sigma}$.}
\jm[I didn't understand susan's comment on this paragraph: ``problem with primes'']{}
 
Fig.~\ref{fig:Adaptation} illustrates the semantics of $\adapt {\sigma'}{\sigma}$. In   $\sigma$ the variable \prg{a} points to an \prg{Account}
 with password \prg{pwd1}, and balance \prg{350};  the variable \prg{a1} points to an \prg{Account}
  with password \prg{pwd2}, and balance \prg{350}; and the continuation is \prg{a1.transfer(}$..,..$\prg{);} \prg{a=a1;}
\prg{a.transfer(}$..,..$\prg{);}. 
We reach $\sigma'$ by executing the first two statements from the continuation.
Thus,  ${\adapt {\sigma'}{\sigma}}\not\models \prg{a.balance==250}$.
Moreover, in $\adapt {\sigma'}{\sigma}$ we introduce the fresh variables \prg{y} and \prg{y1}, and replace \prg{a} and
\prg{a1} by \prg{y} and \prg{y1} in the continuation.
This gives that ${\adapt {\sigma'}{\sigma}} \models \calls {\_} {\prg{a1}} {\prg{transfer}} {...}$ and  ${\adapt {\sigma'}{\sigma}}\not\models \calls {\_} {\prg{a}} {\prg{transfer}} {...}$.

Definition~\ref{d:adapt} \sd{describes} the $\adapt{}{}$ operator in all detail
(it is equivalent to, but not identical to the definition  given in \cite{FASE}).
We introduce fresh variables  $\overline{y}$ -- as many as in the $\sigma'$ top frame variable map
-- $dom(\beta')=\overline{x}$,  and $|\overline{y}| = |\overline{x}|$.  
We extend $\sigma$'s variable map ($\beta$), so that it also maps $\overline{y}$ 
in the way that  $\sigma'$'s variable map ($\beta'$) maps its local variables -- $\beta'' =  \beta[\overline{y} \mapsto  {\beta'(\overline{x})}]$. We rename $\overline{x}$   in $\sigma'$ continuation
to $\overline{y}$ --  $\kappa''=[\overline{y}/\overline{x}]\kappa'$.

\begin{definition}
\label{d:adapt}
For any states $\sigma$, $\sigma'$, heaps $\chi$, $\chi'$, 
variable maps $\beta$, $\beta'$, 
and continuations $\kappa$, $\kappa'$, such that 
$\sigma$=$(\chi,(\beta,\kappa):\psi)$, and $\sigma$=$(\chi',(\beta',\kappa'):\psi')$, we define  
\\
$\strut \ \ \ \ \bullet$  $\adapt{\sigma'}{\sigma} \triangleq (\chi', (\beta'',\kappa'') : \psi')$  
\\
where there exist variables $\overline{y}$ such that \ \ $\beta'' =  \beta[\overline{y} \mapsto  {\beta'(\overline{x})}]$, \ and\  $\kappa''=[\overline{y}/\overline{x}]\kappa'$, and 
$dom(\beta')=\overline{x}$,  and $|\overline{y}| = |\overline{x}|$,\  and\  $\overline{y}$ are fresh in $\beta$ and $\beta'$.
\end{definition}

Strictly speaking, $\adapt {}{}$  does not define one  unique state: Because  variables $\overline{y}$ 
are arbitrarily chosen,   $\adapt {}{}$ describes an infinite set of states. These states satisfy the same assertions
and therefore are   equivalent with each 
other.
This is why it is sound to  use $\adapt {}{}$  as an operator, rather than as a set.

\subsection{Expressiveness}

We discuss expressiveness of \Nec operators, by comparing 
them with one another, with temporal operators, and with other examples from the literature.

\paragraph{Relationship between Necessity Operators}
The three \Nec \ operators
are related by generality. 
 \emph{Only If} ($\onlyIf{A_1}{A_2}{A}$) implies
  \emph{Single-Step Only If} ($\onlyIfSingle{A_1}{A_2}{A}$), since if $A$ is 
a necessary precondition for multiple steps, then it must be a necessary 
precondition for a single step. 
 \emph{Only If} also implies 
an \emph{Only Through}, where the intermediate state is the starting state
of the execution.  There is no further relationship between 
\emph{Single-Step Only If} and \emph{Only Through}.

\paragraph{Relationship with Temporal Logic}
Two of the three \Nec operators can be expressed in traditional
  temporal logic: 
  ${\onlyIf{A_1}{A_2}{A}}$
can be expressed  
 as 
 $A_1\ \wedge\ \Diamond A_2\ \longrightarrow\ A$, and
 $\onlyIfSingle{A_1}{A_2}{A}$
can be expressed  
 as $\ A_1\ \wedge\ \bigcirc A_2\ \longrightarrow\ A$
 (where $\Diamond$ denotes any future state,  and
 $\bigcirc$ denotes the next state).
 Critically, 
$\onlyThrough{A_1}{A_2}{A}$ cannot be encoded in temporal logics
  without ``nominals'' (explicit state references), because the state where $A$ 
 holds must be between the state where $A_1$ holds, and the state
 where $A_2$ holds; and this must be so on \emph{every} execution path
 from $A_1$ to  $A_2$ \cite{hybridLogic2021,nominal-seplogic2020}.
 TLA+, for example, cannot describe ``only through'' conditions
 \cite{tlabook}, but we have found ``only through'' conditions critical
 to our proofs.

\label{s:expressiveness}

\paragraph{The DOM}  
\label{ss:DOM}
This is the motivating example in \cite{dd},
dealing with a tree of DOM nodes: Access to a DOM node
gives access to all its \prg{parent} and \prg{children} nodes, with the ability to
modify the node's \prg{property} -- where  \prg{parent}, \prg{children} and \prg{property}
are fields in class \prg{Node}. Since the top nodes of the tree
usually contain privileged information, while the lower nodes contain
less crucial third-party information, we must be able to limit 
 access given to third parties to only the lower part of the DOM tree. We do this through a \prg{Proxy} class, which has a field \prg{node} pointing to a \prg{Node}, and a field \prg{height}, which restricts the range of \prg{Node}s which may be modified through the use of the particular \prg{Proxy}. Namely, when you hold a \prg{Proxy}  you can modify the \prg{property} of all the descendants of the    \prg{height}-th ancestors of the \prg{node} of that particular \prg{Proxy}.  We say that
\prg{pr} has \emph{modification-capabilities} on \prg{nd}, where \prg{pr} is
a  \prg{Proxy} and \prg{nd} is a \prg{Node}, if the \prg{pr.height}-th  \prg{parent}
of the node at \prg{pr.node} is an ancestor of \prg{nd}.

The specification \prg{DOMSpec} states that the \prg{property} of a node can only change if
some external object presently has 
access to a node of the DOM tree, or to some \prg{Proxy} with modification-capabilties
to the node that was modified.
\begin{lstlisting}[language = Chainmail, mathescape=true, frame=lines]
DOMSpec $\triangleq$ from nd : Node $\wedge$ nd.property = p  to nd.property != p
          onlyIf $\exists$ o.[ $\external {\prg{o}}$ $\wedge$ 
                       $( \  \exists$ nd':Node.[ $\access{\prg{o}}{\prg{nd'}}$ ]  $\vee$ 
                         $\exists$ pr:Proxy,k:$\mathbb{N}$.[$\, \access{\prg{o}}{\prg{pr}}$ $\wedge$ nd.parent$^{\prg{k}}$=pr.node.parent$^{\prg{pr.height}}$ ] $\,$ ) $\,$ ]
\end{lstlisting}

\paragraph{More examples}
In order to investigate \Nec's expressiveness,  
we used it for
examples provided in the literature. 
In Appendix \ref{s:expressiveness:appendix},
we compare with examples proposed by  \citeasnoun{FASE}, and \citeasnoun{VerX}.

\section{Proving Necessity}
\label{s:inference}

In this Section we provide a proof system for constructing 
proofs of the \Nec specifications defined in \S \ref{s:holistic-guarantees}.
As discussed in \S \ref{s:approach},
\sdNr[]{such proofs consist} of 
\jm[]{four parts}: 
\begin{description} 
\item[(Part 1)]
Proving Assertion Encapsulation (\S \ref{s:encaps-proof})
\item[(Part 2)]
Proving Per-Method \Nec specifications for a single internal method from the \funcSpec of that method (\S \ref{s:classical-proof})
\item[(Part 3)]
Proving Per-Step \Nec specifications by combining \sdNr[replaced per-call by per-method, as the prev sentence talks of per-method]{per-method} \Nec specifications (\S \ref{s:module-proof})
\item[(Part 4)]
Raising necessary conditions to construct proofs of \sdN{properties of} emergent behaviour (\S \ref{s:emergent-proof})
\end{description}

\sdr[replace concern by part, and reword]{Part 1 is, to a certain extent, orthogonal to the main aims of our work;
in this paper we propose a simple approach based on the type system, while also acknowledging that 
better solutions are possible.
For Parts 2-4, we  came up with the key ideas outlined in  \S \ref{s:approach}, which we
develop in more detail in \S \ref{s:classical-proof}-\S \ref{s:emergent-proof}.}

\subsection {Assertion Encapsulation}
\label{s:encaps-proof}

{
\sdNr[reword]{\Nec proofs 
often leverage the fact that some assertions cannot be invalidated unless some 
} internal (and thus known)
computation took place. 
{We refer to this property as \emph{Assertion Encapsulation}.}
}
\sdNr[reword, and banned model]{In this work, we define the property $M\ \vDash A'\ \Rightarrow\ \encaps{A}$, which states that
under the conditions described by assertion $A'$, the assertion $A$ is encapsulated by module $M$.
We  do not mandate how this property should be derived -- instead, we rely on a judgment 
$M\ \vdash A'\ \Rightarrow\ \encaps{A}$ provided by some external system. 
Thus, \Nec is parametric over the derivation of the encapsulation
     judgment; in fact, several ways to do that are possible \cite{TAME2003,ownEncaps,objInvars}. In Appendix~\ref{s:encap-proof} and
    Figure~\ref{f:asrt-encap}  we present a 
    rudimentary system that is sufficient to support our example
    proof. }

\subsubsection{Assertion Encapsulation Semantics}

\sdNr[somebody wrote "WHY THE FUCK IS A' the subject and A the aux assertion", and they were right, so I swapped A and A']{}
\sdNr[also, no "models the notion"]{}

\sdNr[]{As we said earlier,  an assertion $A$  is  encapsulated by a module $M$ under condition $A'$,
if in all possible states which arise from execution of module $M$ with any other external module $M'$, and which satisfy $A'$, 
the validity of $A$} 
\sdr[shortened, but all is here]{ can only be changed via computations internal to that module} -- \emph{i.e.},  via a call to
a method from $M$.
In \Loo, that means by
calls to objects defined in $M$ but \jm[removed: that are]{} accessible from the
outside.
\jm[sophia asked if caller should be external, and if that is in coq - it is not in coq, but is implicit given the operational semantics of \Loo and the definition of Arising]{}

\begin{definition}[Assertion Encapsulation]
\label{def:encapsulation}
An assertion $A$ is \emph{encapsulated} by module $M$ and assertion $A'$, written as\ \  $M\ \vDash A'\ \Rightarrow\ \encaps{A}$, \ \ if and only if
for all external modules $M'$, and all states $\sigma$, $\sigma'$
such that $\arising{M'}{M}{\sigma}$:

\begin{tabular}{lr}
$\;\;\;\;$- $\reduction{M'}{M}{\sigma}{\sigma'}$  & \rdelim\}{3}{4mm}[$\;\;\;\Rightarrow\;\;\;$  $\exists x,\ m,\ \overline{z}. (\ \satisfiesA{M}{\sigma}{\calls{\_}{x}{m}{\overline{z}} \wedge\ \internal{x}}\ )$] \\
$\;\;\;\;$- $\satisfiesA{M}{\sigma}{A \wedge  A'}$ \\
$\;\;\;\;$- $\satisfiesA{M}{\sigma' \triangleleft \sigma}{\neg A}$ \\
\end{tabular} 
\end{definition}

\noindent
\sdN{Note that this definition   uses adaptation, 
${\sigma' \triangleleft \sigma}$. The application of the adaptation operator is necessary
because we  interpret the assertion $A$ in the current state, $\sigma$, while we interpret the assertion $\neg A$ in 
the future state, $\sigma' \triangleleft \sigma$.
}

Revisiting the examples from \S~\ref{s:outline}, 
both \ModB and \ModC encapsulate   the \jm[]{equality of the \prg{balance} of an account to some value \prg{bal}: }
\sdr[]{This equality can only be invalidated} through calling  methods on internal objects.
\\
\strut \hspace{1cm}
$\ModB\ \vDash \prg{a}:\prg{Account}\ \Rightarrow\ \encaps{\prg{a.balance}=\prg{bal}}$
\\
\strut \hspace{1cm}
$\ModC\ \vDash \prg{a}:\prg{Account}\ \Rightarrow\ \encaps{\prg{a.balance}=\prg{bal}}$


\noindent 
\sdN{Moreover, the property that an object is only accessible from module-internal objects is encapsulated, that is, for all \prg{o}, and all modules $M$:}
\\
\strut \hspace{1cm}
$M\ \vDash \prg{o}:\prg{Object}\ \Rightarrow\ \encaps{\wrapped{\prg{o}}}$

\noindent
\sdN{This is so because any object which is only internally accessible} can become
  \jm[]{externally accessible} only via an internal call.

\sdN{In general},  code that does not contain 
calls to a \jm[]{given} module is guaranteed not to invalidate any assertions encapsulated by that module.
 Assertion encapsulation has been used in proof systems to {address}   the  {frame} problem
 \cite{objInvars,encaps}. 

\subsubsection{\sdN{Deriving} Assertion Encapsulation}


Our logic does not \sd{deal with, nor} rely on, the specifics of  how   encapsulation
\sdN{is derived}.
\sdN{Instead, it relies} on an encapsulation judgment and expects it to be sound:

\begin{definition}[Encapsulation Soundness]
\label{lem:encap-soundness}
A judgement of the form $\proves{M}{\givenA{A'}{\encaps{A}}}$  is\  \emph{sound}, \ if 
for all modules $M$, and assertions $A$ and $A'$, if \\

$\strut \hspace{1.5cm} \proves{M}{\givenA{A'}{\encaps{A}}}\ \ \ \ $ implies $\ \ \ \ \satisfies{M}{\givenA{A'}{\encaps{A}}}$.
\end{definition}

  \jm[I'm not sure this paragraph should exist. It's key, but does it belong here or add anything?
  \sdN{I modified the paragraph on types}]{}

\paragraph{Types for Assertion Encapsulation}
\label{types}
\sdNr[I have unified the two separate descriptions of the types system.]{}
\sdN{Even though the derivation of assertion encapsulation  is not the focus of this paper, 
for illustrative purposes, we will outline 
now a  very simple type system which supports such derivations:}
We assume that 
field declarations, method arguments
and method results are annotated with class names, and that classes may  
be annotated as \enclosed. A  \enclosed object  
\sdN{is not} accessed by external objects; that is, it is always \inside. 

The type system then checks 
that field assignments, method calls, and method returns adhere to these expectations,
and in particular, that objects of \enclosed type
are never returned from method bodies 
\sdN{-- this is a simplified version of the type system described in \cite{confined}.}
Because the type system is so simple, we do not include its formalization in the paper.
Note however, that the type system has one further implication: modules are typed 
in isolation, thereby implicitly prohibiting
method calls from internal objects to external objects. 

Based on this type system, we define a predicate $\intrnl{e}$, in Appendix \ref{s:encap-proof},
which asserts that any \sdN{objects read} during the evaluation of $e$ are internal.
Thus, any assertion that only involves $\intrnl{\_}$ expressions is encapsulated -- more in Appendix \ref{s:encap-proof}.

\subsection{Per-Method \Nec Specifications}
\label{s:classical-proof}
In this section we detail how we use \funcSpecs
to \jm[]{prove} per-method \Nec specifications  
of the form 
$$\onlyIfSingle{A_1\ \wedge\ x : C\ \wedge\ \calls{\_}{x}{m}{\overline{z}}}{A_2}{A}$$
where $C$ is a class, and $m$ a method in $C$.
\sdNr[removed "Thus, $A$ is a necessary precondition for reaching $A_2$ from $A_1$ via a method call $m$ to an object of class $C$."
we should know that by now]{}

The first key idea in \S \ref{s:approach}
is that if a precondition and a certain statement is \emph{sufficient}
to achieve a particular result, 
then the negation of that precondition
is \emph{necessary} to achieve the negation of the result after executing that statement.
Specifically, 
 \sdN{$\hoare{P}{s}{Q}$ implies} that $\neg P$ is a \emph{necessary precondition} for $\neg Q$ to 
hold following the execution of $\prg{s}$.

For the use in \funcSpecs, we define \emph{Classical assertions}, a subset of \SpecO, comprising only those 
assertions that are commonly present in other specification languages.
\jm[]{They} are restricted to expressions, class assertions, the usual connectives, negation, 
implication, and the usual quantifiers.

\begin{definition}
Classical assertions, $P$, $Q$, are defined as follows 

$
\begin{syntax}
\syntaxElement{P, Q} {} 
		{
		\syntaxline
				{e}
				{e : C}
				{P\ \wedge\ P}
				{P\ \vee\ P}
				{P\ \longrightarrow\ P}
				{\neg P}
				{\forall x.[P]}
				{\exists x.[P]}
		\endsyntaxline
		}
\endSyntaxElement\\
\end{syntax}
$
\label{f:classical-syntax}
\end{definition}

We assume that there exists some
proof system  that \sdN{derives} 
\sdN{functional} specifications of the form  $\proves{M}{\hoare{P}{\prg{s}}{Q}}$.
This implies that we can also have guarantees of  
$$M\ \vdash\ \hoare{P}{\prg{res} = x.m(\overline{z})}{Q}$$
That is,   
 the execution of $x.m(\overline{z})$ 
with the precondition $P$ results in a program state that 
satisfies postcondition $Q$, where the returned value is represented
by \prg{res} in $Q$.
\jm[]{We further assume that such a proof system is sound, i.e. that 
if $\proves{M}{\hoare{P}{\prg{res = x.m($\overline{z}$)}}{Q}}$, then 
for every program state $\sigma$ that satisfies $P$, the execution of the method call \prg{x.m($\overline{z}$)}
results in a program state satisfying $Q$.}
As we have previously discussed (see \S \ref{s:approach}), we build \Nec specifications
on top of \funcSpecs using the fact that 
validity of $\hoare{P}{\prg{res} = x.m(\overline{z})}{Q}$ implies that
$\neg P$ is a necessary pre-condition 
to $\neg Q$ being true after execution of ${\prg{res} = x.m(\overline{z})}$.

Proof  rules for per-method specifications are given in 
Figure \ref{f:classical->singlestep}. \julian{Note that the receiver $x$ in the rules
in \ref{f:classical->singlestep} is implicitly an internal object. This is because 
we only have access to internal code, and thus are only able to prove the validity 
of the associated Hoare triple.}
\sdNr[chopped: "\textsc{If1-Classical} and \textsc{If1-Inside}
raise \funcSpecs to \Nec specifications.
These are rules whose conclusion have the form Single-Step Only If." as it breaks the flow]{}

\begin{figure}[t]
\footnotesize
\begin{mathpar}
\infer
	{
	\proves{M}{\hoare
						{x : C \ \wedge\ P_1\ \wedge\ \neg P}
						{\prg{res} = x.m(\overline{z})}
						{\neg P_2}}
	}
	{
	\proves{M}{\onlyIfSingle
			{P_1\ \wedge\ x : C \wedge\ \calls{\_}{x}{m}{\overline{z}}}
			{P_2}
			{P}}
	}
	\quad(\textsc{If1-Classical})
	\and
\infer
	{
	\proves{M}{\hoare
						{x : C \ \wedge\ \neg P}
						{\prg{res} = x.m(\overline{z})}
						{\prg{res} \neq y}}
	}
	{
	\proves{M}{\onlyIfSingle{\wrapped{y}\ \wedge\ x : C \wedge\ \calls{\_}{x}{m}{\overline{z}}}{\neg \wrapped{y}}{P}}
	}
	\quad(\textsc{If1-Inside})
\end{mathpar}
\caption{Per-Method \Nec specifications}
\label{f:classical->singlestep}
\end{figure}

 \textsc{If1-Classical} states that  
if \sdNr[dropped "by some classical logic" because usually classical logics are first order logics]{}  the execution of $x.m(\overline{z})$, with precondition $P \wedge \neg P_1$,
\sdN{leads} to a state satisfying postcondition $\neg P_2$, then $P_1$ is a \emph{necessary} precondition to the 
resulting state satisfying $P_2$.

\jm[I removed the old bit that was here because I felt we had already explained this several times before.]{}

\jm[]{\textsc{If1-Inside} states} that if  \sdNr[simpliified] {the precondition $\neg P$} guarantees that the result of
the call $x.m(\overline{z})$  
 is not $y$, then $P$ is a necessary pre-condition to invalidate $\wrapped{y}$ by calling
$x.m(\overline{z})$.
This is sound, \sdNr[added details]{because the premise} of \textsc{If1-Inside} implies that  $P$ is
a necessary precondition for \sdN{the call $x.m(\overline{z})$ to return} an object $y$; this, in turn,
implies that    $P$ is a necessary precondition for the call $x.m(\overline{z}$) to 
result in an external object gaining access to $y$.
\sdN{The latter implication is valid} because
\jm[]{the} \sdN{rule is applicable only to external states semantics, which means that}
the call  $x.m(\overline{z})$ is a call from an external object to
some internal object $x$. \sdN{Namely,} there are only four ways
an object $o$ might gain access to another object $o'$: 
\sdNr[replaced $x$ and $y$ by $o$ and $o'$, because $x$ and $y$ are taken from earlier.]{}
(1) $o'$ is created by $o$ as the result of a \prg{new} expression, 
(2) $o'$ is written to some field of $o$, 
(3) $o'$ is passed to $o$ as an argument to a method call on $o$,
or (4) $o'$ is returned to $o$ as the result of a method call from an object $o''$ that has access to $o'$.
\sdN{The rule \textsc{If1-Inside}} is only concerned with 
effects on program state resulting from a method call to some internal object, and thus (1) and (2) need not be considered as 
neither object creation or field writes may result in an external object gaining access \sdNr[was from]{to}  
an
\sdNr[used to say "an internal object", but I think this is too weak]{object that is only internally accessible.}
Since we are only concerned with describing how internal objects grant access to external objects,
our restriction on external method calls within internal code prohibits (3) from occuring. Finally,
(4) is described by \textsc{If1-Inside}.
In further work we plan to weaken the restriction on external method calls, and will  
strengthen this rule.
Note that \textsc{If1-Inside}  is essentially  a specialized version of \textsc{If1-Classical}
for the $\wrapped{\_}$ predicate. Since $\wrapped{\_}$ is not a classical
assertion, we cannot use \funcSpecs to reason about necessary conditions
for invalidating $\wrapped{\_}$.

\subsection{Per-Step \Nec Specifications}
\label{s:module-proof}

\begin{figure}[thb]
\footnotesize
\begin{mathpar}
\infer
	{
	\left[\infer{\textit{for all}\ \ C \in dom(M)\ \ \textit{and}\ \  m \in M(C).\prg{mths}, \\\\
				[\proves{M}{\onlyIfSingle
								{A_1\ \wedge\ x : C\ \wedge\ \calls{\_}{x}{m}{\overline{z}}}
								{A_2}
								{A_3}}]}{}\right]\\
	\proves{M}{A_1\ \longrightarrow\ \neg A_2}\\
	\proves{M}{\givenA{A_1}{\encaps{A_2}}}
	}
	{
	M\ \vdash\ \onlyIfSingle{A_1}{A_2}{A_3}
	}
	\quad(\textsc{If1-Internal})
	\and
\infer
	{
	\proves{M}{A_1 \longrightarrow A_1'}\\
	\proves{M}{A_2 \longrightarrow A_2'}\\
	\proves{M}{A_3' \longrightarrow A_3}\\
	\proves{M}{\onlyIfSingle{A_1'}{A_2'}{A_3'}}
	}
	{\proves{M}{\onlyIfSingle{A_1}{A_2}{A_3}}}
	\quad(\textsc{If1-$\longrightarrow$})
	\and
\infer
	{
	\proves{M}{\onlyIfSingle{A_1}{A_2}{A\ \vee\ A'}} \\
	\proves{M}{\onlyThrough{A'}{A_2}{\prg{false}}}
	}
	{\proves{M}{\onlyIfSingle{A_1}{A_2}{A}}}
	\quad(\textsc{If1-$\vee$E})
	\and
\infer
	{
	\forall y,\; \proves{M}{\onlyIfSingle{([y / x]A_1)}{A_2}{A}}
	}
	{\proves{M}{\onlyIfSingle{\exists x. [A_1]}{A_2}{A}}}
	\quad(\textsc{If1-$\exists_1$})
\end{mathpar}
\caption{Selected rules for Single-Step \emph{Only If}}
\label{f:only-if-single}
\end{figure}

{The second key idea in \S \ref{s:approach}} allows us to
\sdNr[rephrase]{leverage several per-method \Nec specifications 
to obtain one per-step \Nec specification:}
Namely, if an assertion is encapsulated, and all methods within the internal module
\sdN{require} the same condition to the \sdN{invalidation} of that assertion, then 
this condition is a necessary, program-wide, single-step condition   to the invalidation of that assertion.
\sdr[replaced necessary pre-condition by necessary condition]{}

\sdr[] {In this section} we present a selection of the rules whose conclusion is of the form Single Step Only If in Fig. \ref{f:only-if-single}.
 The full rule set can be found in Fig. \ref{f:app:only-if-single}.

\textsc{If1-Internal} 
 lifts a \sdN{set of} per-method \Nec \jm[]{specifications} to a per-step \Nec specification.
Any \Nec specification which is satisfied for \jm[]{all} method
calls sent to any object in a module, is satisfied for \emph{any step}, even
an external step, provided that the effect involved, \ie going from $A_1$ states to
$A_2$ states, is encapsulated.

 The remaining rules are more standard, and are reminiscent of the Hoare logic rule of consequence.
\jm[]{We present a few of the more interesting rules here}:
 
The  rule for implication (\textsc{If1-$\longrightarrow$}) may strengthen
 properties of either the starting or ending state, or 
weaken the necessary precondition. 
%
%
The disjunction
elimination rule (\textsc{IF1-$\vee$E}) mirrors typical disjunction elimination
rules, with a variation stating that if it is not possible  to reach 
the end state from one branch of the disjunction, then we can eliminate 
that branch. 

Two rules support existential elimination on the left hand side. 
\textsc{If1-$\exists_1$} states that if any single step of execution starting
from a state satisfying $[y/x]A_1$ for all possible $y$, reaching some state satisfying
$A_2$ has $A$ as a necessary precondition, it follows that any single step execution
starting in a state where such a $y$ exists, and ending in a state satisfying $A_2$,
must have $A$ as a necessary precondition.  \sdN{The other rule  can be found in Fig. \ref{f:app:only-if-single}.}

\begin{figure}[t]
\footnotesize
\begin{mathpar}
\infer
	{\proves{M}{\onlyIfSingle{A}{\neg A}{A'}}}
	{
	\proves{M}{\onlyThrough{A}{\neg A}{A'}}
	}
	\quad(\textsc{Changes})
	\and
\infer
	{
	\proves{M}{\onlyThrough{A_1}{A_2}{A_3}} \\\\
	\proves{M}{\onlyThrough{A_1}{A_3}{A}}
	}
	{\proves{M}{\onlyThrough{A_1}{A_2}{A}}}
	\quad(\textsc{Trans$_1$})
	\and
\infer
	{
	\proves{M}{\onlyThrough{A_1}{A_2}{A_3}} \\\\
	\proves{M}{\onlyThrough{A_3}{A_2}{A}}
	}
	{\proves{M}{\onlyThrough{A_1}{A_2}{A}}}
	\quad(\textsc{Trans$_2$})
	\and
\infer
	{
	\proves{M}{\onlyIf{A_1}{A_2}{A}}
	}
	{\proves{M}{\onlyThrough{A_1}{A_2}{A}}}
	\quad(\textsc{If})
	\and
\infer
	{}
	{\proves{M}{\onlyThrough{A_1}{A_2}{A_2}}}
	\quad(\textsc{End})
\end{mathpar}
\caption{\scd{Selected rules for} \emph{Only Through} -- the rest can be found in Figure \ref{app:f:only-through-full}}
\label{f:only-through}
\end{figure}
\begin{figure}[t]
\footnotesize
\begin{mathpar}
\infer
	{
	\proves{M}{\onlyThrough{A_1}{A_2}{A_3}} \\
	\proves{M}{\onlyIf{A_1}{A_3}{A}}
	}
	{\proves{M}{\onlyIf{A_1}{A_2}{A}}}
	\quad(\textsc{If-Trans)}
	\and
\infer
	{}
	{\proves{M}{\onlyIf{x\ :\ C}{\neg\ x\ :\ C}{\false}}}
	\quad(\textsc{If-Class})
	\and	
\infer
	{}
	{\proves{M}{\onlyIf{A_1}{A_2}{A_1}}}
	\quad(\textsc{If-Start})
\end{mathpar}
\caption{\scd{Selected rules for} \emph{Only If} -- the rest can be found in Figure \ref{app:f:only-if-full}}
\label{f:only-if}
\end{figure}

%
\subsection{Emergent \Nec Specifications}
\label{s:emergent-proof}

The third key idea in \S \ref{s:approach}  allows us to
\sdNr[rephrase]{leverage several per-step \Nec specifications to 
obtain  multiple-step \Nec specifications, and thus enables the description of the module's emergent behaviour.}
We
\sdN{combine}   per-step \Nec specifications into  
multiple-step \Nec specifications, as well as several  multiple step \Nec specifications into further multiple step \Nec specifications.

 Figure \ref{f:only-through}  {presents} some of the rules with conclusion \emph{Only Through}, while Figure \ref{f:only-if}
provides some of the rules with conclusion \emph{Only If}. 
The \jm[I couldn't read Susan's comment on this]{full rules} can be found in Appendix \ref{a:necSpec}.


\textsc{Changes}, in Figure \ref{f:only-through}, 
{states that 
if   \jm[re:susan's coment, is this a question of primes vs subscripts?]{$A'$} is a necessary condition for the satisfaction of $A$ to change  in \emph{one} step, then
it is also a  necessary condition for the satisfaction of $A$ to change  in \emph{any number of} steps.
This is sound, because if  the satisfaction of some assertion changes over time, then 
 there must be some specific intermediate state where that change occurred.}
\textsc{Changes} is an important 
 enabler for proofs of emergent properties:
\jm[]{Since \Nec specifications}  are concerned with \sdN{necessary conditions for} change,
\sdN{their} proofs typically hinge around such \sdN{necessary conditions} for certain properties  
to change. For example,
under what conditions \sdN{may} our account's balance decrease? 

It \jm[]{might} seem natural \jm[]{that} \textsc{Changes} \jm[]{had} the more
general form:
$$\infer{\proves{M}{\onlyIfSingle{A_1}{A_2}{A_3}}}{\proves{M}{\onlyThrough{A_1}{A_2}{A_3}}}\quad(\textsc{(ChangesUnsound)}$$
\sdN{\textsc{(ChangesUnsound)} is not sound because} the conclusion  of the rule describes 
transitions from a state satisfying $A_1$ to one satisfying $A_2$  \sdN{which may occur} occur over several steps,
\sdN{while the premise describes a transition that takes place} over one single step.
\sdN{Such a concern does not apply to \textsc{(Changes)},}  because 
a change in satisfaction for a specific assertion (\ie $A$ to $\neg A$) can \emph{only} take place in a single step.

\textsc{Trans}$_1$ and \textsc{Trans}$_2$  {are \jm[]{rules about transitivity.}}
\jm[]{They} {state} that necessary conditions to reach intermediate states or 
proceed from intermediate states are themselves necessary intermediate states. 
\jm[moved]{Any \emph{Only If} specification entails the corresponding
 \emph{Only Through} specification (\textsc{If}).}
\jm[]{Finally, \textsc{End} states that the ending condition is 
a necessary intermediate condition.}

\emph{Only If} also includes a rule for transitivity (\textsc{If-Trans}), but 
since the necessary condition must be true in the beginning state,
there is only a single rule. \textsc{If-Class} expresses that
an object's class never changes.
Finally, any starting condition is
itself a necessary precondition (\textsc{If-Start}).

\subsection{Soundness of the \Nec Logic}

\label{s:soundness}

\begin{theorem}[Soundness]
\label{thm:soundness}
Assuming a sound \SpecO proof system, $\proves{M}{A}$, and
a sound encapsulation inference system, $\proves{M}{\givenA{A}{\encaps{A'}}}$,
 and  that on top of these systems we built
 the \Nec logic according to the rules in Figures \ref{f:classical->singlestep},  \ref{f:only-if-single}, 
 \ref{f:only-through},  and \ref{f:only-if},   then, for    all modules $M$, and all \Nec specifications  $S$:
 
 $$\proves{M}{S}\ \ \ \ \ \ \ \mbox{implies}\ \ \ \ \ \  \ \ \ \satisfies{M}{S}$$
\end{theorem}

\begin{proof}
by induction on the derivation of $\proves{M}{S}$.
\end{proof}

Theorem. \ref{thm:soundness} demonstrates 
 that the   \Nec logic is sound with respect to the semantics of \Nec specifications.
 The \Nec logic parametric wrt to the algorithms for proving validity of assertions
 $\proves{M}{A}$, and 
 assertion encapsulation ($\proves{M}{\givenA{A}{\encaps{A'}}}$), and is sound
 provided that these two proof systems are sound.


The mechanized  proof of Theorem \ref{thm:soundness} in Coq 
can be found in the associated artifact \cite{necessityCoq2022}. 
The   Coq formalism deviates slightly from the system as
presented here,  mostly in the formalization of the 
\SpecO language. The Coq version of \SpecO restricts variable usage to expressions, and allows only addresses to 
be used as part of non-expression syntax. 
For example, in the Coq formalism
we can write assertions like $x.f==\prg{this}$ and
$x==\alpha_y$ and  $\access{\alpha_x}{\alpha_y}$, but we cannot write assertions 
like $\access{x}{y}$, where $x$ and $y$ are variables, and $\alpha_x$ and $\alpha_y$ are
addresses.
The reason for this restriction in the Coq formalism is to avoid spending 
significant effort encoding variable
renaming and substitution, a well-known difficulty for languages such as Coq. 
This restriction does not affect the expressiveness of 
our  Coq formalism: we are
able to express assertions such as $\access{x}{y}$, by using addresses and introducing equality expressions 
to connect variables to address, \ie
 $\access{\alpha_x}{\alpha_y} \wedge \alpha_x == x \wedge \alpha_y == y$.
 \jm[]{The Coq formalism makes use of the \prg{CpdtTactics} \cite{chlipala}
 library of tactics to discharge some proofs.}

\section{Proving that \ModC satisifes \SrobustB}
\label{s:examples}
We now revisit our example from  \S  \ref{s:intro} and \S \ref{s:outline},
and outline a proof that \ModC satisfies \SrobustB. 
A {summary} of this proof has already been discussed in \S \ref{s:approach}.
 A more complex variant of this example can be found in Appendix \ref{app:examples}.
 \sdN{It demonstrates dealing with modules consisting of several classes some of which are confined, and
 which use ghost fields
 defined through functions; it also demonstrates proofs of assertion encapsulation of assertions 
 which involve reading the values of 
 several fields.} \julian{Mechanised versions of the proofs in both this Section, and Appendix \ref{app:examples}
 can be found in the associated Coq artifact \cite{necessityCoq2022} in \prg{simple\_bank\_account.v} and \prg{bank\_account.v} respectively.}
 
 \jm[todo: add reference so \prg{simple\_bank\_account.v} \& \prg{bank\_account.v}]{}
 
Recall that an \prg{Account} includes 
 at least a  field  (or ghost field)  called \prg{balance}, and a method called \prg{transfer}. 

 We first rephrase 
\SrobustB to use the $\wrapped{\_}$ predicate.
\begin{lstlisting}[language=Chainmail, mathescape=true, frame=lines]
$\SrobustB$ $\triangleq$ from a:Account $\wedge$ a.balance=bal 
          to a.balance < bal onlyIf $\neg\wrapped{\prg{a.pwd}}$
\end{lstlisting}

We next revisit the   \funcSpec from \S \ref{s:bank} and derive the following 
\prg{PRE}- and \prg{POST}-conditions. The first two pairs of \prg{PRE}-, \prg{POST}-conditions correspond to the first two \prg{ENSURES}
clauses from \S \ref{s:bank}, while the next two pairs correspond to the \prg{MODIFIES}-clause. The current expression in terms
of \prg{PRE}- and \prg{POST}-conditions is weaker than the one in \S \ref{s:bank}, and is not modular, but is
sufficient for proving adherence to  \SrobustB.

\begin{lstlisting}[mathescape=true, frame=lines, language=Chainmail]
$\SclassicP$  $\triangleq$
   method transfer(dest:Account, pwd':Password) -> void  
      (PRE: this.balance$=$bal1 $\wedge$ dest.balance$=$bal2 $\wedge$ this.pwd$=$pwd' $\wedge$ this$\neq$dest
       POST: this.balance=bal1-100 $\wedge$ dest.balance=bal2+100)
      (PRE: this.balance$=$bal1 $\wedge$ dest.balance$=$bal2 $\wedge$ (this.pwd$\neq$pwd' $\vee$ this$=$dest)
       POST: this.balance=bal1 $\wedge$ dest.balance=bal2)
      (PRE: a:Account $\wedge$ a.balance$=$bal $\wedge$ a$\neq$this $\wedge$ a$\neq$dest 
       POST: a.balance=bal)          
      (PRE: a:Account $\wedge$ a.pwd$=$pwd1  
       POST: a.pwd=pwd1)         
\end{lstlisting}



\subsection{Part 1: Assertion Encapsulation}
\label{s:BA-encap}
The first part of the proof demonstrates that the \prg{balance}, \prg{pwd}, and external accessibility to the password are 
encapsulated properties. That is, for the \prg{balance} to change (i.e. for \prg{a.balance = bal} to be invalidated),  
or for the encapsulation of \prg{a.pwd} to be broken (ie for a transition from ${\wrapped{\prg{a,pwd}}}$ to $\neg {\wrapped{\prg{a.pwd}}}$),
internal computation is required. 

We use \sdNr[used to say "a conservative approach to an encapsulation system" but "conservative" here means sound]{a simple encapsulation system}, detailed in \jm[]{Appendix} \ref{s:encap-proof}, 
and provide the proof steps below.
\textbf{\prg{aEnc}} and \textbf{\prg{balanceEnc}} state that 
\jm[]{\prg{a} and \prg{a.balance}} satisfy \sdN{the \textsc{Enc$_e$} predicate. That is, if any objects' contents are to be
looked up during execution of these expressions, then these objects are internal.}
\textsc{Enc}$_e$(\prg{a}) holds because no object's contents is looked up,
while \textsc{Enc}$_e$(\prg{a.balance}) holds because \prg{balance} is a field of \prg{a}, and \prg{a} is
internal.
\\
\begin{figure}[h]
\begin{proofexample}
\proofsteps{\prg{BalEncaps}}
	{\begin{proofexample}
		\proofsteps{\prg{aEnc}}
			{\proofstepwithrule
			{$\proves{\ModC}{\givenA{\prg{a:Account $\wedge$ a.balance=bal}}{\intrnl{\prg{a}}}}$}
				{by \textsc{Enc$_e$-Obj}}
		}
		\endproofsteps
	\end{proofexample}
		}
	{\begin{proofexample}
		\proofsteps{\prg{balanceEnc}}
			{\proofstepwithrule
			{$\proves{\ModC}{\givenA{\prg{a:Account $\wedge$ a.balance=bal}}{\intrnl{\prg{a.balance}}}}$}
				{by \prg{aEnc} and \textsc{Enc-Field}}
		}
		\endproofsteps
	\end{proofexample}
		}
	{\begin{proofexample}
		\proofsteps{\prg{balEnc}}
			{\proofstepwithrule
			{$\proves{\ModC}{\givenA{\prg{a:Account $\wedge$ a.balance=bal}}{\intrnl{\prg{bal}}}}$}
				{by \textsc{Enc$_e$-Int}}
		}
		\endproofsteps
	\end{proofexample}
		}
		{\proofstepwithrule
			{
			$\proves{\ModC}{\givenA{\prg{a:Account $\wedge$ a.balance=bal}}{\encaps{\prg{a.balance=bal}}}}$
			}{by \prg{balanceEnc}, \prg{balEnc}, \textsc{Enc-Eq}, and \textsc{Enc-=}}}
\endproofsteps
\end{proofexample}
\end{figure}

\sdN{Moreover},  \textbf{\prg{balEnc}} states that \prg{bal} satisfies \sdN{the \textsc{Enc}$_e$ predicate}
-- it is an integer, and no object look-up is involved in its calculation.
\textbf{\prg{balanceEnc}} and \textbf{\prg{balEnc}} combine to prove that the assertion \prg{a.balance = bal} is encapsulated --
\sdN{only internal object lookups are involved in the validity of that assertion, and therefore only 
internal computation may cause it to be invalidated.}

\sdN{Using similar reasoning, we} prove that \prg{a.pwd} {is encapsulated} (\textbf{\prg{PwdEncaps}}), and
that \wrapped{\prg{a.pwd}} {is encapsulated} (\textbf{\prg{PwdInsideEncaps}}). 
\sdNr[I chopped the below "That is, if only internal objects have access
to an account's \prg{pwd}, then only internal computation may grant  access to \prg{pwd} 
 to an external object." as it only repeats the definition.]{}

\begin{figure}[h]
\begin{proofexample}
\proofsteps{\prg{PwdEncaps}}
		{\proofstepwithrule
			{
			$\proves{\ModC}{\givenA{\prg{a:Account}}{\encaps{\prg{a.pwd=p}}}}$
			}{by \textsc{Enc$_e$-Obj}, \textsc{Enc-Field}, and \textsc{Enc-Eq}}}
\endproofsteps
\end{proofexample}
\\\begin{proofexample}
\proofsteps{\prg{PwdInsideEncaps}}
		{\proofstepwithrule
			{
			$\proves{\ModC}{\givenA{\prg{a:Account}}{\encaps{\wrapped{\prg{a.balance}}}}}$
			}{by \textsc{Enc-Inside}}}
\endproofsteps
\end{proofexample}
\end{figure}

\sdNr[removing references to Hoare logic.]{}

\subsection{Part 2: Per-Method \Nec Specifications}
\label{s:BA-classical}
Part 2 proves necessary preconditions for each method in 
the module interface. 
\sdN{We employ the rules from  \S \ref{s:classical-proof} which describe how to derive 
necessary preconditions from \funcSpecs.
}

\textbf{\prg{SetBalChange}} uses a 
 \funcSpec and a rule of consequence
to prove that the \prg{set} method in \prg{Account}
never modifies the \prg{balance}. We then use \textsc{If1-Classical}
and our \Nec logic to prove that if it ever did change (a logical absurdity),
then \prg{transfer} must have been called.
\begin{figure}[htb]
{
	\begin{proofexample}
		\proofsteps{SetBalChange}
			{\proofstepwithrule
				{\hoareEx
						{a, a$^\prime$:Account $\wedge$ a$^\prime$.balance=bal}
						{a.set(\_, \_)}
						{a$^\prime$.balance = bal}
						}
					{by \funcSpec}
			}
			{\proofstepwithrule
				{\hoareEx
						{a, a$^\prime$:Account $\wedge$ a$^\prime$.balance = bal $\wedge$ $\neg$ false}
						{a.set(\_, \_)}
						{$\neg$ a$^\prime$.balance < bal }
						}
					{by 
					\sdN{rule of consequence}} 
			}
			{\proofstepwithrule
				{\onlyIfSingleExAlt
						{a, a$^\prime$:Account $\wedge$ a$^\prime$.balance=bal $\wedge$ $\calls{\_}{\prg{a}}{\prg{set}}{\prg{\_, \_}}$}
						{a$^\prime$.balance < bal}
						{false}
						}
					{by \textsc{If1-Classical}}
			}
			{\proofstepwithrule
				{\onlyIfSingleExAlt
						{a, a$^\prime$:Account $\wedge$ a$^\prime$.balance=bal $\wedge$ $\calls{\_}{\prg{a}}{\prg{set}}{\prg{\_, \_}}$}
						{a$^\prime$.balance < bal}
						{$\calls{\_}{\prg{a}^\prime}{\prg{transfer}}{\prg{\_, a$^\prime$.pwd}}$}
						}
					{by \textsc{Absurd} and \textsc{If1-}$\longrightarrow$}
			}
		\endproofsteps
	\end{proofexample}
}
\end{figure}

\sdN{Similarly,} in \textbf{\prg{SetPwdLeak}}  we employ \funcSpecs 
to prove that a method does not leak access to some data (in this case the \prg{pwd}).
Using \textsc{If1-Inside}, we reason that since the return value of \prg{set} is
\prg{void}, and \prg{set} is prohibited from making external method calls,
no call to \prg{set} can result in an object (external or otherwise) gaining access to the \prg{pwd}.
\begin{figure}[htb]
{
	\begin{proofexample}
		\proofsteps{SetPwdLeak}
			{\proofstepwithrule
				{\hoareEx
						{a:Account $\wedge$ a$^\prime$:Account $\wedge$ a.pwd == p}
						{\prg{res}=a$^\prime$.set(\_, \_)}
						{res != pwd}
						}
					{by \funcSpec}
			}
			{\proofstepwithrule
				{\hoareEx
						{a:Account $\wedge$ a$^\prime$:Account $\wedge$ a.pwd == p $\wedge$ $\neg$ false}
						{\prg{res}=a$^\prime$.set(\_, \_)}
						{res != p}
						}
					{by \sdN{rule of consequence}} 
			}
			{\proofstepwithrule
				{\onlyIfSingleExAlt
						{$\wrapped{\prg{pwd}}$ $\wedge$ a, a$^\prime$:Account $\wedge$ a.pwd=p $\wedge$ $\calls{\_}{\prg{a}^\prime}{\prg{set}}{\_, \_}$}
						{$\neg \wrapped{\_}$}
						{false}
						}
					{by \textsc{If1-Inside}}
			}
		\endproofsteps
	\end{proofexample}
	}
	\end{figure}
	
In the same manner as \textbf{\prg{SetBalChange}} and \textbf{\prg{SetPwdLeak}}, we also prove
\textbf{\prg{SetPwdChange}}, \textbf{\prg{TransferBalChange}}, \textbf{\prg{TransferPwdLeak}}, and \textbf{\prg{TransferPwdChange}}. We provide their 
statements, but omit their proofs.
\begin{figure}[htb]
{
	\begin{proofexample}
		\proofsteps{SetPwdChange}
			{\proofstepwithrule
				{\onlyIfSingleExAlt
						{a, a$^\prime$:Account $\wedge$ a$^\prime$.pwd=p $\wedge$ $\calls{\_}{\prg{a}}{\prg{set}}{\prg{\_, \_}}$}
						{$\neg$ a.pwd = p}
						{$\calls{\_}{\prg{a}^\prime}{\prg{set}}{\prg{a$^\prime$.pwd, \_}}$}
						}
					{by \textsc{If1-Classical}}
			}
		\endproofsteps
	\end{proofexample}
}
{
	\begin{proofexample}
		\proofsteps{TransferBalChange}
			{\proofstepwithrule
				{\onlyIfSingleExAlt
						{a, a$^\prime$:Account $\wedge$ a$^\prime$.balance=bal $\wedge$ $\calls{\_}{\prg{a}}{\prg{transfer}}{\prg{\_, \_}}$}
						{a$^\prime$.balance < bal}
						{$\calls{\_}{\prg{a}^\prime}{\prg{transfer}}{\prg{\_, a$^\prime$.pwd}}$}
						}
					{by \textsc{If1-Classical}}
			}
		\endproofsteps
	\end{proofexample}
}
{
	\begin{proofexample}
		\proofsteps{TransferPwdLeak}
			{\proofstepwithrule
				{\onlyIfSingleExAlt
						{$\wrapped{\prg{pwd}}$ $\wedge$ a, a$^\prime$:Account $\wedge$ a.pwd=p $\wedge$ $\calls{\_}{\prg{a}^\prime}{\prg{transfer}}{\_, \_}$}
						{$\neg \wrapped{\_}$}
						{false}
						}
					{by \textsc{If1-Inside}}
			}
		\endproofsteps
	\end{proofexample}
	}
	{
	\begin{proofexample}
		\proofsteps{TransferPwdChange}
			{\proofstepwithrule
				{\onlyIfSingleExAlt
						{a, a$^\prime$:Account $\wedge$ a$^\prime$.pwd=p $\wedge$ $\calls{\_}{\prg{a}}{\prg{transfer}}{\prg{\_, \_}}$}
						{$\neg$ a.pwd = p}
						{$\calls{\_}{\prg{a}^\prime}{\prg{set}}{\prg{a$^\prime$.pwd, \_}}$}
						}
					{by \textsc{If1-Classical}}
			}
		\endproofsteps
	\end{proofexample}
	}
\end{figure}

\subsection{Part 3: Per-Step \Nec Specifications}
Part 3 builds upon the proofs of Parts 1 and 2 to 
construct proofs of necessary preconditions, not for single method execution, 
but \sdN{for} any single execution step. That is, a proof that for
\emph{any} single step in program execution, \sdNr[removed "potentially dangerous", since such proof
steps apply whether the changes are potentially dangerous or not]{} changes
in program state require specific preconditions.
\begin{figure}[htb]
{
	\begin{proofexample}
		\proofsteps{BalanceChange}
			{\proofstepwithrule
				{\onlyIfSingleExAlt
						{a:Account $\wedge$ a.balance=bal}
						{a.balance < bal}
						{$\calls{\_}{\prg{a}}{\prg{transfer}}{\prg{\_, a.pwd}}$}
						}
					{by \textbf{\prg{BalEncaps}}, \textbf{\prg{SetBalChange}}, \textbf{TransferBalChange}, and \textsc{If1-Internal}}
			}
		\endproofsteps
	\end{proofexample}
	}{
	\begin{proofexample}
		\proofsteps{PasswordChange}
			{\proofstepwithrule
				{\onlyIfSingleExAlt
						{a:Account $\wedge$ a.pwd=p}
						{$\neg$ (a.pwd = p)}
						{$\calls{\_}{\prg{a}}{\prg{set}}{\prg{a.pwd, \_}}$}
						}
					{by \textbf{\prg{PwdEncaps}}, \textbf{\prg{SetPwdChange}}, \textbf{TransferPwdChange}, and \textsc{If1-Internal}}
			}
		\endproofsteps
	\end{proofexample}
	}{
	\begin{proofexample}
		\proofsteps{PasswordLeak}
			{\proofstepwithrule
				{\onlyIfSingleExAlt
						{a:Account $\wedge$ a.pwd=p $\wedge$ $\wrapped{\prg{p}}$}
						{$\neg$ $\wrapped{\prg{p}}$}
						{false}
						}
					{by \textbf{\prg{PwdInsideEncaps}}, \textbf{\prg{SetPwdLeak}}, \textbf{TransferPwdLeak}, and \textsc{If1-Internal}}
			}
		\endproofsteps
	\end{proofexample}
}
\end{figure}

\subsection{Part 4: Emergent \Nec Specifications}
Part 4 raises necessary preconditions for single execution steps proven in Part 3 to 
the level of an arbitrary number of execution steps in order to prove specifications of emergent behaviour.
The proof of \SrobustB takes the following form:
\begin{description}
\item [(1)]
If the balance of an account decreases, then
by \prg{BalanceChange} there must have been a call
to \prg{transfer} in \jm[]{\prg{Account}} with the correct password.
\item [(2)]
If there was a call where the \prg{Account}'s password 
was used, then there must have been an intermediate program state
when some external object had access to the password.
\item [(3)]
Either that password was the same password as in the starting 
program state, or it was different:
\begin{description}
\item [(Case A)]
If it is the same as the initial password, then since by \prg{PasswordLeak}
it is impossible to leak the password, it follows that some external object 
must have had access to the password initially.
\item [(Case B)]
If the password is different from the initial password, 
then there must have been an intermediate program state when it 
changed. By \prg{PasswordChange} we know that this must have occurred
by a call to \prg{set} with the correct password. Thus,
there must be a some intermediate program state where the initial
password is known. From here we proceed by the same reasoning 
as \textbf{(Case A)}.
\end{description}
\end{description}
\begin{figure}[htb]
\begin{proofexample}
\proofsteps{\SrobustB}
	{\proofstepwithrule{\onlyThroughExAlt
				{a:Account $\wedge$ a.balance=bal}
				{a.balance < bal}
				{$\calls{\_}{\prg{a}}{\prg{transfer}}{\_, \prg{a.pwd}}$}
				}
			{by \textsc{Changes} and \prg{BalanceChange}}}
	{\proofstepwithrule{\onlyThroughExAlt
				{a:Account $\wedge$ a.balance=bal}
				{b.balance(a) < bal}
				{$\neg$$\wrapped{\prg{a.pwd}}$}
				}
			{by $\longrightarrow$, \textsc{Caller-Ext}, and \textsc{Calls-Args}}}
	{\proofstepwithrule{\onlyThroughEx
				{a:Account $\wedge$ a.balance=bal $\wedge$ a.pwd=p}
				{a.balance < bal}
				{$\neg$$\wrapped{\prg{a.pwd}}$ $\wedge$ (a.pwd=p $\vee$ a.pwd != p)}
				}
			{by $\longrightarrow$ and \textsc{Excluded Middle}}}
	{\proofstepwithrule{\onlyThroughEx
				{a:Account $\wedge$ a.balance=bal $\wedge$ a.pwd=p}
				{a.balance < bal}
				{($\neg$$\wrapped{\prg{a.pwd}}$ $\wedge$ a.pwd=p) $\vee$\\
				($\neg$$\wrapped{\prg{a.pwd}}$ $\wedge$ a.pwd != p)}
				}
			{by $\longrightarrow$}}
	{\proofstepwithrule{\onlyThroughExAlt
				{a:Account $\wedge$ a.balance=bal $\wedge$ a.pwd=p}
				{a.balance < bal}
				{$\neg$$\wrapped{\prg{p}}$ $\vee$
				a.pwd != p}
				}
			{by $\longrightarrow$}}
	{
	\begin{proofexample}
	\proofsteps{Case A ($\neg\wrapped{\prg{p}}$)}
			{\proofstepwithrule
				{\onlyIfExAlt
					{a:Account $\wedge$ a.balance=bal $\wedge$ a.pwd=p}
					{$\neg$$\wrapped{\prg{p}}$}
					{$\wrapped{\prg{p}}\ \vee \neg\wrapped{\prg{p}}$}
					}
				{by \textsc{If-}$\longrightarrow$ and \textsc{Excluded Middle}}}
			{\proofstepwithrule{\onlyIfExAlt
					{a:Account $\wedge$ b:Bank $\wedge$ b.balance(a)=bal $\wedge$ a.password=pwd}
					{$\neg$$\wrapped{\prg{p}}$}
					{$\neg\wrapped{\prg{p}}$}
					}
				{by $\vee$E and \prg{PasswordLeak}}}
	\endproofsteps
	\end{proofexample}
	}
	{
	\begin{proofexample}
	\proofsteps{Case B (\prg{a.pwd != p})}
		{\proofstepwithrule{\onlyThroughExAlt
					{a:Account $\wedge$ b:Bank $\wedge$ b.balance(a)=bal $\wedge$ a.password=pwd}
					{a.pwd != p}
					{$\calls{\_}{\prg{a}}{\prg{set}}{\prg{p}, \_}$}
					}
				{by \textsc{Changes} and \textsc{PasswordChange}}}
		{\proofstepwithrule{\onlyThroughExAlt
					{a:Account $\wedge$ a.balance=bal $\wedge$ a.pwd=p}
					{a.pwd != p}
					{$\neg\wrapped{\prg{p}}$}
					}
				{by $\vee$E and \prg{PasswordLeak}}}
		{\proofstepwithrule{\onlyIfExAlt
					{a:Account $\wedge$ a.balance=bal $\wedge$ a.pwd=p}
					{a.pwd != p}
					{$\neg\wrapped{\prg{p}}$}
					}
				{by \textbf{Case A} and \textsc{Trans}}}
	\endproofsteps
	\end{proofexample}
	}
	{\proofstepwithrule{\onlyIfExAlt
				{a:Account $\wedge$ a.balance=bal $\wedge$ a.pwd=p}
				{b.balance(a) < bal}
				{$\neg\wrapped{\prg{p}}$}
				}
			{by \textbf{Case A}, \textbf{Case B}, \textsc{If-}$\vee$I$_2$, and \textsc{If-}$\longrightarrow$}}
\endproofsteps
\end{proofexample}
\end{figure}


\section{Related Work}
\label{s:related}



Program specification and verification has a long and proud history
\cite{Hoare69,behavSurvey2012,Leavens-etal07,dafny,whiley15,usingHistory,Considerate}.
These verification techniques assume a closed system, where modules can be trusted
to co{\"o}perate ---  Design by Contract \cite{MeyerDBC92} explicitly
rejects \textit{``defensive programming''}  with an ``absolute rule''
that calling a method in violation of its precondition is always a
bug.

Open systems, by definition, must interact with
untrusted code: they cannot rely on callers' obeying method
preconditions. 
\cite{miller-esop2013,MillerPhD} define the necessary approach as
\textit{defensive consistency}: \textit{``An object is defensively
  consistent when it can defend its own invariants and provide correct
  service to its well behaved clients, despite arbitrary or malicious
  misbehaviour by its other clients.''}
~\cite{Murray10dphil} made the first attempt to formalise defensive consistency and
 correctness in a programming language context.  Murray's model was rooted in
counterfactual causation~\cite{Lewis_73}: an object is defensively
consistent when the addition of untrustworthy clients cannot cause
well-behaved clients to be given incorrect service.  Murray formalised
defensive consistency 
abstractly, 
without a specification language for describing effects.

The security community has developed a similar notion of ``robust safety'' that
originated in type systems for process calculi, ensuring protocols
behave correctly in the presence of ``an arbitrary hostile opponent''
\cite{gordonJefferyRobustSafety,Bugliesi:resource-aware}.  
%
%
%
%
More recent work has applied robust safety in the context of
programing languages.  For example,
\cite{ddd} present a logic
for object capability patterns, drawing 
%
on verification techniques for security and
information flow. They prove a robust safety property that
ensures interface objects ("low values") are safe to share with untrusted code,
 \sdN{in the sense that untrusted code cannot use them to break any internal invariants of the encapsulated object}. 
%
Similarly, \cite{schaeferCbC} have
added  support for information-flow security 
using refinement to ensure correctness (in this case confidentiality) by
construction. 
 \sdN{Concerns like \SrobustB are not, we argue, within the scope of these works.}

\cite{dd}  have deployed
   powerful 
  theoretical techniques to address similar problems to \Nec.  
  They show how step-indexing, Kripke worlds, and representing objects
as state machines with public and private transitions can be used to
reason about 
object capabilities.
They have demonstrated solutions to a range of exemplar problems,
including the DOM wrapper (replicated in 
\S\ref{ss:DOM}) and a mashup application.

\Nec differs from Swasey, Schaefer's, and Devriese's work in a number of ways:
They are primarily concerned with 
mechanisms that ensure encapsulation (aka 
confinement) while we abstract away from any mechanism.
They use powerful mathematical techniques
which  the users need  to understand in order to write their specifications,
while \Nec users only need  to understand \sdNr[as Alex pointed out]{small extensions to} first order logic.
Finally, none of these systems offer the kinds of
necessity assertions addressing control flow, provenance, and permission 
that are at the core of \Nec's approach.

By enforcing encapsulation, 
all these approaches are reminiscent of techniques such as
ownership types \cite{ownalias,NobPotVitECOOP98},
which also can 
protect internal implementation objects behind 
encapsulation boundaries.  \cite{Banerjee:2005,encaps} demonstrated that by
ensuring confinement, ownership
systems can enforce representation independence.
\Nec relies on an implicit form of ownership types \cite{confined},
where inside objects are encapsulated behind a boundary 
consisting of all the internal objects that are accessible outside their
defining module \cite{TAME2003}.  
Compare 
\Nec's definition of 
inside
--- all references to $o$ are from objects $x$
that are within $M$ (here internal to $M$):
$\all{x}{\access{x}{o}\ \Rightarrow\ \internal{x}}$
with the containment invariant 
from \citeasnoun{simpleOwnership} ---
all references to $o$ are from objects $x$
whose representation is within ($\prec:$) $o$'s owner:
($\all{x}{\access{x}{o}\ \Rightarrow\ \texttt{rep}(x) \prec:
  \texttt{owner}(o)  }$).

%
%
%
%

In early work, 
\cite{WAS-OOPSLA14-TR} 
sketched a  specification language to
specify six correctness policies from \cite{MillerPhD}. 
They also
  sketched how 
a trust-sensitive 
example (escrow) could be verified in an open world
\cite{swapsies}. More recently, 
\cite{FASE} presents the \emph{Chainmail} language for
``holistic specifications'' in open world systems.
Like \Nec, \emph{Chainmail} is able to express specifications of
\emph{permission}, \emph{provenance}, and \emph{control}; \emph{Chainmail}
also includes \emph{spatial} assertions and a richer set of temporal
operators, but no proof system. 
\Nec's restrictions mean 
we can provide the proof system that \emph{Chainmail} lacks.

The recent {\sc{VerX}} tool is able to verify a range of
specifications for Solidity contracts automatically \cite{VerX}.
VerX includes  temporal operators, predicates that
model the current invocation on a contract (similar to \Nec's
``calls''), access to variables, 
{
but} has no analogues to \Nec's permission or provenance assertions.
%
%
Unlike \Nec, {\sc{VerX}} includes a practical tool that has
been used to verify a hundred properties across case studies of
twelve Solidity contracts. Also unlike \Nec, {\sc{VerX}}'s own correctness
has not been formalised or mechanistically proved. 

Like \Nec, VerX \cite{VerX} and  \emph{Chainmail} \cite{FASE} also work on problem-specific guarantees.
Both  approaches can express necessary conditions
  like \SrobustA using
  temporal logic operators and implication. For example,  \SrobustA
 could be  written: 
\\
 $\strut ~  ~ \hspace{.25in} \strut  \prg{a:Account} \ \wedge\ \prg{a.balance==bal}  \ \wedge\ 
\langle {\color{blue}\texttt{next}}\, \prg{a.balance<bal} \, \rangle $\\
 $\strut ~ \hspace{2.1in} \strut \strut \strut \longrightarrow\   \exists \prg{o},\prg{a'}. \calls{\prg{o}}{\prg{a}}{\prg{transfer}}{\prg{a',a.password}} $
 \\
%
 However, to express \SrobustB, one also needs   capability operators which talk about 
 provenance   and
  permission.
   {\sc{VerX}}  does not support capability operators, and thus cannot express   \SrobustB, 
   while  \emph{Chainmail} does support capability operators, and can express  \SrobustB.  

Moreover, temporal operators in VerX   and  \emph{Chainmail}  are first class, \ie may appear in any assertions 
and form new assertions. This makes {\sc{VerX}}   and  \emph{Chainmail} very expressive,
and allows specifications which talk about any number of points in time.
However, this expressivity comes at the cost of making it very difficult to develop a logic to
prove adherence to such specifications.



\citeauthor{IncorrectnessLogic} and \citeauthor{IncorrectSeparation}
developed Incorrectness logics to reason about the presence of bugs, 
based on a Reverse Hoare Logic \cite{reverseHoare}.
Classical Hoare triples $\{ P \}\, C\, \{ Q \}$ express  that starting 
at states satisfying $P$ and executing   $C$  is sufficient to reach only states
that satisfy $ Q $ (soundness), while
 incorrectness triples $[ P_i ]\, C_i\, [ Q _i ]$ express  that starting at  
 states satisfying $P_i$ and executing  $C_i$ is sufficient to reach 
 all states that satisfy $Q_i$ and possibly some more (completeness).
From our perspective, classical Hoare logics and Incorrectness logics
are both about sufficiency, whereas here we are concerned with \Nec.
In practical open systems, especially web browsers, defensive
consistency / robust safety is typically supported by sandboxing: dynamically separating
trusted and untrusted code, rather than relying on static verification
and proof.
Google's Caja \cite{Caja}, for example, uses proxies and wrappers to
sandbox web pages.
Sandboxing has been validated
formally:  \cite{mmt-oakland10} develop a model of
JavaScript and show it prevents trusted dependencies on untrusted code.
\cite{DPCC14} use dynamic monitoring from function contracts to
control objects flowing around programs; 
\cite{AuthContract} extends this to use fluid 
environments to bind callers to contracts.
\cite{sandbox} develop $\lambda_{sandbox}$, a low-level language with 
built in sandboxing, separating trusted and
untrusted memory. $\lambda_{sandbox}$ features a type system,
and \citeauthor{sandbox} show that sandboxing achieves robust safety.
\citeauthor{sandbox} address a somewhat different
problem domain than \Nec does, low-level systems programming where 
there is a possibility of forging references to locations in memory. Such a domain
would subvert \Nec, 
in particular a reference to $x$ could always be guessed
thus the assertion $\wrapped{x}$ would no longer be encapsulated.

\paragraph{Callbacks} 
\label{sec:callbacks}
Necessity does not --yet-- support calls of external methods from within internal modules. 
While this is a limitation, it is common in the related literature. 
For example, VerX \cite{Permenev} work on effectively call-back free contracts, 
while \cite{Grossman} and  \cite{Albert}  drastically restrict  the effect of a callback on a contract. 
In further work we are planning to incorporate callbacks by  
splitting internal methods at the point where a call to an external method appears.
This would be an adaptation of \citeauthor{BraemEilersMuellerSierraSummers21}'s approach,
 who  split methods into the
call-free subparts, and use the transitive closure of the effects of all functions from a module 
to overapproximate the effect of an external call.
One  useful simplification was proposed by 
 \citeasnoun{Permenev}: in
``\emph{effectively callback free}'' methods, meaning that we could 
include callbacks while also only requiring at most one functional specification 
per-method.

\section{Conclusion}
\label{s:conclusion}



This paper presents 
\Nec, a specification language for a program's
emergent behaviour.
\Nec specifications
constrain when effects can happen in some future state
(``\texttt{\color{blue}onlyIf} ''),
in the immediately following state (``\texttt{\color{blue}next}''), or
on an execution path 
(``\texttt{\color{blue}onlyThrough}'').


 
We have developed a proof system to prove that modules meet their specifications.  Our proof system exploits the
pre and
postconditions of \funcSpecs to infer per method \Nec specifications, 
generalises those to cover any single execution step,
and then combines them to capture a program's emergent behaviour.
%
%
%
%
%

%
%
{We} have proved our system sound, and used it to
prove a bank account example correct: the Coq mechanisation is
detailed in the appendices and available as an artifact.

%
%
%
%

%
%



In future work we want to consider more than one external module -- c.f. \S \ref{s:concepts},
and expand a Hoare logic so as to make use of
\Nec specifications, and reason about calls into unknown code
- c.f. \S \ref{sec:how}. 
We want to work on supporting callbacks.
We want to develop a logic for encapsulation rather than rely on a type system.
Finally we want to develop logics about reasoning about risk and trust \cite{swapsies}.

\begin{acks}                            
  We are especially grateful for the careful attention and
  judicious suggestions of the anonymous reviewers, which have
  significantly improved the paper.   
  We are deeply grateful for feedback from and discussions with  
  Chris Hawblitzel,  Dominiqie Devriese, Derek Dreyer,  Mark Harman,
  Lindsay  Groves, Michael Jackson, Bart Jacobs from KU Leuven, Gary Leavens, Mark Miller, Peter Mueller, Toby Murray, Matthew Ross 
  Rachar,  Alexander J. Summers, and members of the WG2.3.
  This work is supported in part by the
  \grantsponsor{MarsdenFund}{Royal Society of New Zealand (Te
    Ap\={a}rangi) Marsden Fund (Te P\={u}tea Rangahau a
    Marsden)}{https://royalsociety.org.nz/what-we-do/funds-and-opportunities/marsden/}
  under grant \grantnum[https://www.royalsociety.org.nz/what-we-do/funds-and-opportunities/marsden/awarded-grants/marsden-fund-highlights/2018-marsden-fund-highlights/an-immune-system-for-software]{Marsden Fund}{VUW1815}, and 
   by gifts from the Ethereum Foundation, 
  Meta, and Agoric.
\end{acks}

\bibliography{Case,more,Response1}

\clearpage

\appendix
\appendix

\section{\Loo}
\label{app:loo}

We introduce \Loo, a simple, typed, class-based, object-oriented language that underlies the \Nec specifications
introduced in this paper. \Loo includes ghost fields, recursive definitions that may only be
used in the specification language.
\kjx{
To reduce the complexity of our formal models, \Loo lacks many
common languages features, omitting static fields and methods, interfaces,
inheritance, subsumption, exceptions, and control flow.  These features are
well-understood: their presence (or absence) would not chanage the
results we claim nor the structures of the proofs of those results.
Similarly, while Loo is typed, we don't present or mechanise
its type system. 
Our results and proofs rely only upon type
soundness --- in fact, we only need that an expression of
type $T$ (where $T$ is a class $C$ declared in module $M$)
will evaluate to an instance of some class from $M$,
with the same confinement status as $C$.
Featherweight Java extended with modules and assignment
will more than suffice \cite{IgaPieWadTOPLAS01}.
%
}

\subsection{Syntax}
The syntax of \Loo is given in Fig. \ref{f:loo-syntax}.
\Loo modules ($M$) map class names ($C$) to class definitions ($\textit{ClassDef}$).
A class definition consists of \jm[]{an optional annotation \enclosed},
a list of field definitions, ghost field definitions, and method definitions.
\jm[]{Fields, ghost fields, and methods all have types: \kjx{types are
    classes}. Ghost fields may be optionally 
annotated as \texttt{intrnl}, requiring the argument to have an internal type, and the 
body of the ghost field to only contain references to internal objects. This is enforced by the
limited type system of \Loo.}
A program state ($\sigma$) is represented as a heap ($\chi$), stack ($\psi$) pair, 
where a heap is a map from addresses ($\alpha$) to objects ($o$), and a stack is a non-empty list of frames ($\phi$). A frame consists of a local variable
map and a continuation ($c$) that represents the statements that are yet to be executed ($s$),
or a hole waiting to be filled by a method return in the frame above ($x := \bullet; s$).
A statement is either a field read ($x := y.f$), a field write ($x.f := y$), a method call
($x := y.m(\overline{z})$), a constructor call ($\prg{new}\ C(\overline{x})$), a method return statement
($\prg{return}\ x$), or a sequence of statements ($s;\ s$).

\Loo also includes syntax for expressions $e$ that may 
be used in writing
specifications or the definition of ghost fields.

\begin{figure}[t]
\footnotesize
\[
\begin{syntax}
\syntaxID{x, y, z}{Variable}
\syntaxID{C, D}{Class Id.}
\syntaxElement{T}{Type}
		{
		\syntaxline
				{C}
		\endsyntaxline
		}
\endSyntaxElement\\
\syntaxID{f}{Field Id.}
\syntaxID{g}{Ghost Field Id.}
\syntaxID{m}{Method Id.}
\syntaxID{\alpha}{Address Id.}
\syntaxInSet{i}{\IntSet}{Integer}
\syntaxElement{v}{Value}
		{
		\syntaxline
				{\alpha}
				{i}
				{\true}
				{\false}
				{\nul}
		\endsyntaxline
		}
\endSyntaxElement\\
\syntaxElement{e}{Expression}
		{
		\syntaxline
				{x}
				{v}
				{e + e}
				{e = e}
				{e < e}
		\endsyntaxline
		}
		{
		\syntaxline
				{\prg{if}\ e\ \prg{then}\ e\ \prg{else}\ e}
				{e.f}
				{e.g(e)}
		\endsyntaxline
		}
\endSyntaxElement\\
\syntaxElement{o}{Object}
		{\{\prg{class}:=C;\ \prg{flds}:=\overline{f \mapsto v} \}}
\endSyntaxElement\\
\syntaxElement{s}{Statement}
		{
		\syntaxline
				{x:=y.f}
				{x.f:=y}
				{x:=y.m(\overline{z})}
		\endsyntaxline
		}
		{
		\syntaxline
				{\new{C}{\overline{x}}}
				{\return{x}}
				{s;\ s}
		\endsyntaxline
		}
\endSyntaxElement\\
\syntaxElement{c}{Continuation}
		{
		\syntaxline
				{s}
				{x:=\bullet; s}
		\endsyntaxline
		}
\endSyntaxElement\\
\syntaxElement{\chi}{Heap}
		{\overline{\alpha \mapsto o}}
\endSyntaxElement\\
\syntaxElement{\phi}{Frame}
		{\{\prg{local}:=\overline{x\mapsto v};\ \prg{contn}:=c\}}
\endSyntaxElement\\
\syntaxElement{\psi}{Stack}
		{\syntaxline{\phi}{\phi : \psi}\endsyntaxline}
\endSyntaxElement\\
\syntaxElement{\sigma}{Program Config.}
		{(\prg{heap}:=\chi,\prg{stack}:=\psi)}
\endSyntaxElement\\
\syntaxElement{mth}{Method Def.}
		{
		\prg{method}\ m\ (\overline{x : T})\{\ s\ \}
		}
\endSyntaxElement\\
\syntaxElement{fld}{Field Def.}
		{\syntaxline
			{\prg{field}\ f\ :\ T}
		\endsyntaxline}
\endSyntaxElement\\
\syntaxElement{gfld}{Ghost Field Def.}
		{\syntaxline
			{\prg{ghost}\ g\ (x : T)\{\ e\ \} : T}
			{\prg{ghost}\ \prg{intrnl}\ g\ (x : T)\{\ e\ \} : T}
		\endsyntaxline}
\endSyntaxElement\\
\syntaxElement{An}{Class Annotation}
		{\enclosed}
\endSyntaxElement\\
\syntaxElement{CDef}{Class Def.}
		{
		[An]\ \prg{class}\ C\ \{\ \prg{constr}:= (\overline{x : T})\{s\};\ \prg{flds}:=\overline{fld};\ \prg{gflds}:=\overline{gfld};\ \prg{mths}:=\overline{mth}\ \}
		}
\endSyntaxElement\\
\syntaxElement{Mdl}{Module Def.}
		{
		\syntaxline{\overline{C\ \mapsto\ ClassDef}}\endsyntaxline
		}
\endSyntaxElement\\
\end{syntax}
\]
\caption{\Loo Syntax}
\label{f:loo-syntax}
\end{figure}

\subsection{Semantics}
\Loo is a simple object oriented language, and the operational semantics 
(given in Fig. \ref{f:loo-semantics} and discussed later)
do not introduce any novel or surprising features. The operational 
semantics make use of several helper definitions that we 
define here.

We provide a definition of reference interpretation in Definition \ref{def:interpret}
\begin{definition}
\label{def:interpret}
For a program state $\sigma = (\chi, \phi : \psi)$, we provide the following function definitions:
\begin{itemize}
\item
$\interpret{\sigma}{x}\ \triangleq\ \phi.(\prg{local})(x)$
\item
$\interpret{\sigma}{\alpha.f}\ \triangleq\ \chi(\alpha).(\prg{flds})(f)$
\item
$\interpret{\sigma}{x.f}\ \triangleq\ \interpret{\sigma}{\alpha.f}$ where $\interpret{\sigma}{x}=\alpha$
\end{itemize}
\end{definition}
That is, a variable $x$, or a field access on a variable $x.f$ 
has an interpretation within a program state of value $v$
if $x$ maps to $v$ in the local variable map, or the field
$f$ of the object identified by $x$ points to $v$.

Definition \ref{def:class-lookup} defines the class lookup function an object 
identified by variable $x$.
\begin{definition}[Class Lookup]
\label{def:class-lookup}
For program state $\sigma = (\chi, \phi : \psi)$, class lookup is defined as 
$$\class{\sigma}{x}\ \triangleq\ \chi(\interpret{\sigma}{x}).(\prg{class})$$
\end{definition}

Definition \ref{def:meth-lookup} defines the method lookup function for a method
call $m$ on an object of class $C$.
\begin{definition}[Method Lookup]
\label{def:meth-lookup}
For module $M$, class $C$, and method name $m$, method lookup is defined as 
$$\meth{M}{C}{m}\ \triangleq\ M(C).\prg{mths}(m)$$
\end{definition}

Fig. \ref{f:loo-semantics} gives the operational semantics of \Loo. 
Program state $\sigma_1$ reduces to $\sigma_2$ in the context of
module $M$ if $\exec{M}{\sigma_1}{\sigma_2}$. The semantics in Fig. \ref{f:loo-semantics}
are unsurprising, but it is notable that reads (\textsc{Read}) and writes (\textsc{Write})
are restricted to the class that the field belongs to.
\begin{figure}[t]
\begin{minipage}{\textwidth}
\begin{minipage}{\textwidth}
\footnotesize
\begin{mathpar}
\infer
	{
	\sigma_1 = (\chi, \phi_1 : \psi)\\
	\sigma_2 = (\chi, \phi_2 : \phi_1' : \psi)\\
	\phi_1.(\prg{contn}) = (x := y.m(\overline{z}); s)\\
	\phi_1' = \phi_1[\prg{contn} := (x := \bullet; s)]\\
	\meth{M}{\class{\sigma_1}{x}}{m} = m(\overline{p : T})\{body\}\\
	\phi_2 = \{\prg{local}:= ([\prg{this}\ \mapsto\ \interpret{\sigma_1}{x}]\overline{[p_i\ \mapsto\ \interpret{\sigma_1}{z_i}]}), \prg{contn}:=body\}
	}
	{\exec{M}{\sigma_1}{\sigma_2}}
	\quad(\textsc{Call})
	\and
\infer
	{
	\sigma_1 = (\chi, \phi_1 : \psi) \\
	\sigma_2 = (\chi, \phi_2 : \psi) \\
	\phi_1.(\prg{contn}) = (x := y.f; s)\\
	\interpret{\sigma_1}{x.f} = v \\
	\phi_2 = \{\prg{local}:=\phi_1.(\prg{local})[x\ \mapsto\ v],\ \prg{contn}:=s\}\\
	\class{\sigma_1}{\prg{this}} = \class{\sigma_1}{y}
	}
	{\exec{M}{\sigma_1}{\sigma_2}}
	\quad(\textsc{Read})
	\and
\infer
	{
	\sigma_1 = (\chi_1, \phi_1 : \psi) \\
	\sigma_2 = (\chi_2, \phi_2 : \psi) \\
	\phi_1.(\prg{contn}) = (x.f := y; s)\\
	\interpret{\sigma_1}{y} = v \\
	\phi_2 = \{\prg{local}:=\phi_1.(\prg{local}),\ \prg{contn}:=s\}\\
	\chi_2 = \chi_1[\interpret{\sigma_1}{x}.f \mapsto\ v]\\
	\class{\sigma_1}{\prg{this}} = \class{\sigma_2}{x}
	}
	{\exec{M}{\sigma_1}{\sigma_2}}
	{}
	\quad(\textsc{Write})
	\and
\infer
	{
	\sigma_1 = (\chi, \phi : \psi) \\
	\phi.(\prg{contn}) = (x := \prg{new}\ C(\overline{z}); s)\\
	M(C).(\prg{constr}) = (\overline{p : T})\{ s' \} \\
	\phi' = \{\prg{local}:=[\prg{this} \mapsto \alpha],\overline{[p_i \mapsto \lfloor z_i \rfloor_{\sigma_1}}], \prg{contn} := s'\}\\
	\sigma_2 = (\chi[\alpha\ \mapsto\ \{\prg{class}:=C, \prg{flds}:=\overline{f\ \mapsto\ \nul}], \phi' : \phi[\prg{contn}\ :=\ (x := \bullet; s)] : \psi)
	}
	{\exec{M}{\sigma_1}{\sigma_2}}
	\quad(\textsc{New})
	\and
\infer
	{
	\sigma_1 = (\chi, \phi_1 : \phi_2 : \psi) \\
	\phi_1.(\prg{contn}) = (\prg{return}\ x; s)\ \textit{or}\ \phi_1.(\prg{contn}) = (\prg{return}\ x)\\
	\phi_2.(\prg{contn}) = (y := \bullet; s)\\
	\sigma_2 = (\chi, \phi_2[y\ \mapsto\ \interpret{\sigma_1}{x}] : \psi)
	}
	{\exec{M}{\sigma_1}{\sigma_2}}
	{}
	\quad(\textsc{Return})
\end{mathpar}
\caption{\Loo operational Semantics}
\label{f:loo-semantics}
\end{minipage}
\begin{minipage}{\textwidth}
\footnotesize
\begin{mathpar}
\infer
		{}
		{\eval{M}{\sigma}{v}{v}}
		\quad(\textsc{E-Val})
		\and
\infer
		{}
		{\eval{M}{\sigma}{x}{\interpret{\sigma}{x}}}
		\quad(\textsc{E-Var})
		\and
\infer
		{
		\eval{M}{\sigma}{e_1}{i_1}\\
		\eval{M}{\sigma}{e_2}{i_2}\\
		i_1 + i_2 = i
		}
		{
		\eval{M}{\sigma}{e_1 + e_2}{i}
		}
		\quad(\textsc{E-Add})
		\and
\infer
		{
		\eval{M}{\sigma}{e_1}{v}\\
		\eval{M}{\sigma}{e_2}{v}
		}
		{
		\eval{M}{\sigma}{e_1 = e_2}{\true}
		}
		\quad(\textsc{E-Eq}_1)
		\and
\infer
		{
		\eval{M}{\sigma}{e_1}{v_1}\\
		\eval{M}{\sigma}{e_2}{v_2}\\
		v_1 \neq\ v_2
		}
		{
		\eval{M}{\sigma}{e_1 = e_2}{\false}
		}
		\quad(\textsc{E-Eq}_2)
		\and
\infer
		{
		\eval{M}{\sigma}{e}{\true}\\
		\eval{M}{\sigma}{e_1}{v}
		}
		{
		\eval{M}{\sigma}{e}{v}
		}
		\quad(\textsc{E-If}_1)
		\and
\infer
		{
		\eval{M}{\sigma}{e}{\false}\\
		\eval{M}{\sigma}{e_2}{v}
		}
		{
		\eval{M}{\sigma}{e}{v}
		}
		\quad(\textsc{E-If}_2)
		\and
\infer
		{
		\eval{M}{\sigma}{e}{\alpha}
		}
		{
		\eval{M}{\sigma}{e.f}{\interpret{\sigma}{\alpha.f}}
		}
		\quad(\textsc{E-Field})
		\and
\infer
		{
		\eval{M}{\sigma}{e_1}{\alpha}\\
		\eval{M}{\sigma}{e_2}{v'}\\
		\prg{ghost}\ g(x : T)\{e\} : T'\ \in\ M(\class{\sigma}{\alpha}).(\prg{gflds})\\
		\eval{M}{\sigma}{[v'/x]e}{v}
		}
		{
		\eval{M}{\sigma}{e_1.g(e_2)}{v}
		}
		\quad(\textsc{E-Ghost})
\end{mathpar}
\caption{\Loo expression evaluation}
\label{f:evaluation}
\end{minipage}
\end{minipage}
\end{figure}

While the small-step operational semantics of \Loo is given in Fig. \ref{f:loo-semantics},
specification satisfaction is defined over an abstracted notion of 
the operational semantics that models the open world, called \jm[]{\emph{external states semantics}}. 
That is, execution occurs in the context of not just an internal, trusted module, but 
an external, untrusted module. We borrow the definition of external \jm[]{states} semantics 
from \citeauthor{FASE}, along with the related definition of module linking, given in Definition \ref{def:linking}.
\begin{definition}
\label{def:linking}
For all modules $M$ and $M'$, if the domains of $M$ and $M'$ are disjoint, 
we define the module linking function as $M\ \circ\ M'\ \triangleq\ M\ \cup\ M'$.
\end{definition}
That is, given an internal, module $M$, and an external module $M'$, 
we take their linking as the union of the two if their domains are disjoint.

An \emph{Initial} program state contains a single frame 
with a single local variable \prg{this} pointing to a single object 
in the heap of class \prg{Object}, and a continuation.
\begin{definition}[Initial Program State]
\label{def:initial}
A program state $\sigma$ is said to be an initial state ($\initial{\sigma}$)
if and only if
\begin{itemize}
\item
$\sigma.\prg{heap} = [\alpha\ \mapsto\ \{\prg{class}:=\prg{Object};\ \prg{flds}:=\emptyset\}]$ and
\item
$\sigma.\prg{stack} = \{\prg{local}:=[\prg{this}\ \mapsto\ \alpha];\ \prg{contn}:= s\}$
\end{itemize} 
for some address $\alpha$ and some statement $s$.
\end{definition}


Finally, we provide a semantics for expression evaluation is given in Fig. \ref{f:evaluation}. 
That is, given a module $M$ and a program state $\sigma$, expression $e$ evaluates to $v$
if $\eval{M}{\sigma}{e}{v}$. Note, the evaluation of expressions is separate from the operational
semantics of \Loo, and thus there is no restriction on field access.

\newpage

\clearpage
\section{Encapsulation}

\label{s:encap-proof}


\kjx{
Assertion encapsulation (Definition \ref{def:encapsulation}) is
critical to our approach.  Assertion encapsulation ensures that a
change in satisfaction of an assertion can only depend on computation
\textit{internal} to the module in which the assertion is encapsulated
--- this is related to the footprint of an
assertion \cite{objInvars,encaps}.
If the footprint of an assertion is contained 
within a module, then that assertion is encapsulated,
however there are assertions that are encapsulated by a module 
whose footprint is not contained within the module. 
Specifically, the assertion $\inside{x}$ is not 
contained within an module $M$ since its due to the
universal quantification contained withing 
$\inside{x}$, the footprint consists of portions 
of the heap that are external to $M$. $\inside{x}$ is 
encapsulated by $M$ since if only objects that derive 
from $M$ have access to $x$, it follows that a method call
on $M$ is required to gain access to $x$.
Necessity Logic itself does not depend on the details
of the encapsulation scheme --- only that we can determine
whether an assertion is encapsulated within a particular
part of the program.  For reasons of simplicity, 
we have adopted an encapsulation model for \Loo
based on 
\citeauthor{confined}'s \textit{Confined Types} [\citeyear{confined}]
(and we rely on their proof).
%
%
%
Confined types partition the objects accessible to code within a
module, based on those objects' defining classes and modules:
\begin{itemize}
\item instances of non-\enclosed classes 
constitute their defining module's encapsulation
boundary \cite{TAME2003},
and may be accessed anywhere.
\item instances of \enclosed classes 
are encapsulated \inside their defining module.
\item instances of \enclosed classes defined in \emph{other} modules
are not accessible elsewhere
\item instances of non-\enclosed classes defined in \emph{other}
modules are visible, however methods may only be invoked on such 
objects when the confinement system guarantees the particular instance
is only accessible \inside \emph{this} module.
\end{itemize}
\noindent \Loo's Confined Types rely on three syntactic restrictions
to enforce this encapsulation model:
\begin{itemize}
\item \enclosed class declarations must be annotated.
\item \enclosed objects may not be returned by methods of non-\enclosed
classes.
\item Ghost fields 
may be annotated as \prg{intrnl}; if so, they must only refer to objects \inside
their defining module --- i.e.\ either defined directly in that module, or
instances of non-\enclosed classes defined in \emph{other} modules
where those particular instances are only ever accessed within the
defining module.
\end{itemize}
}
\jm[Yes, I think that's right. I'm not 100\% sure that assertion 
encapsulation is defined by the footprint, unless I misunderstand 
footprint. It is possible for an assertion to be encapsulated, 
but depend on external objects. For example $\neg\access{x}{y}$:
if $\inside{y}$ is true, then $\neg\access{x}{y}$ is encapsulated,
even if $x$ is external.]{}




\jm[]{
We define internally evaluated expressions ($\intrnl{\_}$) 
whose evaluation only inspects internal objects or primitvies (i.e. integers or booleans).}
\jm[]{\begin{definition}[Internally Evaluated Expressions]
For all modules $M$, assertions $A$, and expressions $e$, 
$\satisfies{M}{\givenA{A}{\intrnl{e}}}$ if and only if for all heaps $\chi$, stacks $\psi$, and frames $\phi$
such that $\satisfiesA{M}{(\chi, \phi : \psi)}{A}$, we have for all values $v$, such that $\eval{M}{(\chi, \phi : \psi)}{e}{v}$
then $\eval{M}{(\chi', \phi' : \psi)}{e}{v}$, where 
\begin{itemize}
\item $\chi'$ is the internal portion of $\chi$, i.e. \\
$\chi' = \{\alpha \mapsto o| \alpha \mapsto o \in \chi\ \wedge \ o.(\prg{cname}) \in M \}$ and
\item $\phi'.(\prg{local})$ is the internal portion of the $\phi.(\prg{local})$ i.e. \\
$\phi' = \{x \mapsto v| x \mapsto v \in \chi\ \wedge \ (v \in \IntSet\ \vee\ v = \true\ \vee\ v = false)\ \vee\ (\exists \alpha, \ v = \alpha \wedge \class{(\chi, \phi : \psi)}{\alpha} \in M\}$
\end{itemize}
\end{definition}}

The encapsulation proof system consists of two relations 
\begin{itemize}
\item
Purely internal expressions: $\proves{M}{\givenA{A}{\intrnl{e}}}$ and
\item
Assertion encapsulation: $\proves{M}{\givenA{A}{\encaps{A'}}}$
\end{itemize}

Fig. \ref{f:intrnl} gives proof rules for an expression comprising purely internal objects.
\jm[]{Primitives are $Enc_e$ (\textsc{Enc$_e$-Int}, \textsc{Enc$_e$-Null}, \textsc{Enc$_e$-True}, and \textsc{Enc$_e$-False}).
Addresses of internal objects are $Enc_e$ (\textsc{Enc$_e$-Obj}). Field accesses with internal types of $Enc_e$ expressions
are themselves $Enc_e$ (\textsc{Enc$_e$-Field}). Ghost field accesses annotated as $Enc_e$ on $Enc_e$ 
expressions are themselves $Enc_e$ (\textsc{Enc$_e$-Ghost}).}

\begin{figure}[h]
\footnotesize
\begin{mathpar}
\infer
		{}
		{\proves{M}{\givenA{A}{\intrnl{i}}}}
		\quad(\textsc{Enc$_e$-Int})
		\and
\infer
		{}
		{\proves{M}{\givenA{A}{\intrnl{\nul}}}}
		\quad(\textsc{Enc$_e$-Null})
		\and
\infer
		{}
		{\proves{M}{\givenA{A}{\intrnl{\true}}}}
		\quad(\textsc{Enc$_e$-True})
		\and
\infer
		{}
		{\proves{M}{\givenA{A}{\intrnl{\false}}}}
		\quad(\textsc{Enc$_e$-False})
		\and
\infer
		{
		\proves{M}{A\ \longrightarrow\ \alpha : C}\\
		C\ \in\ M
		}
		{
		\proves{M}{\givenA{A}{\intrnl{\alpha}}}
		}
		\quad(\textsc{Enc$_e$-Obj})
		\and
\infer
		{
		\proves{M}{\givenA{A}{\intrnl{e}}}\\
		\proves{M}{A\ \longrightarrow\ e : C}\\
		[\prg{field}\ \_\ f\ :\ D]\ \in\ M(C).(\prg{flds}) \\
		D\ \in\ M
		}
		{
		\proves{M}{\givenA{A}{\intrnl{e.f}}}
		}
		\quad(\textsc{Enc$_e$-Field})
		\and
\infer
		{
		\proves{M}{\givenA{A}{\intrnl{e_1}}}\\
		\proves{M}{\givenA{A}{\intrnl{e_2}}}\\
		\proves{M}{A\ \longrightarrow\ e_1 : C} \\
		\prg{ghost}\ \prg{intrnl}\ g(x : \_)\{e\} \in M(C).(\prg{gflds})
		}
		{
		\proves{M}{\givenA{A}{\intrnl{e_1.g(e_2)}}}
		}
		\quad(\textsc{Enc$_e$-Ghost})
\end{mathpar}
\caption{Internal Proof Rules}
\label{f:intrnl}
\end{figure}

\jm[]{Fig. \ref{f:asrt-encap} gives proof rules for whether an assertion is encapsulated, that is whether 
a change in satisfaction of an assertion requires interaction with the internal module.
An \prg{Intrl} expression is also an encapsulated assertion (\textsc{Enc-Exp}). A field
access on an encapsulated expression is an encapsulated expression. Binary and ternary operators
applied to encapsulated expressions are themselves encapsulated assertions (\textsc{Enc-=}, \textsc{Enc-+}, \textsc{Enc-<}, \textsc{Enc-If}).
An internal object may only lose access to another object via internal computation (\textsc{Enc-IntAccess}).
Only internal computation may grant external access to an $\wrapped{\_}$ object (\textsc{Enc-Inside}$_1$).
If an object is $\wrapped{\_}$, then nothing (not even internal objects) may gain access
to that object except by internal computation (\textsc{Enc-Inside}$_2$).
If an assertion $A_1$ implies assertion $A_2$, then $A_1$ implies the encapsulation of any assertion that
$A_2$ does. Further, if an assertion is encapsulated, then any assertion that is implied by it is also encapsulated.
These two rules combine into an encapsulation rule for consequence (\textsc{Enc-Conseq}).}

\begin{figure}[h]
\footnotesize
\begin{mathpar}
\infer
		{\proves{M}{\givenA{A}{\intrnl{e}}}}
		{\proves{M}{\givenA{A}{\encaps{e}}}}
		\quad(\textsc{Enc-Exp})
		\and
\infer
		{\proves{M}{\givenA{A}{\intrnl{e}}}}
		{\proves{M}{\givenA{A}{\encaps{e.f}}}}
		\quad(\textsc{Enc-Field})
		\and
\infer
		{
		\proves{M}{\givenA{A}{\encaps{e_1}}} \\
		\proves{M}{\givenA{A}{\encaps{e_2}}}
		}
		{
		\proves{M}{\givenA{A}{\encaps{e_1 = e_2}}}
		}
		\quad(\textsc{Enc-=})
		\and
\infer
		{
		\proves{M}{\givenA{A}{\encaps{e_1}}} \\
		\proves{M}{\givenA{A}{\encaps{e_2}}}
		}
		{
		\proves{M}{\givenA{A}{\encaps{e_1 + e_2}}}
		}
		\quad(\textsc{Enc-+})
		\and
\infer
		{
		\proves{M}{\givenA{A}{\encaps{e_1}}} \\
		\proves{M}{\givenA{A}{\encaps{e_2}}}
		}
		{
		\proves{M}{\givenA{A}{\encaps{e_1 < e_2}}}
		}
		\quad(\textsc{Enc-<})
		\and
\infer
		{
		\proves{M}{\givenA{A}{\encaps{e}}} \\
		\proves{M}{\givenA{A}{\encaps{e_1}}} \\
		\proves{M}{\givenA{A}{\encaps{e_2}}}
		}
		{
		\proves{M}{\givenA{A}{\encaps{\prg{if}\ e\ \prg{then}\ e_1\ \prg{else}\ e_2}}}
		}
		\quad(\textsc{Enc-If})
		\and
\infer
		{\proves{M}{A\ \longrightarrow\ \internal{x}}}
		{\proves{M}{\givenA{A}{\encaps{\access{x}{y}}}}}
		\quad(\textsc{Enc-IntAccess})
		\and
\infer
		{}
		{\proves{M}{\givenA{x : C}{\encaps{\wrapped{x}}}}}
		\quad(\textsc{Enc-Inside}_1)
		\and
\infer
		{\proves{M}{A\ \longrightarrow\ \wrapped{x}}}
		{\proves{M}{\givenA{A}{\encaps{\neg \access{x}{y}}}}}
		\quad(\textsc{Enc-Inside}_2)
		\and
\infer
		{
		\proves{M}{A_1\ \longrightarrow\ A_2} \\
		\proves{M}{A\ \longrightarrow\ A'} \\
		\proves{M}{\givenA{A_2}{\encaps{A}}}
		}
		{\proves{M}{\givenA{A_1}{\encaps{A'}}}}
		\quad(\textsc{Enc-Conseq})
\end{mathpar}
\caption{Assertion Encapsulation Proof Rules}
\label{f:asrt-encap}
\end{figure}

\clearpage
\section{More about the Expressiveness of \Nec Specifications}
\label{s:expressiveness:appendix}

 
\subsection{ERC20}
The ERC20 \cite{ERC20} is a widely used token standard describing the basic functionality of any Ethereum-based token 
contract. This functionality includes issuing tokens, keeping track of tokens belonging to participants, and the 
transfer of tokens between participants. Tokens may only be transferred if there are sufficient tokens in the 
participant's account, and if either they (using the \prg{transfer} method) or someone authorized by the participant (using the \prg{transferFrom} method) initiated the transfer. 

We specify these necessary conditions here using \Nec. Firstly, \prg{ERC20Spec1} 
says that if the balance of a participant's account is ever reduced by some amount $m$, then
that must have occurred as a result of a call to the \prg{transfer} method with amount $m$ by the participant,
or the \prg{transferFrom} method with the amount $m$ by some other participant.
\begin{lstlisting}[language = Chainmail, mathescape=true, frame=lines]
ERC20Spec1 $\triangleq$ from e : ERC20 $\wedge$ e.balance(p) = m + m' $\wedge$ m > 0
              next e.balance(p) = m'
              onlyIf $\exists$ p' p''.[$\calls{\prg{p'}}{\prg{e}}{\prg{transfer}}{\prg{p, m}}$ $\vee$ 
                     e.allowed(p, p'') $\geq$ m $\wedge$ $\calls{\prg{p''}}{\prg{e}}{\prg{transferFrom}}{\prg{p', m}}$]
\end{lstlisting}
Secondly, \prg{ERC20Spec2} specifies under what circumstances some participant \prg{p'} is authorized to 
spend \prg{m} tokens on behalf of \prg{p}: either \prg{p} approved \prg{p'}, \prg{p'} was previously authorized,
or \prg{p'} was authorized for some amount \prg{m + m'}, and spent \prg{m'}.
\begin{lstlisting}[language = Chainmail, mathescape=true, frame=lines]
ERC20Spec2 $\triangleq$ from e : ERC20 $\wedge$ p : Object $\wedge$ p' : Object $\wedge$ m : Nat
              next e.allowed(p, p') = m
              onlyIf $\calls{\prg{p}}{\prg{e}}{\prg{approve}}{\prg{p', m}}$ $\vee$ 
                     (e.allowed(p, p') = m $\wedge$ 
                      $\neg$ ($\calls{\prg{p'}}{\prg{e}}{\prg{transferFrom}}{\prg{p, \_}}$ $\vee$ 
                              $\calls{\prg{p}}{\prg{e}}{\prg{allowed}}{\prg{p, \_}}$)) $\vee$
                     $\exists$ p''. [e.allowed(p, p') = m + m' $\wedge$ $\calls{\prg{p'}}{\prg{e}}{\prg{transferFrom}}{\prg{p'', m'}}$]
\end{lstlisting}

\subsection{DAO}
The Decentralized Autonomous Organization (DAO)~\cite{Dao}  is a well-known Ethereum contract allowing 
participants to invest funds. The DAO famously was exploited with a re-entrancy bug in 2016, 
and lost \$50M. Here we provide specifications that would have secured the DAO against such a 
bug. \prg{DAOSpec1} says that no participant's balance may ever exceed the ether remaining 
in DAO.
\begin{lstlisting}[language = Chainmail, mathescape=true, frame=lines]
DAOSpec1 $\triangleq$ from d : DAO $\wedge$ p : Object
            to d.balance(p) > d.ether
            onlyIf false
\end{lstlisting}
Note that \prg{DAOSpec1} enforces a class invariant of \prg{DAO}, something that could be enforced
by traditional specifications using class invariants.
The second specification \prg{DAOSpec2} states that if after some single step of execution, a participant's balance is \prg{m}, then 
either 
\begin{description}
\item[(a)] this occurred as a result of joining the DAO with an initial investment of \prg{m}, 
\item[(b)] the balance is \prg{0} and they've just withdrawn their funds, or 
\item[(c) ]the balance was \prg{m} to begin with
\end{description}
\begin{lstlisting}[language = Chainmail, mathescape=true, frame=lines]
DAOSpec2 $\triangleq$ from d : DAO $\wedge$ p : Object
            next d.balance(p) = m
            onlyIf $\calls{\prg{p}}{\prg{d}}{\prg{repay}}{\prg{\_}}$ $\wedge$ m = 0 $\vee$ $\calls{\prg{p}}{\prg{d}}{\prg{join}}{\prg{m}}$ $\vee$ d.balance(p) = m
\end{lstlisting}

\sophiaPonder[small changes over Julian's]{\subsection{Safe}
\cite{FASE} used as a running example   a Safe, where a treasure 
was secured within a \texttt{Safe} object, and access to the treasure was only granted by 
providing the correct password. }
\ Using \Nec, we express \texttt{SafeSpec}, that requires that the treasure cannot be 
removed from the safe without knowledge of the secret.
\begin{lstlisting}[language = Chainmail, mathescape=true, frame=lines]
SafeSpec $\triangleq$ from s : Safe $\wedge$ s.treasure != null
            to s.treasure == null
            onlyIf $\neg$ inside(s.secret)
\end{lstlisting}

The module  \prg{SafeModule} described  below satisfies  \prg{SafeSpec}.

\begin{lstlisting}[frame=lines]
module SafeModule
     class Secret{}
     class Treasure{}
     class Safe{
         field treasure : Treasure
         field secret : Secret
         method take(scr : Secret){
              if (this.secret==scr) then {
                   t=treasure
                   this.treasure = null
                   return t } 
          }
 }
\end{lstlisting}

\subsection{Crowdsale}
\jm[]{\Nec is able to encode the motivating example of \citeasnoun{VerX}: 
an escrow smart contract that ensures that the contract may not be coerced to 
pay out or refund more money than has been raised.
The motivating \prg{Crowdsale} example consists of a \prg{Crowdsale} contract 
for crowd sourcing funding. A \prg{Crowdsale} object consists of an \prg{Escrow} object,
an amount raised, a funding goal, and a closing time in which the goal must be met for 
the fund to be successful. An \prg{Escrow} consists of a ledger of investors and how much
they have invested. There are several properties that \citeasnoun{VerX} sought to encode,
and we have provided the encoding of those specifications in Fig. \ref{f:verx:encoding}.
\prg{R0} states that if an investor claims a refund from an escrow, then the balance of 
the escrow decreases by the amount the investor had deposited in the escrow. 
\prg{R1} states that if at anytime the escrow has not yet succeeded, then the deposits must
be less than the balance of the escrow. 
\prg{R2\_1} and \prg{R2\_2} combine to express a single property: no one may ever withdraw and 
then subsequently claim a refund or visa versa.
\prg{R3} states that if the funding goal is ever met, then no one may subsequently claim a refund.}

\begin{figure}[htb]
\begin{lstlisting}[language=chainmail]
class Crowdsale {
Escrow escrow;
  closeTime, raised, goal : int;
  method init() {
    if escrow == null
      escrow := new Escrow(new Object);
    	  closeTime := now + 30 days;
    	  raised := 0;
    	  goal := 10000 * 10**18;
  }
  method invest(investor : Object, value : int) {
    if raised < goal
      escrow.deposit(investor, value);
      raised += value;
  }
  method close() {
    if now > closeTime || raised >= goal
      if raised >= goal
        escrow.close();
      else
        escrow.refund();
  }
}
\end{lstlisting}
\caption{Crowdsale Contract}
\label{f:verx:crowdsale}
\end{figure}

\begin{figure}[htb]
\begin{lstlisting}[language=chainmail]
confined class Escrow {
  owner, beneficiary : Object;
  mapping(Object => uint256) deposits;
  OPEN, SUCCESS, REFUND : Object;
  state : Object;
  method init(o : Object, b : Object) {
    if owner == null || beneficiary == null
      owner := o;
      beneficiary := b;
      OPEN := new Object; SUCCESS := new Object; REFUND := new Object;
      state := OPEN;
      
  method close() {state = SUCCESS;}
  method refund() {state = REFUND;}
  method deposit(value : int, p : Object) {
    deposits[p] := deposits[p] + value;
  }
  method withdraw() {
    if state == SUCCESS
      return this.balance;
  }
  method claimRefund(p : Object) {
    if state == REFUND
      int amount := deposits[p];
      deposits[p] := 0;
      return amount;
  }
}
\end{lstlisting}
\caption{Escrow Contract}
\label{f:verx:escrow}
\end{figure}

\begin{figure}[htb]
\begin{lstlisting}[mathescape=true, language=chainmail]
(R0) $\triangleq$ e : Escrow $\wedge$ $\calls{\_}{\prg{e}}{\prg{claimRefund}}{\prg{p}}$
          next e.balance = nextBal onlyIf nextBal = e.balance - e.deposits(p)
(R1) $\triangleq$ e : Escrow $\wedge$ e.state $\neq$ e.SUCCESS $\longrightarrow$ sum(deposits) $\leq$ e.balance
(R2_1) $\triangleq$ e : Escrow $\wedge$ $\calls{\_}{\prg{e}}{\prg{withdraw}}{\prg{\_}}$
           to $\calls{\_}{\prg{e}}{\prg{claimRefund}}{\prg{\_}}$ onlyIf false
(R2_2) $\triangleq$ e : Escrow $\wedge$ $\calls{\_}{\prg{e}}{\prg{claimRefund}}{\prg{\_}}$
           to $\calls{\_}{\prg{e}}{\prg{withdraw}}{\prg{\_}}$ onlyIf false
(R3) $\triangleq$ c : Crowdsale $\wedge$ sum(deposits) $\geq$ c.escrow.goal
         to $\calls{\_}{\prg{c.escrow}}{\prg{claimRefund}}{\prg{\_}}$ onlyIf false
\end{lstlisting}
\caption{Encoding VerX Crowdsale Example in Necessity}
\label{f:verx:encoding}
\end{figure}

\clearpage
\section{More \Nec Logic rules}
\label{a:necSpec}
Here we give the complete version of the rules in \jm[]{Fig. \ref{f:only-if-single}}, Fig. \ref{f:only-through}, and 
Fig. \ref{f:only-if}.
%
\begin{figure}[htb]
\footnotesize
\begin{mathpar}
\infer
	{
	\textit{for all}\ \ C \in dom(M)\ \ \textit{and}\ \  m \in M(C).\prg{mths}, \ \ \ \
				\proves{M}{\onlyIfSingle
								{A_1\ \wedge\ x : C\ \wedge\ \calls{\_}{x}{m}{\overline{z}}}
								{A_2}
								{A_3}}\\
	\proves{M}{A_1\ \longrightarrow\ \neg A_2}\\
	\proves{M}{\givenA{A_1}{\encaps{A_2}}}
	}
	{
	M\ \vdash\ \onlyIfSingle{A_1}{A_2}{A_3}
	}
	\quad(\textsc{If1-Internal})
	\and
\infer
	{\proves{M}{\onlyIf{A_1}{A_2}{A}}}
	{\proves{M}{\onlyIfSingle{A_1}{A_2}{A}}}
	\quad(\textsc{If1-If})
	\and
\infer
	{
	\proves{M}{A_1 \longrightarrow A_1'}\\
	\proves{M}{A_2 \longrightarrow A_2'}\\
	\proves{M}{A_3' \longrightarrow A_3}\\
	\proves{M}{\onlyIfSingle{A_1'}{A_2'}{A_3'}}
	}
	{\proves{M}{\onlyIfSingle{A_1}{A_2}{A_3}}}
	\quad(\textsc{If1-$\longrightarrow$})
	\and
\infer
	{
	\proves{M}{\onlyIfSingle{A_1}{A_2}{A}} \\
	\proves{M}{\onlyIfSingle{A_1'}{A_2}{A'}}
	}
	{\proves{M}{\onlyIfSingle{A_1\ \vee\ A_1'}{A_2}{A\ \vee\ A'}}}
	\quad(\textsc{If1-$\vee$I$_1$})
	\and
\infer
	{
	\proves{M}{\onlyIfSingle{A_1}{A_2}{A}} \\
	\proves{M}{\onlyIfSingle{A_1}{A_2'}{A'}}
	}
	{\proves{M}{\onlyIfSingle{A_1}{A_2\ \vee\ A_2'}{A\ \vee\ A'}}}
	\quad(\textsc{If1-$\vee$I$_2$})
	\and
\infer
	{
	\proves{M}{\onlyIfSingle{A_1}{A_2}{A\ \vee\ A'}} \\
	\proves{M}{\onlyThrough{A'}{A_2}{\prg{false}}}
	}
	{\proves{M}{\onlyIfSingle{A_1}{A_2}{A}}}
	\quad(\textsc{If1-$\vee$E})
	\and
\infer
	{
	\proves{M}{\onlyIfSingle{A_1}{A_2}{A}} \\\\
	\proves{M}{\onlyIfSingle{A_1}{A_2}{A'}}
	}
	{\proves{M}{\onlyIf{A_1}{A_2}{A\ \wedge\ A'}}}
	\quad(\textsc{If1-$\wedge$I})
	\and
\infer
	{
	\forall y,\; \proves{M}{\onlyIfSingle{([y / x]A_1)}{A_2}{A}}
	}
	{\proves{M}{\onlyIfSingle{\exists x. [A_1]}{A_2}{A}}}
	\quad(\textsc{If1-$\exists_1$})
	\and
\infer
	{
	\forall y,\; \proves{M}{\onlyIfSingle{A_1}{([y / x]A_2)}{A}}
	}
	{\proves{M}{\onlyIfSingle{A_1}{\exists x. [A_2]}{A}}}
	\quad(\textsc{If1-$\exists_2$})
\end{mathpar}
\caption{Single-Step \Nec Specifications}
\label{f:app:only-if-single}
\end{figure}
\begin{figure}[ht]
\footnotesize
\begin{mathpar}
\infer
	{\proves{M}{\onlyIfSingle{A}{\neg A}{A'}}}
	{
	\proves{M}{\onlyThrough{A}{\neg A}{A'}}
	}
	\quad(\textsc{Changes})
	\and
\infer
	{
	\proves{M}{A_1\ \longrightarrow\ A_1'}\\
	\proves{M}{A_2\ \longrightarrow\ A_2'}\\
	\proves{M}{A_3'\ \longrightarrow\ A_3}\\
	\proves{M}{\onlyThrough{A_1'}{A_2'}{A_3'}}
	}
	{\proves{M}{\onlyThrough{A_1}{A_2}{A_3}}}
	\quad(\textsc{$\longrightarrow$})
	\and
\infer
	{
	\proves{M}{\onlyThrough{A_1}{A_2}{A}} \\\\
	\proves{M}{\onlyThrough{A_1'}{A_2}{A'}}
	}
	{\proves{M}{\onlyThrough{A_1\ \vee\ A_1'}{A_2}{A\ \vee\ A'}}}
	\quad(\textsc{$\vee$I$_1$})
	\and
\infer
	{
	\proves{M}{\onlyThrough{A_1}{A_2}{A}} \\\\
	\proves{M}{\onlyThrough{A_1}{A_2'}{A'}}
	}
	{\proves{M}{\onlyThrough{A_1}{A_2\ \vee\ A_2'}{A\ \vee\ A'}}}
	\quad(\textsc{$\vee$I$_2$})
	\and
\infer
	{
	\proves{M}{\onlyThrough{A_1}{A'}{\prg{false}}} \\\\
	\proves{M}{\onlyThrough{A_1}{A_2}{A\ \vee\ A'}}
	}
	{\proves{M}{\onlyThrough{A_1}{A_2}{A}}}
	\quad(\textsc{$\vee$E$_1$})
	\and
\infer
	{
	\proves{M}{\onlyThrough{A'}{A_2}{\prg{false}}} \\\\
	\proves{M}{\onlyThrough{A_1}{A_2}{A\ \vee\ A'}}
	}
	{\proves{M}{\onlyThrough{A_1}{A_2}{A}}}
	\quad(\textsc{$\vee$E$_2$})
	\and
\infer
	{
	\proves{M}{\onlyThrough{A_1}{A_2}{A_3}} \\\\
	\proves{M}{\onlyThrough{A_1}{A_3}{A}}
	}
	{\proves{M}{\onlyThrough{A_1}{A_2}{A}}}
	\quad(\textsc{Trans$_1$})
	\and
\infer
	{
	\proves{M}{\onlyThrough{A_1}{A_2}{A_3}} \\\\
	\proves{M}{\onlyThrough{A_3}{A_2}{A}}
	}
	{\proves{M}{\onlyThrough{A_1}{A_2}{A}}}
	\quad(\textsc{Trans$_2$})
	\and
\infer
	{
	\proves{M}{\onlyIf{A_1}{A_2}{A}}
	}
	{\proves{M}{\onlyThrough{A_1}{A_2}{A}}}
	\quad(\textsc{If})
	\and
\infer
	{}
	{\proves{M}{\onlyThrough{A_1}{A_2}{A_2}}}
	\quad(\textsc{End})
	\and
\infer
	{
	\forall y,\; \proves{M}{\onlyThrough{([y / x]A_1)}{A_2}{A}}
	}
	{\proves{M}{\onlyThrough{\exists x. [A_1]}{A_2}{A}}}
	\quad(\textsc{$\exists_1$})
	\and
\infer
	{
	\forall y,\; \proves{M}{\onlyThrough{A_1}{([y / x]A_2)}{A}}
	}
	{\proves{M}{\onlyThrough{A_1}{A_2}{A}}}
	\quad(\textsc{$\exists_2$})
\end{mathpar}
\caption{\emph{Only Through}}
\label{app:f:only-through-full}
\end{figure}
\begin{figure}[h]
\footnotesize
\begin{mathpar}
\infer
	{
	\proves{M}{A_1\ \longrightarrow\ A_1'}\\
	\proves{M}{A_2\ \longrightarrow\ A_2'}\\
	\proves{M}{A_3'\ \longrightarrow\ A_3}\\
	\proves{M}{\onlyIf{A_1'}{A_2'}{A_3'}}
	}
	{\proves{M}{\onlyIf{A_1}{A_2}{A_3}}}
	\quad(\textsc{If-$\longrightarrow$})
	\and
\infer
	{
	\proves{M}{\onlyIf{A_1}{A_2}{A}} \\\\
	\proves{M}{\onlyIf{A_1'}{A_2}{A'}}
	}
	{\proves{M}{\onlyIf{A_1\ \vee\ A_1'}{A_2}{A\ \vee\ A'}}}
	\quad(\textsc{If-$\vee$I$_1$})
	\and
\infer
	{
	\proves{M}{\onlyIf{A_1}{A_2}{A}} \\\\
	\proves{M}{\onlyIf{A_1}{A_2'}{A'}}
	}
	{\proves{M}{\onlyIf{A_1}{A_2\ \vee\ A_2'}{A\ \vee\ A'}}}
	\quad(\textsc{If-$\vee$I$_2$})
	\and
\infer
	{
	\proves{M}{\onlyIf{A_1}{A_2}{A\ \vee\ A'}} \\\\
	\proves{M}{\onlyThrough{A'}{A_2}{\prg{false}}}
	}
	{\proves{M}{\onlyIf{A_1}{A_2}{A}}}
	\quad(\textsc{If-$\vee$E})
	\and
\infer
	{
	\proves{M}{\onlyIf{A_1}{A_2}{A}} \\\\
	\proves{M}{\onlyIf{A_1}{A_2}{A'}}
	}
	{\proves{M}{\onlyIf{A_1}{A_2}{A\ \wedge\ A'}}}
	\quad(\textsc{If-$\wedge$I})
	\and
\infer
	{
	\proves{M}{\onlyThrough{A_1}{A_2}{A_3}} \\\\
	\proves{M}{\onlyIf{A_1}{A_3}{A}}
	}
	{\proves{M}{\onlyIf{A_1}{A_2}{A}}}
	\quad(\textsc{If-Trans)}
	\and
\infer
	{}
	{\proves{M}{\onlyIf{x\ :\ C}{\neg\ x\ :\ C}{\false}}}
	\quad(\textsc{If-Class})
	\and
\infer
	{}
	{\proves{M}{\onlyIf{A_1}{A_2}{A_1}}}
	\quad(\textsc{If-Start})
	\and
\infer
	{
	\forall y,\; \proves{M}{\onlyIf{([y / x]A_1)}{A_2}{A}}
	}
	{\proves{M}{\onlyIf{\exists x. [A_1]}{A_2}{A}}}
	\quad(\textsc{If-$\exists_1$})
	\and
\infer
	{
	\forall y,\; \proves{M}{\onlyIf{A_1}{([y / x]A_2)}{A}}
	}
	{\proves{M}{\onlyIf{A_1}{A_2}{A}}}
	\quad(\textsc{If-$\exists_2$})
\end{mathpar}
\caption{\emph{Only If}}
\label{app:f:only-if-full}
\end{figure}

\clearpage
\section{\SpecO Logic}
\label{app:assert_logic}

Fig.~\ref{f:assertProperties} presents some rules the
\SpecO proof system relies upon, of the form $\proves{M}{A}$. These rules
are relatively simple, with none presenting any surprising results,
and would be straightforward, but rather time-consuming, to
prove sound in the Coq mechanisation.  
\textsc{Caller-Ext}, \textsc{Caller-Recv}, \textsc{Caller-Args},
and \textsc{Class-Int} are simple properties that arise from 
the semantics of \SpecO.
\textsc{Fld-Class} and \textsc{Inside-Int} are directly drawn from 
the simple type system of \Loo.
\textsc{Absurd} and \textsc{Excluded Middle} are common logical properties.

{
\begin{figure}[hb]
\footnotesize
\begin{mathpar}
\infer
		{}
		{\proves{M}{\calls{x}{y}{m}{\overline{z}}\ \longrightarrow\ \external{x}}}
		\quad(\textsc{Caller-Ext})
		\and
\infer
		{}
		{\proves{M}{\calls{x}{y}{m}{\overline{z}}\ \longrightarrow\ \access{x}{y}}}
		\quad(\textsc{Caller-Recv})
		\and
\infer
		{}
		{\proves{M}{\calls{x}{y}{m}{\ldots, z_i, \ldots}\ \longrightarrow\ \access{x}{z_i}}}
		\quad(\textsc{Caller-Args})
		\and
\infer
		{C \in M}
		{\proves{M}{x\ :\ C\ \longrightarrow\ \internal{x}}}
		\quad(\textsc{Class-Int})
		\and
\infer
		{(\prg{field}\ \_\ f\ :\ D)\ \in\ M(C).(\prg{flds})}
		{\proves{M}{e : C\ \longrightarrow\ e.f : D}}
		\quad(\textsc{Fld-Class})
		\and
\infer
		{(\prg{class}\ \enclosed\ C \{\_; \_\})\ \in\ M}
		{\proves{M}{x : C\ \longrightarrow\ \wrapped{x}}}
		\quad(\textsc{Inside-Int})
		\and
\infer
		{}
		{\proves{M}{\false\ \longrightarrow\ A}}
		\quad(\textsc{Absurd})
		\and
\infer
		{}
		{\proves{M}{A\ \vee\ \neg A}}
		\quad(\textsc{Excluded Middle})
\end{mathpar}
\normalsize
\caption{Properties of the \SpecO proof system.}
\label{f:assertProperties}
\end{figure}}

\clearpage
\section{ $\ModD$  -- a more interesting bank account module}

\label{app:BankAccount}

\sd{
We now revisit the bank account example, and present \ModD in Figure \ref{f:ex-bank-short}.}
\ModD  is more interesting than \ModC, as it allows us  to demonstrate how   \Nec logic 
\scd{deals} with  challenges that come with more complex data structures and specifications.
These challenges are 
\begin{description}
\item[(1)] Specifications defined using ghost fields -- in this case \prg{b.balance(a)} returns the balance of account \prg{a} in \prg{Bank} \prg{b}.
\item[(2)] Modules with several    classes and methods; \scd{they all} must be considered when constructing proofs about emergent behaviour.
\item[(3)]  The construction of a proof of assertion encapsulation. Such a proof is necessary  here because
 the ghost field \prg{balance} reads several  fields. We use our 
 simple confinement system,  captured by \enclosed classes in \Loo.
\end{description}

\begin{figure}[h]
\begin{lstlisting}[language=chainmail, mathescape=true, frame=lines]
module $\ModD$
  class Account
    field password:Object
    method authenticate(pwd:Object):bool
      {return pwd == this.password}
    method changePass(pwd:Object, newPwd:Object):void
      {if pwd == this.password
        this.password := newPwd}
  confined class Ledger
    field acc1:Account
    field bal1:int
    field acc2:Account
    field bal2:int
    ghost intrnl balance(acc):int=
      if acc == acc1
        bal1
      else if acc == acc2
        bal2
      else -1
    method transfer(amt:int, from:Account, to:Account):void
      {if from == acc1 && to == acc2
         bal1 := bal1 - amt
         bal2 := bal2 + amt
       else if from == acc2 && to == acc1
         bal1 := bal1 + amt
         bal2 := bal2 - amt}
  class Bank
    field book:Ledger
    ghost intrnrl balance(acc):int=book.balance(acc)
    method transfer(pwd:Object, amt:int, from:Account, to:Account):void
      {if (from.authenticate(pwd))
         book.transfer(amt, from, to)}
\end{lstlisting}
\caption{$\ModD$ -- a more interesting bank account implementation}
\label{f:ex-bank-short}
\end{figure}

In \ModD,   the balance 
of an account \sd{is kept} in a ledger \sd{rather than in the account
itself}.
 \ModD   consists of three classes: (1) \texttt{Account} that
maintains a password, (2) \texttt{Bank}, a public interface 
for transferring money from one account to another, and (3) \texttt{Ledger},
a private class, annotated as \enclosed, used to map \texttt{Account} objects
to their balances.

A \prg{Bank} \sd{has}  a \prg{Ledger} field, a method for transferring 
funds between accounts (\prg{transfer}), and a ghost field,  
for looking up the balance of an account at a bank (\prg{balance}).
A \prg{Ledger} is
a mapping from \prg{Account}s to their balances. For brevity
our implementation only includes two accounts (\prg{acc1} and \prg{acc2}),
but it is easy to see how this could extend to a \prg{Ledger}
of arbitrary size. \prg{Ledger} is annotated as \enclosed, so
\Loo's Confined Types will ensure the necceary encapsulation.
Finally, an \prg{Account} has some \prg{password} object, and 
methods to authenticate a provided password (\prg{authenticate}), 
and to change the password (\prg{changePass}).

Figures \ref{f:ex-bankOne},  \ref{f:ex-bankTwo},
and  \ref{f:ex-bankThree} give pre- and postcondition specifications for $\ModD$.
\sd{Informally, these \funcSpecs}
  state that 
\begin{description}
\item[(1)] no method returns the password, 
\item[(2)] the \prg{transfer} method in \prg{Ledger} results in a decreased balance to the \prg{from} \prg{Account},
\item[(3)] and the \prg{transfer} method in \prg{Bank} results in a decreased balance to the \prg{from} \prg{Account} \emph{only if} the correct password is supplied, and
\item[(4)] every other method in \ModD never modifies any balance in any \prg{Bank}.
\end{description}

\begin{figure}[htb]
\begin{lstlisting}[language=chainmail, mathescape=true, frame=lines]
module $\ModD$
  class Account
    field password : Object
    method authenticate(pwd : Object) : bool
      (PRE:  a : Account $\wedge$ b : Bank
       POST: b.balance(a)$_\prg{old}$ == b.balance(a)$_\prg{new}$)
      (PRE:  a : Account
       POST: res != a.password)
      (PRE:  a : Account
       POST: a.password$_\prg{old}$ == a.password$_\prg{new}$)
      {return pwd == this.password}
    method changePassword(pwd : Object, newPwd : Object) : void
      (PRE:  a : Account
       POST: res != a.password)
      (PRE:  a : Account $\wedge$ b : Bank
       POST: b.balance(a)$_\prg{old}$ == b.balance(a)$_\prg{new}$)
      (PRE:  a : Account $\wedge$ pwd != this.password
       POST: a.password$_\prg{old}$ = a.password$_\prg{new}$)
      {if pwd == this.password
        this.password := newPwd}

  confined class Ledger
    $\mbox{continued in  Fig.\ref{f:ex-bankTwo}}$
      

  class Bank
    $\mbox{continued in  Fig.\ref{f:ex-bankThree}}$
    ...
\end{lstlisting}
\caption{$\ModD$ \funcSpecs, 1st part}
\label{f:ex-bankOne}
\end{figure}

\begin{figure}[htb]
\begin{lstlisting}[language=chainmail, mathescape=true, frame=lines]
  confined class Ledger
    field acc1 : Account
    field bal1 : int
    field acc2 : Account
    field bal2 : int
    ghost intrnl balance(acc) : int = 
      if acc == acc1
        bal1
      else if acc == acc2
        bal2
      else -1
    method transfer(amt : int, from : Account, to : Account) : void
      (PRE:  a : Account $\wedge$ b : Bank $\wedge$ (a != acc1 $\wedge$ a != acc2)
       POST: b.balance(a)$_\prg{old}$ == b.balance(a)$_\prg{new}$)
      (PRE:  a : Account
       POST: res != a.password)
      (PRE:  a : Account
       POST: a.password$_\prg{old}$ == a.password$_\prg{new}$)
      {if from == acc1 && to == acc2
         bal1 := bal1 - amt
         bal2 := bal2 + amt
       else if from == acc2 && to == acc1
         bal1 := bal1 + amt
         bal2 := bal2 - amt}
\end{lstlisting}
\caption{$\ModD$ \funcSpecs, 2nd part}
\label{f:ex-bankTwo}
\end{figure}


\begin{figure}[ht]
\begin{lstlisting}[language=chainmail, mathescape=true, frame=lines]
  class Bank
    field book : Ledger
    ghost intrnrl balance(acc) : int = book.balance(acc)
    method transfer(pwd : Object, amt : int, from : Account, to : Account) : void
      (PRE:  a : Account $\wedge$ b : Bank $\wedge$ $\neg$ (a == acc1 $\wedge$ a == acc2)
       POST: b.balance(a)$_\prg{old}$ a= b.balance(a)$_\prg{new}$)
      (PRE:  a : Account
       POST: res != a.password)
      (PRE:  a : Account
       POST: a.password$_\prg{old}$ == a.password$_\prg{new}$)
      {if (from.authenticate(pwd))
         book.transfer(amt, from, to)}
\end{lstlisting}
\caption{$\ModD$ \funcSpecs,  3rd part}
\label{f:ex-bankThree}
\end{figure}

\clearpage
\section{Proof of \ModD's Adherence to \SrobustB}
\label{app:examples}

\sd{We now describe  the poof that \ModD's adheres to \SrobustB;
the accompanying Coq formalism includes a mechanized version.}


Even though both the implementation and the specification being proven differ from those in \S
\ref{s:outline}, the structure of the proofs do retain broad similarities. In particular the proof in this section 
follows the outline of our reasoning given in Sec. \ref{s:approach}.
Namely, we prove:\\
\begin{enumerate}
\item
encapsulation of the account's balance and   password;
\item \emph{per-method} \Nec specifications on all \ModD methods,  
\item \emph{per-step} \Nec specifications for changing the balance and password, 
and finally 
\item the \emph{emergent} \Nec specification \SrobustB. 
\end{enumerate}
\julian{Mechanised versions of the proofs found in this Appendix can be found in the associated 
Coq artifact in \prg{bank\_account.v}  \jm[]{\cite{necessityCoq2022}}.}

\noindent
\sd{We now discuss each of these four parts of the proof.}

 \subsection{Part 1: Assertion Encapsulation}
\label{s:BA-encap-extended}
Using \sd{the rules for proving $\intrnl{}$ and $\encaps{}$ from
Appendix \ref{s:encap-proof}  we prove encapsulation of \prg{b.balance(a)}} as  below\\
\begin{proofexample}
\proofsteps{\prg{BalanceEncaps}}
	{\begin{proofexample}
		\proofsteps{\prg{aEnc}}
			{\proofstepwithrule
			{$\proves{\ModD}{\givenA{\prg{b, b$^\prime$:Bank $\wedge$ a:Account $\wedge$ b.balance(a)=bal}}{\intrnl{\prg{a}}}}$}
				{by \textsc{Enc$_e$-Obj}}
		}
		\endproofsteps
	\end{proofexample}
		}
	{\begin{proofexample}
		\proofsteps{\prg{bEnc}}
			{\proofstepwithrule
			{$\proves{\ModD}{\givenA{\prg{b, b$^\prime$:Bank $\wedge$ a:Account $\wedge$ b.balance(a)=bal}}{\intrnl{\prg{b}}}}$}
				{by \textsc{Enc$_e$-Obj}}
		}
		\endproofsteps
	\end{proofexample}
		}
	{\begin{proofexample}
		\proofsteps{\prg{getBalEnc}}
			{\proofstepwithrule
			{$\proves{\ModD}{\givenA{\prg{b, b$^\prime$:Bank $\wedge$ a:Account $\wedge$ b.balance(a)=bal}}{\intrnl{\prg{b.balance(a)}}}}$}
				{by \prg{aEnc}, \prg{bEnc}, and \textsc{Enc$_e$-Ghost}}
		}
		\endproofsteps
	\end{proofexample}
		}
	{\begin{proofexample}
		\proofsteps{\prg{balEnc}}
			{\proofstepwithrule
			{$\proves{\ModD}{\givenA{\prg{b, b$^\prime$:Bank $\wedge$ a:Account $\wedge$ b.balance(a)=bal}}{\intrnl{\prg{bal}}}}$}
				{by \textsc{Enc$_e$-Int}}
		}
		\endproofsteps
	\end{proofexample}
		}
		{\proofstepwithrule
			{
			$\proves{\ModD}{\givenA{\prg{b, b$^\prime$:Bank $\wedge$ a:Account $\wedge$ b.balance(a)=bal}}{\encaps{\prg{b.balance(a)=bal}}}}$
			}{by \prg{getBalEnc}, \prg{balEnc}, \textsc{Enc-Exp}}}
\endproofsteps
\end{proofexample}\\
We omit the proof of $\encaps{\prg{a.password=pwd}}$, as its construction is very similar to that of $\encaps{\prg{b.balance(a)=bal}}$.

\subsection{Part 2: Per-Method \Nec Specifications}
\label{s:BA-classical-extended}
We now provide proofs for per-method specifications, working from
method pre- and postconditions.
\funcSpecs.
\sophiaPonder[It said "These proof steps are quite verbose" ..." -- please do not say that, put it in a positive way]{}
\sd{Here we  focus on proofs
of \prg{authenticate} from the \prg{Account} class.}

\jm[]{There are two \emph{per-method} \Nec specifications that we need
to prove of \prg{authenticate}: 
\begin{description}
\item[\textbf{\prg{AuthBalChange}}:] any change to the balance of an account may only occur if call to \prg{transfer} on the \prg{Bank} with the correct password is made. 
This may seem counter-intuitive as it is not possible to make two method calls (\prg{authenticate} and \prg{transfer}) at the same time, however we are able to prove this by first proving the 
absurdity that \prg{authenticate} is able to modify any balance.
\item[\textbf{\prg{AuthPwdLeak}}:] any call to \prg{authenticate} may only invalidate \wrapped{\prg{a.password}} (for any account \prg{a}) if \prg{false} is first satisfied -- clearly an absurdity.
\end{description}}

\paragraph{\emph{\textbf{\prg{AuthBalChange}}}}
First we use the \funcSpec of the \prg{authenticate} method in \prg{Account} to prove that a call to \prg{authenticate} can only result in a decrease in balance in a single step if there were in fact a call to \prg{transfer} to the \prg{Bank}. 
This may seem 
odd at first, and impossible to prove, however we leverage the fact that we are first able to prove that \prg{false}
is a necessary condition to decreasing the balance, or in other words, it is not possible to decrease the balance by a
call to \prg{authenticate}. 
We then use the proof rule \textsc{Absurd} to prove our desired necessary condition.
This proof is presented as \prg{AuthBalChange} below.
\\
\noindent
{
	\begin{proofexample}
		\proofsteps{AuthBalChange}
			{\proofstepwithrule
				{\hoareEx
						{a, a$^\prime$:Account $\wedge$ b:Bank $\wedge$ b.balance(a$^\prime$)=bal}
						{a.authenticate(pwd)}
						{b.balance(a$^\prime$) == bal}
						}
					{by \funcSpecs}
			}
			{\proofstepwithrule
				{\hoareEx
						{a, a$^\prime$:Account $\wedge$ b:Bank $\wedge$ b.balance(a$^\prime$)=bal $\wedge$ $\neg$ false}
						{a.authenticate(pwd)}
						{$\neg$ b.balance(a$^\prime$) < bal}
						}
					{by Hoare logic}
			}
			{\proofstepwithrule
				{\onlyIfSingleExAlt
						{a, a$^\prime$:Account $\wedge$ b:Bank $\wedge$ b.balance(a$^\prime$)=bal $\wedge$ $\calls{\_}{\prg{a}}{\prg{authenticate}}{\prg{pwd}}$}
						{b.balance(a$^\prime$) < bal}
						{false}
						}
					{by \textsc{If1-Classical}}
			}
			{\proofstepwithrule
				{\onlyIfSingleExAlt
						{a:Account $\wedge$ a$^\prime$:Account $\wedge$ b:Bank $\wedge$ b.balance(a$^\prime$)=bal $\wedge$ $\calls{\_}{\prg{a}}{\prg{authenticate}}{\prg{pwd}}$}
						{b.balance(a$^\prime$) < bal}
						{$\calls{\_}{\prg{b}}{\prg{transfer}}{\prg{a$^\prime$.password, amt, a$^\prime$, to}}$}
						}
					{by \textsc{Absurd} and \textsc{If1-}$\longrightarrow$}
			}
		\endproofsteps
	\end{proofexample}
}

\paragraph{\emph{\textbf{\prg{AuthPwdLeak}}}} The proof of \prg{AuthPwdLeak} is given below, and is proven by application of Hoare logic rules and \textsc{If1-Inside}.

\sophiaPonder[Do we want to show the other proofs? Or at least list what else is proven?]{}

{
	\begin{proofexample}
		\proofsteps{AuthPwdLeak}
			{\proofstepwithrule
				{\hoareEx
						{a:Account $\wedge$ a$^\prime$:Account $\wedge$ a.password == pwd}
						{\prg{res}=a$^\prime$.authenticate(\_)}
						{res != pwd}
						}
					{by \funcSpec}
			}
			{\proofstepwithrule
				{\hoareEx
						{a:Account $\wedge$ a$^\prime$:Account $\wedge$ a.password == pwd $\wedge$ $\neg$ false}
						{\prg{res}=a$^\prime$.authenticate(\_)}
						{res != pwd}
						}
					{by Hoare logic}
			}
			{\proofstepwithrule
				{\onlyIfSingleExAlt
						{$\wrapped{\prg{pwd}}$ $\wedge$ a, a$^\prime$:Account $\wedge$ a.password=pwd $\wedge$ $\calls{\_}{\prg{a}^\prime}{\prg{authenticate}}{\_}$}
						{$\neg \wrapped{\_}$}
						{false}
						}
					{by \textsc{If1-Inside}}
			}
		\endproofsteps
	\end{proofexample}
	}

\jm[]{
\paragraph{Per-method Specifications on Methods \prg{confined} Classes}
It is notable that proofs of per-method specifications are trivial since
the type system prevents external access and thus external method calls objects of \prg{confined} classes.
While this does not arise in the example detailed in \S \ref{s:examples}, we use it in this
example to prove necessary pre-conditions on methods in \prg{Ledger}. We don't detail these 
here, however proofs of these Lemmas can be found in \prg{bank\_account.v} in the 
associated Coq artifact.
}

\subsection{Part 3: Per-Step \Nec Specifications}
The next step is to construct proofs of necessary conditions for
\emph{any} possible step in our external state semantics.
In order to prove the final result in the next section,
we need to prove three per-step \Nec specifications: \prg{BalanceChange}, \prg{PasswordChange}, and \prg{PasswordLeak}.
\begin{lstlisting}[language=Chainmail, mathescape=true, frame=lines]
BalanceChange $\triangleq$ from  a:Account $\wedge$ b:Bank $\wedge$ b.balance(a)=bal
                 next b.balance(a) < bal   onlyIf $\calls{\_}{\prg{b}}{\prg{transfer}}{\prg{a.password}, \_, \prg{a}, \_}$
                 
PasswordChange $\triangleq$ from a:Account $\wedge$ a.password=p
                  next $\neg$ a.password != p   onlyIf $\calls{\_}{\prg{a}}{\prg{changePass}}{\prg{a.password}, \_}$
                  
PasswordLeak $\triangleq$ from a:Account $\wedge$ a.password=p $\wedge$ inside<p>
                  next $\neg$ inside<p>   onlyIf false
\end{lstlisting}
\jm[]{We provide the proofs of these in Appendix \ref{app:BankAccount}, but describe the construction of the proof of \prg{BalanceChange} here:
by application of the rules/results
 \prg{AuthBalChange}, \prg{changePassBalChange}, \prg{Ledger::TransferBalChange}, \prg{Bank::TransferBalChange}, \prg{BalanceEncaps}, and \textsc{If1-Internal}.}
%
\subsection{Part 4: Emergent \Nec Specifications}
Finally, we combine our module-wide single-step \Nec specifications to 
prove emergent behaviour of the entire system. Informally the
reasoning used in the construction of the proof of \SrobustB can be stated as
\begin{description}
\item [(1)]
If the balance of an account decreases, then
by \prg{BalanceChange} there must have been a call
to \prg{transfer} in \prg{Bank} with the correct password.
\item [(2)]
If there was a call where the \prg{Account}'s password 
was used, then there must have been an intermediate program state
when some external object had access to the password.
\item [(3)]
Either that password was the same password as in the {starting} 
program state, or it was different:
\begin{description}
\item [(Case A)]
If it is the same as the initial password, then since by \prg{PasswordLeak}
it is impossible to leak the password, it follows that some external object 
must have had access to the password initially.
\item [(Case B)]
If the password is different from the initial password, 
then there must have been an  {intermediate} program state when it 
changed. By \prg{PasswordChange} we know that this must have occurred
by a call to \prg{changePassword} with the correct password. Thus,
there must be a some  {intermediate} program state where the initial
password is known. From here we proceed by the same reasoning 
as \textbf{(Case A)}.
\end{description}
\end{description}
\begin{proofexample}
\proofsteps{\SrobustB}
	{\proofstepwithrule{\onlyThroughExAlt
				{a:Account $\wedge$ b:Bank $\wedge$ b.balance(a)=bal}
				{b.balance(a) < bal}
				{$\calls{\_}{\prg{b}}{\prg{transfer}}{\prg{a.password}, \_, \prg{a}, \_}$}
				}
			{by \textsc{Changes} and \prg{BalanceChange}}}
	{\proofstepwithrule{\onlyThroughExAlt
				{a:Account $\wedge$ b:Bank $\wedge$ b.balance(a)=bal}
				{b.balance(a) < bal}
				{$\exists$ o.[$\external{\prg{o}}$ $\wedge$ $\access{\prg{o}}{\prg{a.password}}$]}
				}
			{by $\longrightarrow$, \textsc{Caller-Ext}, and \textsc{Calls-Args}}}
	{\proofstepwithrule{\onlyThroughExAlt
				{a:Account $\wedge$ b:Bank $\wedge$ b.balance(a)=bal $\wedge$ a.password=pwd}
				{b.balance(a) < bal}
				{$\neg$$\wrapped{\prg{a.password}}$}
				}
			{by $\longrightarrow$}}
	{\proofstepwithrule{\onlyThroughEx
				{a:Account $\wedge$ b:Bank $\wedge$ b.balance(a)=bal $\wedge$ a.password=pwd}
				{b.balance(a) < bal}
				{$\neg$$\wrapped{\prg{a.password}}$ $\wedge$ (a.password=pwd $\vee$ a.password != pwd)}
				}
			{by $\longrightarrow$ and \textsc{Excluded Middle}}}
	{\proofstepwithrule{\onlyThroughEx
				{a:Account $\wedge$ b:Bank $\wedge$ b.balance(a)=bal $\wedge$ a.password=pwd}
				{b.balance(a) < bal}
				{($\neg$$\wrapped{\prg{a.password}}$ $\wedge$ a.password=pwd) $\vee$\\
				($\neg$$\wrapped{\prg{a.password}}$ $\wedge$ a.password != pwd)}
				}
			{by $\longrightarrow$}}
	{\proofstepwithrule{\onlyThroughExAlt
				{a:Account $\wedge$ b:Bank $\wedge$ b.balance(a)=bal $\wedge$ a.password=pwd}
				{b.balance(a) < bal}
				{$\neg$$\wrapped{\prg{pwd}}$ $\vee$
				a.password != pwd}
				}
			{by $\longrightarrow$}}
	{
	\begin{proofexample}
	\proofsteps{Case A ($\neg\wrapped{\prg{pwd}}$)}
			{\proofstepwithrule
				{\onlyIfExAlt
					{a:Account $\wedge$ b:Bank $\wedge$ b.balance(a)=bal $\wedge$ a.password=pwd}
					{$\neg$$\wrapped{\prg{pwd}}$}
					{$\wrapped{\prg{pwd}}\ \vee \neg\wrapped{\prg{pwd}}$}
					}
				{by \textsc{If-}$\longrightarrow$ and \textsc{Excluded Middle}}}
			{\proofstepwithrule{\onlyIfExAlt
					{a:Account $\wedge$ b:Bank $\wedge$ b.balance(a)=bal $\wedge$ a.password=pwd}
					{$\neg$$\wrapped{\prg{pwd}}$}
					{$\neg\wrapped{\prg{pwd}}$}
					}
				{by $\vee$E and \prg{PasswordLeak}}}
	\endproofsteps
	\end{proofexample}
	}
	{
	\begin{proofexample}
	\proofsteps{Case B (\prg{a.password != pwd})}
		{\proofstepwithrule{\onlyThroughExAlt
					{a:Account $\wedge$ b:Bank $\wedge$ b.balance(a)=bal $\wedge$ a.password=pwd}
					{a.password != pwd}
					{$\calls{\_}{\prg{a}}{\prg{changePass}}{\prg{pwd}, \_}$}
					}
				{by \textsc{Changes} and \textsc{PasswordChange}}}
		{\proofstepwithrule{\onlyThroughExAlt
					{a:Account $\wedge$ b:Bank $\wedge$ b.balance(a)=bal $\wedge$ a.password=pwd}
					{a.password != pwd}
					{$\neg\wrapped{\prg{pwd}}$}
					}
				{by $\vee$E and \prg{PasswordLeak}}}
		{\proofstepwithrule{\onlyIfExAlt
					{a:Account $\wedge$ b:Bank $\wedge$ b.balance(a)=bal $\wedge$ a.password=pwd}
					{a.password != pwd}
					{$\neg\wrapped{\prg{pwd}}$}
					}
				{by \textbf{Case A} and \textsc{Trans}}}
	\endproofsteps
	\end{proofexample}
	}
	{\proofstepwithrule{\onlyIfExAlt
				{a:Account $\wedge$ b:Bank $\wedge$ b.balance(a)=bal $\wedge$ a.password=pwd}
				{b.balance(a) < bal}
				{$\neg\wrapped{\prg{pwd}}$}
				}
			{by \textbf{Case A}, \textbf{Case B}, \textsc{If-}$\vee$I$_2$, and \textsc{If-}$\longrightarrow$}}
\endproofsteps
\end{proofexample}

\clearpage
\section{Proof of Guarantee of Safety in \S\ref{sec:how}}
\label{app:safety}

In this section we provide a proof sketch that \SrobustB ensures our balance
does not decrease in contexts with no access to our password. This 
property is expressed in \S\ref{sec:how}, and the example is repeated below.

\begin{lstlisting}[mathescape=true, language=chainmail, frame=lines]
module $\ModParam{1}$
     ...
    method cautious(untrusted:Object)
        a = new Account
        p = new Password
        a.set(null,p)
        ...
        untrusted.make_payment(a)
        ...
\end{lstlisting}
\jm[]{
The guarantee for the above code snippet is that  as long as 
\prg{untrusted} does not have external access (whether transitive or direct)
to \sdN{\prg{a.pwd}} before the call on line 7, then \prg{a.balance} will not decrease during the 
execution of line 8. This property is expressed and proven in Theorem \ref{thm:safety}.
}

\begin{theorem}[\SrobustB Guarantees Account Safety]
\label{thm:safety}
Let \prg{BankMdl} be some module that satisfies \SrobustB, let 
$M$ be any external module, and $\sigma_1 = (\chi_1, \phi_1 : \psi_1)$ be some \textit{Arising} program state,
\sdN{$\arising{M}{\prg{BankMdl}}{\sigma_1}$}.
\\
If
\begin{enumerate}
\item
the continuation of $\sdN{\phi_1}$ is
\begin{verbatim}
    a = new Account; 
    p = new Password; 
    a.set(null,p); 
    s; 
    untrusted.make_payment(a, z1, ..., zn); ...
\end{verbatim}
\item
$\sigma_2 = (\chi_2, \phi_2 : \psi_2)$ is the program state immediately preceding the execution of \prg{s}
\item
$\sigma_3 = (\chi_3, \phi_3 : \psi_3)$ is the program state immediately following the execution of \prg{s}
\item
$\sigma_4 = (\chi_4, \phi_4 : \psi_4)$ is the program state immediately following the execution of\\ \prg{untrusted}\prg{.make\_payment}\prg{(a, z1, ..., zn)} 
\item
for all objects $o \in \chi_3$ which are transitively accessible (i.e. the transitive closure of $\access{\_}{\_}$) from \prg{untrusted}
\sdN{or from \prg{z1},...\prg{zn}}:\\
 $\strut \ \ \ \  \ \ \satisfies{\prg{BankMdl}; \sigma_3}{\access{o}{\sdN{\prg{a.pwd}}}}$, \ \ \
implies \ \ \  $\satisfies{\prg{BankMdl}; \sigma_3}{\internal{o}}$, 
\item
$\satisfies{\prg{BankMdl}; \sigma_3}{\prg{a.balance} = b}$
\end{enumerate}
then 
\begin{itemize}
\item
$\satisfies{\prg{BankMdl};\sigma_4}{\prg{a.balance} \geq b}$.
\end{itemize}
\end{theorem}

\vspace{.1in}

\noindent
{\bf{Proof Idea}}

\noindent
\sdS{We would like to apply \SrobustB in state $\sigma_3$, and argue that since by (5) no external object transitively accessible from \prg{a},  
\prg{untrusted}, \prg{z1}, ... \prg{zn} has access to the password, the balance in $\sigma_4$ will not decrease over what it was in $\sigma_3$.}
\sdS{However, the challenge is that the premise of \SrobustB is stronger than what we have in (5). Namely the premise of \SrobustB requires
that no external object has (direct) access to the password, but this
 requirement might not hold in  $\sigma_3$: depending on the contents of the code in \prg{s},} there may exist  external 
objects that have access to \prg{a.password}.  For example, if \prg{s} 
is the empty code, then $\sigma_1(\prg{this})$ has access to \prg{a}. 

\sdN{To address this challenge}, we will   create 
a program state, say \sdN{$\sigma_3'$}. In the new program state $\sigma_3'$ there will be no external access to \prg{a.password}. 
Also, $\sigma_3'$  must be similar enough to 
$\sigma_3$ so  that the execution of \prg{untrusted.make\_payment(a, z1, ..., zn)} \sdS{starting from state $\sigma_3$ is effectively
 equivalent to the execution of \prg{untrusted.make\_payment(a, z1, ..., zn)}  when starting from  $\sigma_3'$.}
 \sdN{Moreover,}  \sdN{$\sigma_3'$}, must also be \textit{Arising} for us to apply the \Nec specification 
\SrobustB to it.

 \sdN{This throws up a new challenge: $\sigma_3'$ is not necessarily \textit{Arising} in \prg{BankMdl} and $M$. We address the latter challenge by creating a new module, $M'$, such that  $\arising{M'}{\prg{BankMdl}}{\sigma_3'}$.}

\vspace{.1in}

\noindent
{\bf{Proof Sketch}}

\noindent
We construct $M'$ from $M$ by 1) modifying  all methods in all 
classes in $M$  so that all methods are duplicated: a) the original version, and b) 
a version almost identical  to that in $M$ with the addition that it keeps track of all the objects which contain fields pointing to
any objects of the \prg{Password} class, 
2) We add to all classes in $M$ a method called \prg{nullify}
that compares the contents of each of its fields with the method's argument, and if they are
equal overwrites the field with \prg{null}, 
3) all method calls  are replaced by those in part 1a, except of the body of \prg{make\_payment},
4) we modify the code in \prg{s} (and any methods called from it) so that it also keeps track of the current value of
\prg{a.pwd}, 
5) after \prg{s} and before the call \prg{untrusted.make\_payment(a, z1, ..., zn)} we insert  code which
runs through the list created in part 1, and calls \prg{nullify} with the current value of \prg{a.pwd} by \prg{null} as its argument.

By staring with the same initial configuration which reached $\sigma_3$,
 but now using $M'$ as the external module, we reach $\sigma_3'$, 
that is, $\arising{M'}{\prg{BankMdl}}{\sigma_3'}$. Moreover,  $\sigma_3'$ 
satisfies the premise of  \SrobustB. 
We execute $\prg{untrusted}\prg{.make\_payment}\prg{(a, z1, ..., zn)}$ in the context of   $\sigma_3'$ 
and reach $\sigma_4'$. We apply  \SrobustB, and obtain that $\satisfies{\prg{BankMdl};\sigma_4'}{\prg{a.balance} \geq b}$.

We use the latter fact, to conclude that $\satisfies{\prg{BankMdl};\sigma_4}{\prg{a.balance} \geq b}$.
Namely, $\sigma_3$ and $\sigma_3'$ are equivalent -- up to renaming of addresses  -- for all all the objects which are
reachable from \prg{o}, \prg{z1}, ... \prg{zn}, and for all objects from 
$\prg{BankMdl}$. Therefore, the execution of \prg{make\_payment} in $M;\prg{BankMdl}$ and $\sigma_3$
will be "equivalent" to that in $M';\prg{BankMdl}$ and $\sigma_3'$. Therefore, $\sigma_4$ and $\sigma_4'$ are equivalent -- up to renaming of addresses  -- for all all the objects which are
reachable from \prg{o}, \prg{z1}, ... \prg{zn} and for all objects from 
$\prg{BankMdl}$.
This gives us that  $\satisfies{\prg{BankMdl};\sigma_4}{\prg{a.balance} \geq b}$.

%
%
%


\end{document}